%% file: AmplitudeRecursions.tex
\documentclass[cntpempty]{ipart}

\startlocaldefs
\usepackage{hyphenat}
\usepackage[pdfencoding=auto,bookmarks=true,hyperfigures=true]{hyperref}%
\PassOptionsToPackage{unicode}{hyperref}

\usepackage{xparse}

\usepackage[usenames,dvipsnames,table]{xcolor}
\definecolor{mygreen}{rgb}{0,0.4,0}
\definecolor{myblue}{rgb}{0,0.0,0.4}
\definecolor{refrcolor}{rgb}{0,0.4,0}
\definecolor{cgreen}{rgb}{0,0.7,0}
\definecolor{ecolor}{rgb}{.52,.03,.06}
\definecolor{bgcolor}{rgb}{.96,.95,.80}
\definecolor{bgcolordark}{rgb}{.80,.80,.67}
\definecolor{faint}{rgb}{.80,.80,.80}

\newcommand{\cb}{\cellcolor{MidnightBlue!20}}
\usepackage[compile=false,fonts]{mpostinl}
\mpostsetup{template={AmpRec-#1.mps}}

\newcommand{\eqn}[1]{eq.~\eqref{#1}}
\newcommand{\Eqn}[1]{Equation~\eqref{#1}}
\newcommand{\eqns}[2]{eqs.~\eqref{#1} and~\eqref{#2}}

\newcommand{\rcite}[1]{ref.~\cite{#1}}
\newcommand{\rcites}[1]{refs.~\cite{#1}}

\theoremstyle{plain}
\newtheorem{theorem}{Theorem}
\newtheorem{proposition}[theorem]{Proposition}

\newtheorem{definition}[theorem]{Definition}

\makeatletter
\newcommand{\namedref}[2]{#1~\hyperref[#2]{\ref*{#2}}}
\newcommand{\secref}{\@ifstar{\namedref{Section}}{\namedref{section}}}
\newcommand{\subsecref}{\@ifstar{\namedref{Subsection}}{\namedref{subsection}}}
\newcommand{\appref}{\@ifstar{\namedref{Appendix}}{\namedref{appendix}}}
\newcommand{\tabref}{\@ifstar{\namedref{Table}}{\namedref{table}}}
\newcommand{\figref}{\@ifstar{\namedref{Figure}}{\namedref{figure}}}
\makeatother

\newcommand{\ap}{\alpha'}
\newcommand{\SL}{\mathrm{SL}}
\newcommand{\SI}[1]{\Sel[#1]}
\newcommand{\El}{\tau}
\newcommand{\pd}{\partial}
\newcommand{\nol}{N}
\newcommand{\snol}{L}
\newcommand{\wmax}{{w_\text{max}}}

\ExplSyntaxOn
\NewDocumentCommand{\Gtargz}{m m}
{
 \Gt\left(\begin{smallmatrix}
 \Gtargz_print:n {#1} \\
 \Gtargz_print:n {#2}
 \end{smallmatrix};z\right)
}
\seq_new:N \l_Gtargz_list_seq
\cs_new_protected:Npn \Gtargz_print:n #1
{
  \seq_set_split:Nnn \l_Gtargz_list_seq { , } { #1 }
  \seq_use:Nn \l_Gtargz_list_seq { , & }
}
\ExplSyntaxOff

\ExplSyntaxOn
\NewDocumentCommand{\SIE}{m m}
{
\SelE\!\Big[\begin{smallmatrix}
 \SI_print:n {#1} \\
 \SI_print:n {#2}
 \end{smallmatrix}\Big]
}
\seq_new:N \l_SI_list_seq
\cs_new_protected:Npn \SI_print:n #1
{
  \seq_set_split:Nnn \l_SI_list_seq { , } { #1 }
  \seq_use:Nn \l_SI_list_seq { , & }
}
\ExplSyntaxOff

\ExplSyntaxOn
\NewDocumentCommand{\Gtargzt}{m m}
{
 \Gt\left(\begin{smallmatrix}
 \Gtargzt_print:n {#1} \\
 \Gtargzt_print:n {#2}
 \end{smallmatrix};z,\tau\right)
}
\seq_new:N \l_Gtargzt_list_seq
\cs_new_protected:Npn \Gtargzt_print:n #1
{
  \seq_set_split:Nnn \l_Gtargzt_list_seq { , } { #1 }
  \seq_use:Nn \l_Gtargzt_list_seq { , & }
}
\ExplSyntaxOff

\makeatletter
\providecommand*{\shuffle}{%
  \mathbin{\mathpalette\shuffle@{}}%
}
\newcommand*{\shuffle@}[2]{%
  \sbox0{$#1\vcenter{}$}%
  \kern .15\ht0 
  \rlap{\vrule height .25\ht0 depth 0pt width 2.5\ht0}%
  \raise.1\ht0\hbox to 2.5\ht0{%
    \vrule height 1.75\ht0 depth -.1\ht0 width .17\ht0 %
    \hfill
    \vrule height 1.75\ht0 depth -.1\ht0 width .17\ht0 %
    \hfill
    \vrule height 1.75\ht0 depth -.1\ht0 width .17\ht0 %
  }%
  \kern .15\ht0 
}
\makeatother

\newcommand{\nnl}{\nonumber\\}

\newcommand{\colvec}[1]{\begin{pmatrix}#1\end{pmatrix}}

\DeclareMathOperator{\Gt}{\tilde{\Gamma}}
\DeclareMathOperator{\KN}{KN}
\DeclareMathOperator{\KNE}{KN^{\El}}
\DeclareMathOperator{\Sel}{S}
\DeclareMathOperator{\SelE}{S^{\El}}
\DeclareMathOperator{\Selbld}{\mathbf{S}}
\DeclareMathOperator{\SelbldE}{\mathbf{S}^{\El}}
\DeclareMathOperator{\SelbldEw}{\mathbf{S}^{\El}_{\mathit{w}}}
\DeclareMathOperator{\bC}{\mathbf{C}}

\DeclareMathOperator{\bF}{\boldsymbol{\mathrm{F}}}

\DeclareMathOperator{\bhF}{{\bf{\hat F}}}
\DeclareMathOperator{\LL}{L}
\DeclareMathOperator{\FF}{F}

\newcommand{\dz}{{dz}}

\newcommand{\ZC}{\mathbb C}

\newcommand{\ZN}{\mathbb N}

\newcommand{\ZR}{\mathbb R}
\newcommand{\ZZ}{\mathbb Z}

\newcommand{\CB}{\mathcal{B}}

\newcommand{\CF}{\mathcal{F}}       
\newcommand{\CG}{\mathcal{G}}

\newcommand{\CN}{\mathcal{N}}      
      
\newcommand{\CO}{\mathcal{O}}

\newcommand{\CV}{\mathcal{V}}

\def\tree{\text{tree}}
\def\oneloop{\text{1-loop}}
\def\mmin{\text{min}}
\def\mmax{\text{max}}

\def\YM{\textrm{YM}}

\def\open{\textrm{open}}

\def\reg{\text{reg}}

\endlocaldefs

\pubyear{\ }
\volume{\ }
\issue{\ }
\firstpage{0}
\lastpage{0}
\arxiv{1912.09927}

\begin{document}

\begin{frontmatter}

\title{Amplitude recursions with an extra marked point\protect\thanksref{T1}
}
\thankstext{T1}{AK would like to thank the IMPRS for Mathematical and Physical Aspects of Gravitation, Cosmology and Quantum Field Theory, of which he was a member and which rendered his studies possible.  Furthermore, AK is supported by the Swiss Studies Foundation, to which he would like to express his gratitude.}

\begin{aug}

\author{\fnms{Johannes} \snm{Broedel}\ead[label=e1]{jbroedel@ethz.ch}}
\address{ETH Zurich, Wolfgang-Pauli-Str.~27, 8093 Zurich, Switzerland\\\printead{e1}}
\and
\author{\fnms{Andre} \snm{Kaderli}\ead[label=e2]{andre.kaderli@gmail.com}}
\address{Humboldt-University Berlin, Unter den Linden 6, 10099 Berlin, Germany \\\printead{e2}}
\end{aug}

\begin{abstract}
The recursive calculation of Selberg integrals by Aomoto and Terasoma using the Knizh\-nik--Za\-molod\-chi\-kov equation and the Drinfeld associator makes use of an auxiliary point and facilitates the recursive evaluation of string amplitudes at genus zero: open-string $N$-point amplitudes can be obtained from those at $N{-}1$ points.\\ We establish a similar formalism at genus one, which allows the recursive calculation of genus-one Selberg integrals using an extra marked point in a differential equation of Knizhnik--Zamolodchikov--Bernard type. Hereby genus-one Selberg integrals are related to genus-zero Selberg integrals.  Accordingly, $N$-point open-string amplitudes at one loop can be obtained from $(N{+}2)$-point open-string amplitudes at tree level.   The construction is related to and in accordance with various recent results in intersection theory and string theory.
\end{abstract}

%


\end{frontmatter}


\input{AmplitudeRecursionsmp}

\section{Introduction}
\label{sec:introduction}

\subsection{Recursion for open-string amplitudes at genus zero}
Scattering amplitudes in open superstring theories at tree level are correlation functions of vertex operators inserted on the boundary of a disk. When evaluating those conformal correlators, the properties of the particular string theory in question can be straightforwardly incorporated in the so-called polarization part. What remains is the evaluation of the so-called configuration\hyp{}space integrals such as the four-point Veneziano amplitude \cite{Veneziano:1968yb}
\begin{equation}\label{eqn:ex4PtGenus0}
	\int_{0}^1 d x_3\, x_3^{s_{13}}(x_3-1)^{s_{23}} \frac{s_{13}}{x_3}=\frac{\Gamma(1+s_{13})\Gamma(1+s_{23})}{\Gamma(1+s_{13}+s_{23})}\,.
\end{equation}
The complex parameters 
\begin{equation}
	\label{eqn:Mandelstam}
	s_{i_1...i_r}=\ap(k_{i_1}+\ldots+k_{i_r})^2
\end{equation}
are Mandelstam variables built from the momenta $k_{i_p}$ of the external particles. Throughout this article, these variables are assumed to meet a sufficient condition for the convergence of the integrals to be considered, such as $\Re(s_{i_1...i_r})>0$ for consecutive insertion points $x_{i_1}{<}\cdots{<}x_{i_r}$ \cite{Mandelstam:1974fq,Brown:2019wna}.

In the case of $N$-point interactions and for appropriately fixed $s_{ij}$, the integrands of the configuration-space integrals are defined on the configuration space\footnote{This is the real moduli space $\mathcal{M}_{0,N}=\mathcal{M}_{0,N}(\ZR)=\CF_{N,3}$. Below, we will introduce more general configuration spaces $\CF_{L+1,k+1}$, which is why we rather use the notation $\CF_{N,3}$ than $\mathcal{M}_{0,N}(\ZR)$.} $\CF_{N,3}$ of $N{-}3$ insertion points $x_i$ on $\ZR\setminus\{0,1\}$: these are the unfixed insertion points on the real line, which parametrises the boundary of the disk, formed by the tree-level worldsheet, embedded into the Riemann sphere. The $\SL(2)$-symmetry of the Riemann sphere is used to fix three of the $N$ punctures at zero, one and infinity. The configuration-space integrals are obtained from iteratively integrating these integrands over the $N{-}3$ variables of $\CF_{N,3}$, i.e.\ the unfixed insertion points, on the unit interval. Finally, $\ap$ serves as counting parameter and will be identified with the inverse string tension, when considering actual string scattering amplitudes.

The $N$-point configuration\hyp{}space integrals in genus-zero open-string amplitudes are examples of Selberg integrals \cite{Selberg44}, which we denote and construct as follows: consider the $(L{+}1)$-punctured Riemann sphere with 
\begin{align}\label{eqn:fixedPunctures}
(x_1,x_2,x_{L+1})&=(0,1,\infty)
\end{align}
fixed by the $\SL(2)$-symmetry of the Riemann sphere. Writing
\begin{align}
x_{ij}&=x_{i,j}=x_i-x_j\,,
\end{align}
the corresponding integrals of Selberg type are iteratively defined by
\begin{align}
	\label{eqn:SelbergzeroIntro}
	\SI{i_{k+1},\dots,i_L}(x_1,\dots,x_k)&=\int_0^{x_k}\frac{dx_{k+1}}{x_{k+1,i_{k+1}}}\SI{i_{k+2},\dots,i_L}(x_1,\dots,x_{k+1})\,,
\end{align}
and the empty Selberg integral (or Selberg seed) is defined as\footnote{We use the notation $\prod_{x_a\leq x_i<x_j\leq x_b}=\prod_{i,j\in \{1,2,\dots,L\}:\,x_a\leq x_i<x_j\leq x_b}$. This notation ensures that all the differences $x_{ji}$ appearing in the product from \eqn{eqn:SelbergzeroSeedIntro} are positive and real. In agreement with the standard notation in string theory, we will include the absolute values in the definitions of the Koba--Nielsen factors and propagators, and usually write the Mandelstam variables $s_{ij}=s_{ji}$ as $s_{ij}$ with $i<j$.}
\begin{equation}
	\label{eqn:SelbergzeroSeedIntro}
	\SI{}(x_1,\dots,x_L)
	=\prod_{0\leq x_i<x_j\leq 1}x_{ji}^{s_{ij}}\,. 
\end{equation}
The definition \eqref{eqn:SelbergzeroIntro} presumes that the so-called admissibility condition
\begin{equation}\label{eqn:admissibilityIntro}
	1\leq i_{p}< p\qquad\forall p\in\lbrace k+1,\ldots, L\rbrace
\end{equation}
is met. The integral in \eqn{eqn:SelbergzeroIntro} is said to be of type $(k,L{+}1)$ and is, for fixed $s_{ij}$, a function on $\CF_{k+1,3}$.

Aomoto \cite{aomoto1987} and Terasoma \cite{Terasoma} showed that Selberg integrals of type $(2,L)$ can be obtained algebraically from those of type $(2,L{-}1)$: one starts from a basis vector $\Selbld(x_3)$ for Selberg integrals of type $(3,L{+}1)$, which contain an auxiliary point $x_3$ in contrast to the integrals of type $(2,L)$ and $(2,L{-}1)$, respectively. Taking the derivative with respect to $x_3$ leads to an equation of Knizhnik--Zamolodchikov (KZ) type \cite{Knizhnik:1984nr}
\begin{equation}
\label{eqn:KZexampleIntro}
\frac{d}{d x_3}\Selbld(x_3) = \Big(\frac{e_0}{x_3}+\frac{e_1}{x_3-1}\Big)\Selbld(x_3)\,,
\end{equation}
where the (braid) matrices $e_0$ and $e_1$ have entries which are homogeneous polynomials of degree one in the parameters $s_{ij}$.  The regularized boundary values 
\begin{align}\label{eqn:regLimitGenusZero}
\bC_0 = \lim_{x_3 \rightarrow 0} x^{-e_0}\Selbld(x_3) \ , \ \ \ \bC_1 = \lim_{x_3\rightarrow 1} (1-x_3)^{-e_1} \Selbld(x_3)\,.
\end{align}
of the differential equation \eqref{eqn:KZexample} can be shown to be related by the Drinfeld associator \cite{Drinfeld:1989st,Drinfeld2}
\begin{align}
\label{eqn:genusZeroAssociatorEqIntro}
\bC_1&=\mathbf{\Phi}(e_0,e_1)\, \bC_0.
\end{align}
What makes this construction useful for physicists is the fact that the $(N{-}1)$-point and the $N$-point configuration\hyp{}space integrals at genus zero can be identified (upon proper assignment of the Mandelstam variables) as linear combinations of the components of $\bC_0$ and $\bC_1$ respectively, where $N{=}L$. This relationship has been used to derive a recursive construction for all configuration\hyp{}space integrals on genus zero: it provides an analogue of the Parke--Taylor formula \cite{Parke:1986gb} for string theory \cite{Broedel:2013aza}. 

\subsection{Open-string scattering at genus one}
For a long time physicists have tried to find a similar recursive algorithm at genus one. In this article, we are going to establish such a construction.  One-loop open-string amplitudes are calculated on an annulus: again, there is a polarization part and configuration\hyp{}space integrals. For simplicity we are going to stick to those configuration\hyp{}space integrals where points are inserted on one boundary exclusively. Upon embedding the annulus into a torus, the relevant boundary is identified with the $A$-cycle and parametrised by the unit interval.  In the two-point case, the open-string one-loop configuration\hyp{}space integral is of the form
\begin{equation}\label{eqn:ex2PtGenus1}
   \int_0^{1}dz_2\, e^{s_{12}\Gt_{21}} g_{21}^{(0)}\,.
\end{equation}
The functions $g_{ij}^{(0)}$ and $\Gt_{ij}= \int^{z_{ij}} dz\,g^{(1)}(z,\tau) $ (cf.~\eqn{sec:eMPL:DefReg}) are defined by an infinite class of functions $g_{ij}^{(n)}=g^{(n)}(z_{ij},\tau)$ and integrals thereof, where $n$ is a non-negative integer, $z_1=0$ and $z_{ij}=z_i-z_j$ is the difference of insertion points on the $A$-cycle of the torus. These functions are suitable genus-one analogues of the fractions in $\frac{dx_{k+1}}{x_{k+1,i_{k+1}}}$ from \eqn{eqn:SelbergzeroIntro} and the genus-zero propagator $\log x_{ji}$ appearing in \eqn{eqn:SelbergzeroSeedIntro} in exponentiated form, respectively. They are defined by the Eisenstein--Kronecker series $F(z,\eta,\tau)$ \cite{Kronecker,BrownLev}
\begin{equation}
F(z,\eta,\tau) =
\frac{\theta_1'(0,\tau)\theta_1(z+\eta,\tau)}{\theta_1(z,\tau)\theta_1(\eta,\tau)} \ ,
\label{alt1}
\end{equation}
where $\theta_1$ is the odd Jacobi function and $'$ denotes a derivative with
respect to the first argument:  expanding in the second complex argument $\eta$, the function $g^{(n)}$ is the coefficient of $\eta^{n-1}$, i.e.\
\begin{equation}
\label{eqn:differentials}
\eta F(z,\eta,\tau) = \sum_{n=0}^{\infty}g^{(n)}(z,\tau)\eta^{n}\,.
\end{equation}
Various properties of these functions will be discussed thoroughly in \subsecref{subsec:eMPLeMZV}. 

Considering the similarity between genus-zero configuration\hyp{}space integrals such as the four-point example in \eqn{eqn:ex4PtGenus0} and genus-zero Selberg integrals \eqref{eqn:SelbergzeroIntro}, it is very natural to define a suitable genus-one analogue of Selberg integrals:
\begin{definition}
	Let $L\geq 2$, $0=z_1<z_L<...<z_2<1$ and $\tau$ the modular parameter of the torus $\ZC/(\ZZ+\tau\ZZ)$. Let the empty genus-one Selberg integral (or genus-one Selberg seed) be
	\begin{equation}
	\label{eqn:SelbergSeedEIntro}
	\SelE=\SIE{}{}(z_1,\dots,z_L)=\prod_{0=z_1\leq z_i< z_j\leq z_2}\exp\left(s_{ij}\Gt_{ji}\right)\,.
	\end{equation}
	Genus-one Selberg integrals are then defined recursively by
	\begin{align}
	\label{eqn:SelbergIntro}
	&\SIE{n_{k+1},\dots,n_L}{i_{k+1},\dots,i_L}(z_1,\dots,z_k)\nnl
	&\phantom{bbbb}=\int_0^{z_k}dz_{k+1}\, g^{(n_{k+1})}_{k+1,i_{k+1}}\SIE{n_{k+2},\dots,n_L}{i_{k+2},\dots,i_L}(z_1,\dots,z_{k+1})\,,
	\end{align}
	where $1\leq i_p<p$ for $k+1\leq p\leq L$ and $n_{k{+}1},\dots, n_L$ are non-negative integers.
\end{definition}
The successful concept to calculate open-string configuration-space integrals from Selberg integrals, which worked for genus zero, will be extended here: starting from genus-one Selberg integrals of type $(1,L{-}1)$, which contain the genus-one open-string configuration-space integrals, one can introduce an auxiliary point $z_2$ leading to genus-one Selberg integrals of type $(2,L)$. Given a class of type-$(2,L)$ Selberg integrals of fixed weight $w=\sum_{i=k+1}^L n_i$, one can find a vector $\Selbld^{\El}_w(z_2)$ of basis elements with respect to Fay identities (a genus-one generalization of partial fractioning) and integration by parts. Concatenating all those basis vectors into an infinitely long vector $\SelbldE(z_2)$, one has constructed the genus-one analogue of the (finite-length) vector $\Selbld(x_3)$ from above. This article is devoted to the construction of these integrals and to proving the following theorem:

\begin{theorem}{(Elliptic KZB-system)}
	Let $\SelbldE(z_2)$ be the vector of genus-one Selberg integrals of type $(2,L)$ with
	auxiliary point $z_2$. The derivative with respect to the auxiliary point $z_2$ can be written in the form 
	\begin{align}
	\label{eqn:KZBz2Intro}
	\frac{\pd}{\pd z_{2}} \SelbldE(z_2)
	&=\sum_{n\geq 0}g^{(n)}_{21}x^{(n)} \SelbldE(z_2)\,,
	\end{align}
	which is a system of elliptic KZB-type.
	The non-vanishing entries of the matrices $x^{(n)}$ are $\ZZ$-linear combinations of the parameters $s_{ij}$.
\end{theorem}

While showing the closure of the above system of differential equations is elaborate, two regularized boundary values can be easily associated to each other following the statements in the next proposition.
\begin{proposition}
	The regularized boundary values
	\begin{align}
		\bC_1^\El&=\lim_{z_2\rightarrow 1}(-2 \pi i (1{-}z_2))^{-x^{(1)}}\SelbldE(z_2)\text{ and }
	\bC_0^\El&=\lim_{z_2\rightarrow 0}(-2 \pi i z_2)^{-x^{(1)}}\SelbldE(z_2)
	\end{align}
	are related by the A-cycle component $\Phi(x^{(0)},x^{(1)},x^{(2)},...)$ of the KZB associator via
	\begin{equation}\label{eqn:assocEqIntro}
	\bC_1^\El=\Phi(x^{(0)},x^{(1)},x^{(2)},...)\bC_0^\El\,.
	\end{equation}
	The regularized boundary value $\bC_1^\El$ contains $(L{-}1)$-point configuration\hyp{}space integrals at genus one whereas $\bC_0^\El$ contains $(L{+}1)$-point configuration\hyp{}space integrals at genus zero.
\end{proposition}
Therefore, the $N$-point configuration\hyp{}space integrals appearing in open-string amplitudes at genus one can be calculated from the $(N{+}2)$-point integrals at genus zero via \eqn{eqn:assocEqIntro}, with $N{=}L{-}1$.
As examples we will consider the construction suitable for \mbox{two-,} three- and four-point configuration\hyp{}space integrals at genus one, which allow the determination of the (planar) \mbox{two-,} three- and four-point one-loop scattering amplitudes in open string theory.  
\subsection{Contents} In \secref{sec:genuszero} we are going to review the recursive evaluation of Selberg integrals at genus zero. We will apply the technique to genus-zero open-string amplitudes in a way equivalent to the approach in \rcite{Broedel:2013aza}. We are going to develop the genus-one formalism in \secref{sec:genusone}, prove the main theorem there and discuss the relation between genus-one objects and those at genus zero. Three examples are provided in \secref{sec:examples}. In \secref{sec:conclusion} we conclude and point out several open questions.
\subsection{Acknowledgements}
We are grateful to Rob Klabbers, Sebastian Mizera, Oliver Schlotterer and Federico Zerbini for various discussions on the content of this article and related subjects.  We would like to thank Carlos Mafra and Oliver Schlotterer for coordination of two similar projects in rather different languages.   

\section{Genus zero (tree level)}
\label{sec:genuszero}

In this section we are going to review the recursive evaluation of genus-zero Selberg integrals of Aomoto and Terasoma \cite{Aomoto,Terasoma} and relate it to the formalism for calculating open-string tree-level configuration\hyp{}space integrals put forward in \rcite{Broedel:2013aza}.  

\subsection{Singularities, iterated integrals and multiple zeta values}

Con\-fi\-gu\-ration-space integrals for open-string tree-level amplitudes are defined on the boundary of a disk,
on which the integration parameters $x_i$ are located. Integrating over all configurations of the parameters while keeping their ordering along the boundary intact leads to iterated integrals.  

Whithin the context of open-string tree-level configuration\hyp{}space integrals, all integrations can either be performed trivially or can be traced back to iterated integrals of the following differential form on the Riemann sphere with a simple pole at the fixed insertion points \mbox{$a_j\in \{x_1,x_2\}=\{0,1\}$:}
\begin{equation}
	\label{eqn:abeliangenus0}
	\frac{dx_i}{x_i-a_j}\,.
\end{equation}
Accordingly, the canonical form of iterated integrals appearing in the $\alpha'$-expansion of genus-zero configuration\hyp{}space integrals are multiple polylogarithms
\begin{equation}
	\label{eqn:GPolylog}
	G(a_1,a_2,\ldots,a_r;x)=\int_0^xdx'\,\frac{1}{x'-a_1}G(a_2,\ldots,a_r;x'),\quad G(;x)=1\,,
\end{equation}
with $a_i\in\lbrace0,1\rbrace$ and $a_r\neq 0$. Below, it will be useful to write this subclass of Goncharov polylogarithms\cite{GONCHAROV1995197,Goncharov:2001iea} indexed by words of the form
\begin{equation}
w=e_0^{n_r-1}e_1\dots e_0^{n_1-1}e_1\,,
\end{equation}
where $n_i\geq 1$:
\begin{equation}\label{eqn:MPLdef}
G_w(x)=G(\underbrace{0,\dots,0}_{n_r -1},1,\dots,\underbrace{0,\dots,0}_{n_1 -1},1;x)\,.
\end{equation}
Evaluating the above multiple polylogarithms with $n_r>1$ at $x=1$ leads to multiple zeta values
\begin{equation}
 \label{eqn:mzv}
 \zeta_w=(-1)^r G_w(1)=\sum_{1\leq k_1<\dots<k_r}\frac{1}{k_1^{n_1}\dots k_r^{n_r}}\,,
\end{equation}
which are the transcendental numbers appearing in the $\alpha'$-expansion of open-string configuration\hyp{}space integrals at genus zero.

In the above definitions of multiple polylogarithms and multiple zeta values divergent situations have been excluded. However, using the definitions
\begin{align}\label{eqn:regGenusZero}
G_{e_0}(x)&=G(0;x)=\log(x)\nonumber\\
\zeta_{e_1}&=\zeta_{e_0}=0
\end{align}
and shuffle relations between iterated integrals extended to all multiple polylogarithms and multiple zeta values
\begin{equation}
G_{w'} (x) G_{w''}(x)=G_{w'\shuffle \, w''}(x),\,\zeta_{w'} \zeta_{w''}=\zeta_{w'\shuffle \, w''}\quad w',w''\in\lbrace e_0,e_1\rbrace^\times\,,
\end{equation}
one can extend the definition to all words from $\lbrace e_0,e_1\rbrace^\times$. This regularization scheme is referred to as \textit{shuffle regularization} (or \textit{tangential basepoint regularization}), see e.g.\ \rcites{Deligne89,Brown:ICM14}.

\subsection{Selberg Integrals}
\label{subsec:SelbergIntegrals}
In comparison to the iterated integrals defined in \eqn{eqn:GPolylog} above, genus-zero configuration\hyp{}space integrals have one more ingredient: the empty integral to be iteratively integrated with integration kernels $1/(x{-}a_i)$ is not one, but the so-called Koba--Nielsen factor. It contains the open-string tree-level Green function, also known as \textit{propagator}, weighted by Mandelstam variables $s_{ij}$ defined in \eqn{eqn:Mandelstam}. 
The Green function is the integral over the differential form from \eqn{eqn:abeliangenus0}: 
$\log |x_{ij}| = G(0,|x_{ij}|) = \int_1^{|x_{ij}|}\frac{dx}{x}$, where $x_{ij}=x_i-x_j$ is the difference of two insertion points. 

The class of integrals accommodating the above features are \textit{Selberg integrals} \cite{Selberg44,aomoto1987,Terasoma}. Consider $L$ points on the unit interval with the ordering
\begin{equation}\label{eqn:genus0ordering}
0=x_1<x_L<x_{L-1}<\dots<x_3<x_2=1
\end{equation}
and define the empty
Selberg integral or \textit{Selberg seed}
\begin{equation}
	\label{eqn:SelbergzeroSeed}
	\Sel=\SI{}(x_1,\dots,x_L)=\prod_{0\leq x_i<x_j\leq 1}\exp\left(s_{ij}\log x_{ji}\right)=\prod_{0\leq x_i<x_j\leq 1}x_{ji}^{s_{ij}}\,,
\end{equation}
with\footnote{The empty integral $\SI{}{}$ is called the \textit{Koba--Nielsen factor} if the parameters $s_{ij}$ are identified with the momenta of the corresponding external states of a scattering amplitude. Note that according to \eqn{eqn:genus0ordering} all the differences $x_{ji}$ in definition \eqref{eqn:SelbergzeroSeed} are positive and real. This does apply to the genus-zero propagator mentioned above as well.} complex parameters $s_{ij}$. The Selberg seed is integrated over various integration kernels of the form $1/x_{ij}$ which lead to functions denoted by
\begin{align}
	\label{eqn:Selbergzero}
	\SI{i_{k+1},\dots,i_L}(x_1,\dots,x_k)&=\int_0^{x_k}\frac{dx_{k+1}}{x_{k+1,i_{k+1}}}\SI{i_{k+2},\dots,i_L}(x_1,\dots,x_{k+1})\,,
\end{align}
where 
\begin{equation}\label{eqn:admissibility}
	1\leq i_{p}< p\qquad\forall p\in\lbrace k+1,\ldots, L\rbrace\,.
\end{equation}
The above \textit{admissibility condition} motivates the definition of \textit{admissible} iterated integrals: the integration kernel $1/x_{k+1,i_{k+1}}$ in \eqn{eqn:Selbergzero} can not depend on variables which have already been integrated out. In accordance with \rcite{aomoto1987}, this property is called \textit{admissibility} and an integral with an integrand of the form $S \prod_k 1/x_{k,i_k}$ satisfying \eqn{eqn:admissibility} \textit{admissible}.  As argued in \subsecref{subsec:genus0boundaryValues} and \subsecref{subsec:Relate}, Selberg integrals of length $L-3$
\begin{align}
\label{eqn:relevantSelberg}
\SI{i_{4},\dots,i_L}(x_1,x_2,x_3)&=\int_0^{x_3}\frac{dx_{4}}{x_{4,i_{4}}}\SI{i_{5},\dots,i_L}(x_1,\dots,x_{4})\nnl
&= \int_{\mathcal{C}(x_3)}\prod_{i=4}^L dx_i\, \Sel \prod_{k=4}^L \frac{1}{x_{k,i_k}}\,,
\end{align}
where $\mathcal{C}(x_3)$ is the region of integration denoted by
\begin{equation}
\label{eqn:integrationRegion}
\mathcal{C}(x_i)=\{0=x_1<x_L<x_{L-1}<\dots<x_i\}
\end{equation}
for $x_i\leq x_2=1$, include in the limit of merging punctures $x_3\to 1=x_2$ all integrals appearing in the calculation of $L$-point open-string tree-level scattering amplitudes. 

For appropriately fixed $s_{ij}$ and fixed unintegrated insertion points $x_{1}=0,x_2=1,x_3,\dots,x_{k}$ and $x_{L+1}=\infty$, the integrands in the Selberg integrals $ \SI{i_{k+1},\dots,i_L}(x_1,\dots,x_k)$ are functions defined on the configuration space of the $L-k$ insertion points $x_{k+1},\dots,x_{L}$ on the $k$-punctured real line \mbox{$\ZR\setminus \{x_{1},\dots,x_{k}\}$:} 
\begin{align}
&\mathcal{F}_{\snol+1,k+1}=\nnl
&\quad\{(x_{k+1},x_{k+2},\dots,x_{\snol})\in (\ZR\setminus \{x_{1},\dots,x_{k}\})^{\snol-k}|\forall i\neq j: x_i\neq x_j\}\,.
\end{align}
The differential forms
\begin{equation}\label{eqn:form}
\bigwedge_{p=k+1}^L \frac{dx_p}{x_{p,i_p}}\,,
\end{equation}
where $1\leq i_{p}< p$, appearing in the integrands of the Selberg integrals in \eqn{eqn:Selbergzero}, represent elements of a basis of the twisted de Rham cohomology of the ($L{+}1$)-punctured Riemann sphere with $k{+}1$ fixed coordinates: their twisted cohomology classes defined by the connection $d+ d\log S$ pulled back to $\mathcal{F}_{\snol+1,k+1}$ appear in such a basis \cite{aomoto1987}, see also \rcite{Mizera:2019gea}. 

The Selberg integrals $ \SI{i_{k+1},\dots,i_L}(x_1,\dots,x_k)$ in turn, depend on the unintegrated variables $x_1=0,x_2=1, x_3,\dots,x_k$ and $x_{L+1}=\infty$ with \mbox{$x_i\neq x_j$}, and are therefore functions defined on the configuration space $\mathcal{F}_{k+1,3}$. In particular, the Selberg seed from \eqn{eqn:SelbergzeroSeed} is defined on $\mathcal{F}_{L+1,3}$. A configuration of the form \eqref{eqn:genus0ordering} in $\mathcal{F}_{L+1,3}$ can be depicted on the real line plus infinity embedded into a circle on the Riemann sphere as follows:
\begin{equation}\label{fig:config}
	\phantom{mmmm}\mpostuse{genuszerogeneral}\,.
\end{equation}

\subsection{KZ equation for an auxiliary point}

The open-string configuration\hyp{}space integrals at genus zero are recovered from the integrals
\begin{align}
\label{eqn:interestingSelbergzero}
\SI{i_{4},\dots,i_\snol}(x_1=0,x_2=1,x_3)
\end{align}
defined in \eqn{eqn:relevantSelberg} in the following two regularized limits:
\begin{itemize}
	\item in the limit $x_3\to x_2=1$, it is merged with the point $x_2$ and one fixed puncture is removed. The integrands of the Selberg integrals defined on $\mathcal{F}_{L+1,4}$ degenerate to integrands on $\mathcal{F}_{L,3}$ of the integrals $ \SI{i_{4},\dots,i_L}(x_1{=}0,x_2{=}1,x_3{=}x_2)$. The space $\mathcal{F}_{L,3}$ is the configuration space known from open-string calculations with three fixed coordinates on which $L$-point tree-level amplitudes are defined. Indeed, as shown below, they will be recovered in this limit from the Selberg integrals.
	\item the merging of $x_3\to x_1=0$ is slightly more involved and will lead to the $(L{-}1)$-point integrals in a certain soft limit, which leads to an additional degeneration of the integrands to functions defined on the configuration space $\mathcal{F}_{L-1,3}$ relevant for $L{-}1$-point tree-level amplitudes.  
\end{itemize}
Thus, for $x_3\in (0,1)$ the puncture interpolates between the $L$- and $(L{-}1)$-point open-string configuration\hyp{}space integrals. These two boundary values can be related by a recursive procedure involving matrix operations \cite{aomoto1987,Terasoma}, which leads to the genus-zero string recursion in \rcite{Broedel:2013aza}. The main idea hereby is the use of $x_3$ as an \textit{auxiliary} insertion point, such that differentiating with respect to $x_3$ leads to a KZ equation \eqref{eqn:KZexampleIntro}, whose regularised boundary values in the above limits can be related via the Drinfeld associator according to \eqn{eqn:genusZeroAssociatorEqIntro}.
In the remainder of this subsection, we will review differential equations for the Selberg integrals, while limits/boundary values of the differential equation are discussed in \subsecref{subsec:genus0boundaryValues} below. Attached to the point $x_3$ there is an auxiliary external momentum $k_3$ with associated Mandelstam variables $s_{3i}$, $i\in\lbrace 1,2,4,5,\ldots\snol\rbrace$. For the moment we are not imposing any conditions like the momentum conservation and consider the variables $s_{ij}=s_{ji}$ as independent parameters whose interpretation as Mandelstam variables in a scattering amplitude context will become clear when considering the limits $x_3\to0$ and $x_3\to1$ below. 

Therefore, let us explore differential equations with respect to the
auxiliary point $x_3$ acting on the Selberg integrals
\eqref{eqn:interestingSelbergzero}:
\begin{equation}
	\label{eqn:Selbergderivative}
	\frac{d}{d x_3}\SI{i_4,i_5,\ldots,i_\snol}(0,1,x_3)=\frac{d}{d x_3} \int_{\mathcal{C}(x_3)}\prod_{i=4}^\snol dx_i\, \Sel \prod_{k=4}^\snol \frac{1}{x_{k,i_k}}\,.
\end{equation}
Noting that the Selberg seed \eqref{eqn:SelbergzeroSeed} vanishes for $x_i=x_j$ and $\Re(s_{ij})>0$
\begin{equation}
\label{eqn:genus0Svanishing}
\Sel\!|_{x_i=x_j}=0\,,
\end{equation}
it follows that the derivative in \eqn{eqn:Selbergderivative} only acts non-trivially on the integrand and not on the integration domain. The identity
\begin{equation}
\frac{\partial}{\partial x_i}\frac{1}{x_{ij}} = -\frac{\partial}{\partial x_j}\frac{1}{x_{ij}}
\end{equation}
and integration by parts may be used to let partial derivatives act on the
Selberg seed only:
\begin{equation}
	\label{eqn:SelbergderivativeIbP}
	\frac{d}{d x_3}\SI{i_4,i_5,\ldots,i_\snol}(0,1,x_3)= \int_{\mathcal{C}(x_3)}\prod_{i=4}^\snol dx_i\, \left(\sum_{j\in U_3}\frac{\partial}{\partial x_j}\Sel\right) \prod_{k=4}^\snol \frac{1}{x_{k i_k}}\,.
\end{equation}
The set $U_3$ in the previous equation can be stated explicitly: 
\begin{align}
\label{eqn:U3}
U_3&=\Bigg\lbrace j\in\{3,4,\dots,L\}\,\Big|\,j=3\text{ or there exist labels } 3=j_1,j_2,\dots,j_m=j \nnl
   &\phantom{bbbbbbbbbbbb}\text{ such that } \prod_{i=1}^{m-1}\frac{1}{x_{j_{i+1},j_i}} \text{ is a factor of } \prod_{k=4}^\snol \frac{1}{x_{k i_k}}\Bigg\rbrace\,.
\end{align}
Partial derivatives of the Selberg seed yield factors of $s_{jl}/x_{jl}$
\begin{equation}
	\label{eqn:partialSeed}
\frac{\partial}{\partial x_j}\Sel=\sum_{l\neq j} \frac{s_{jl}}{x_{jl}}\Sel\,,
\end{equation}
such that
\begin{equation}
\label{eqn:SelbergderivativeNoDerivative}
\frac{d}{d x_3}\SI{i_4,i_5,\ldots,i_\snol}(0,1,x_3)= \int_{\mathcal{C}(x_3)}\prod_{i=4}^\snol dx_i\, \Sel \sum_{j\in U_3}\sum_{l\not\in U_3}\frac{s_{jl}}{x_{jl}} \prod_{k=4}^\snol \frac{1}{x_{k i_k}}\,.
\end{equation}
Admissibility of $\SI{i_4,i_5,\ldots,i_\snol}(0,1,x_3)$ implies that upon consecutive applications of partial fractioning
\begin{equation}
\label{eqn:partialFractioning}
\frac{1}{x_{k,l}}\frac{1}{x_{k,m}}=\left(\frac{1}{x_{k,l}}-\frac{1}{x_{k,m}}\right)\frac{1}{x_{l,m}}\,,
\end{equation}
where $k>l>m$, we will again find (admissible) Selberg integrals,
however, with different labels $i_k$.

All integrals on the right-hand side of \eqn{eqn:Selbergderivative} will contain a prefactor of the form
\begin{equation}
\frac{s_{ij}}{x_{31}}=\frac{s_{ij}}{x_{3}}\quad \text{or}\quad \frac{s_{ij}}{x_{32}}=\frac{s_{ij}}{x_{3}-1}\,,
\end{equation}
since the indices in $x_{31}$ and $x_{32}$ can no longer be reduced by partial fractioning. Accordingly, if we consider the vector of all admissible integrals 
\begin{equation}
\label{eqn:defSeL}
\Selbld(x_3)=\colvec{\SI{i_4,i_5,\ldots,i_\snol}(0,1,x_3)}_{1\leq i_k<k},
\end{equation}
its derivative with respect to $x_3$ can be phrased in terms of a vector equation
\begin{equation}
\label{eqn:KZexample}
\frac{d}{d x_3}\Selbld(x_3) = \Big(\frac{e_0}{x_3}+\frac{e_1}{x_3-1}\Big)\Selbld(x_3)\,,
\end{equation}
where the entries of the $(\snol{-}1)!/2\times (\snol{-}1)!/2$ matrices $e_0$ and $e_1$ either vanish or are $\ZZ$-linear combinations of the parameters $s_{ij}$.  In an amplitude context later, this implies (cf.~\eqn{eqn:Mandelstam}) that $e_0$ and $e_1$ are proportional to $\ap$.

The fact that the derivative of $ \SI{i_4,i_5,\ldots,i_\snol}(0,1,x_3)$ is expressible as a linear combination of iterated integrals $\SI{i_4,i_5,\ldots,i_\snol}(0,1,x_3)$ 
originates in the property mentioned below \eqn{eqn:form} of the differential forms appearing in the integrand in \eqn{eqn:Selbergderivative}: they are the building blocks for the so-called fibration basis of the twisted de Rham cohomology of the $(L{+}1)$-punctured Riemann sphere with four fixed coordinates \cite{aomoto1987,Mizera:2019gea}. Note that for each $4\leq k\leq L$, one can get rid of one particular index $1\leq i'_k<k$ by partial fractioning and integration by parts. Thus, one can identify a suitable basis of the iterated integrals  $\SI{i_4,i_5,\ldots,i_\snol}(0,1,x_3)$ as 
\begin{equation}
	\label{eqn:basisGenusZero}
	\mathcal{B}_{i'_4,i'_5,\dots,i'_\snol}=\lbrace  \SI{i_4,i_5,\ldots,i_\snol}(0,1,x_3)|1\leq i_k<k, i_k\neq i'_k \rbrace\,,
\end{equation}
that is, the ticked indices do \textit{not} appear as labels. Accordingly, one can reduce the vector of all admissible integrals to 
\begin{equation}
\Selbld(x_3)|_{\mathcal{B}_{i'_4,i'_5,\dots,i'_\snol}}
\end{equation}
for which one finds the differential equation
\begin{equation}
\label{eqn:KZexampleBasis}
	\frac{d}{d x_3}\Selbld(x_3)|_{\mathcal{B}_{i'_4,i'_5,\dots,i'_\snol}} = \Big(\frac{e_0}{x_3}+\frac{e_1}{x_3-1}\Big)\Selbld(x_3)|_{\mathcal{B}_{i'_4,i'_5,\dots,i'_\snol}}\,.
\end{equation}
While the matrices $e_0$ and $e_1$ contain entries from the same class as for the matrices in \eqn{eqn:KZexample}, they are now of dimension $(\snol-2)!\times (\snol-2)!$.  These matrices turn out to be braid matrices, that is, representations of the braid group of $\snol{+}1$ distinguishable strands with three strands held fixed. It is well known how to obtain these matrices recursively \cite{aomoto1987,Terasoma,Mizera:2019gea}.

Of course, the choice of the basis is a priori arbitrary. However, depending on the intended use, certain choices turn out to be much more beneficial than others in practice. For example, the recursive definition of the matrices in $e_0$ and $e_1$ in \rcite{Mizera:2019gea} are constructed for the choice $\mathcal{B}_{1,1,\dots,1}$, i.e.\ $2\leq i_k<k$. On the other hand, the limits considered in \subsecref{subsec:genus0boundaryValues} will conveniently be formulated in the basis $\mathcal{B}_{2,2,\dots,2}$.\\[4pt]
\noindent\textbf{Example.} Let us consider the simplest example $\snol=4$ and show the above calculational steps explicitly for the basis $\mathcal{B}_2=\{\Sel[1](0,1,x_3),\Sel[3](0,1,x_3)\, \}$,
where 
\begin{equation}
\Sel[i_4](0,1,x_3)=\int_0^{x_3}dx_4\, \Sel \frac{1}{x_{4,i_4}}\,,\qquad \Sel=x_{41}^{s_{14}}x_{31}^{s_{13}}x_{34}^{s_{34}}x_{21}^{s_{12}}x_{24}^{s_{24}}x_{23}^{s_{23}}\,.
\end{equation}
The integrands are functions defined on $\mathcal{F}_{5,4}=\{x_4\in \ZR| x_4\neq x_1,x_2,x_3,x_5\}$ and we consider the following order of the punctures:
\begin{equation}
0=x_1<x_4<x_3<x_2=1<x_5=\infty\,.
\end{equation}
The set $\mathcal{B}_2$ is indeed a basis, since the only remaining Selberg integral $\Sel[2](0,1,x_3)$ can be expressed in terms of elements in $\mathcal{B}_2$ using
\begin{equation}
\label{eqn:ibpBasis}
s_{14}\Sel[1](0,1,x_3)+s_{24}\Sel[2](0,1,x_3)+s_{34}\Sel[3](0,1,x_3)=0\,.
\end{equation}
Now, let us calculate the derivatives of $\Selbld(x_3)|_{\mathcal{B}_2}$ in order to recover the KZ equation \eqref{eqn:KZexampleBasis}: for the first basis element, we find $U_3(\SI{1}(0,1,x_3))=\{3\}$, such that according to \eqn{eqn:SelbergderivativeNoDerivative}
\begin{align}
\frac{d}{d x_3}\SI{1}(0,1,x_3)&= \int_{0}^{x_3} dx_4\, \Sel \left(\frac{s_{13}}{x_{31}}+\frac{s_{34}}{x_{34}}+\frac{s_{23}}{x_{32}} \right)\frac{1}{x_{4 1}}\nnl
&=\frac{s_{13}}{x_3}\SI{1}(0,1,x_3)+\frac{s_{23}}{x_3-1}\SI{1}(0,1,x_3)\nnl
&\quad+\frac{s_{34}}{x_{3}}\left(\SI{1}(0,1,x_3)-\SI{3}(0,1,x_3)\right)\,
\end{align}
where we have used the partial fractioning identity \eqref{eqn:partialFractioning} for the third equality. Similarly, for the second basis element we find
$U_3(\SI{3}(0,1,x_3))=\{3,4\}$, such that
\begin{align}
\frac{d}{d x_3}\SI{3}(0,1,x_3)&= \int_{0}^{x_3} dx_4\, \Sel \left(\frac{s_{13}}{x_{31}}+\frac{s_{23}}{x_{32}}+\frac{s_{14}}{x_{41}}+\frac{s_{24}}{x_{42}} \right)\frac{1}{x_{4 3}}\nnl
&=\frac{s_{13}}{x_3}\SI{3}(0,1,x_3)+\frac{s_{23}}{x_3-1}\SI{3}(0,1,x_3)\nnl
&\quad+\frac{s_{14}}{x_3}\left(\SI{3}(0,1,x_3)-\SI{1}(0,1,x_3)\right)\nnl
&\phantom{=}
+\frac{1}{x_3-1}\left((s_{24}+s_{34})\SI{3}(0,1,x_3)+s_{14}\SI{1}(0,1,x_3)\right)\,,
\end{align}
where we have again used partial fractioning \eqref{eqn:partialFractioning} and
integration by parts \eqref{eqn:ibpBasis} for the fourth equality. 
Overall we find the differential equation
\begin{equation}
\frac{d}{d x_3} \Selbld(x_3)|_{\mathcal{B}_2}=\left(\frac{e_0}{x_3}
+\frac{e_1}{x_3-1}
\right)
\Selbld(x_3)|_{\mathcal{B}_2}\,,
\end{equation}
which is indeed of the form of the KZ equation \eqref{eqn:KZexampleBasis} with
the matrices
\begin{equation}
\label{eqn:Genus0e1Example}
e_0=\begin{pmatrix}
s_{13}+s_{34}&-s_{34}\\
-s_{14}&s_{13}+s_{14}
\end{pmatrix}\,,\qquad e_1=\begin{pmatrix}
s_{23}&0\\
s_{14}&s_{234}
\end{pmatrix}
\end{equation}
given by the braid matrices from \rcite{Mizera:2019gea}.

\subsection{Boundary values for the KZ equation}\label{subsec:genus0boundaryValues}

\Eqn{eqn:KZexample} is of KZ type \cite{Knizhnik:1984nr}.  Well known from \rcites{Drinfeld:1989st,Drinfeld2}, we provide a brief summary of the relation between the two boundary values $z_3\to 0,1$ in \appref{app:KZassociator}.

Let us consider the regularized limits \eqref{eqn:regLimitGenusZero} when taking the auxiliary point $x_3$ to either zero or one in \eqref{eqn:KZexample}\footnote{The following paragraph is closely related to the original analysis of Selberg integrals in \rcite{Terasoma}, which serves as the prime reference for our investigation and led to the formulation of the amplitude recursion in \rcite{Broedel:2013aza}.}:
\begin{align}
\bC_0 = \lim_{x_3 \rightarrow 0} x_3^{-e_0}\Selbld(x_3) \ , \ \ \ \bC_1 = \lim_{x_3\rightarrow 1} (1-x_3)^{-e_1} \Selbld(x_3)\,.
\end{align}
\textbf{Boundary value $\bC^\El_1$:} Let us start by considering the limit
$x_3\to x_2=1$, which is depicted in the following figure:
\begin{equation}
  \label{fig:disk0}
  \phantom{mmmm}\mpostuse{genuszerox3to1}\,.
\end{equation}
The relevant integrals in the amplitude recursion in this limit turn out to be the Selberg integrals in $\mathcal{B}_{2,2,\dots,2}\cap \mathcal{B}_{3,3,\dots,3}$ with integrands defined on the configuration space $\mathcal{F}_{\snol+1,4}$ with $1\leq i_k<k$ and $i_k\neq 2,3$. For these integrals, the action of the prefactor $(1-x_3)^{-e_1}$ is particularly simple: on the one hand, the set $U_3$ in \eqn{eqn:SelbergderivativeNoDerivative} is simply $U_3=\{3\}$.  On the other hand, the only appearance of the insertion point $x_2$ in the integral $\Sel[i_4,i_5,\dots,i_L](x_1=0,x_2=1,x_3)$ with $i_k\neq 2$ is in the Selberg seed. Therefore using partial fractioning to obtain the KZ form from \eqn{eqn:SelbergderivativeNoDerivative} does not introduce any factor of $1/x_{32}$ other than $s_{23}/x_{32}$ obtained from differentiating the Selberg seed. Thus, for the basis $ \mathcal{B}_{2,2,\dots,2}$, the representation $e_1$ in the KZ equation \eqref{eqn:KZexampleBasis} is of the form
\begin{equation}\label{eqn:e1B222}
e_1=\,\begin{pmatrix}s_{23}\mathbb{I}_{(L-3)!\times (L-3)!}&0_{ (L-3)\times (L-3)!}\\
A_{(L-3)!\times (L-3)}& B_{(L-3)\times (L-3)}\end{pmatrix}\,,
\end{equation}
where the upper left block proportional to the identity corresponds to the integrals in $ \mathcal{B}_{2,2,\dots,2}\cap \mathcal{B}_{3,3,\dots,3}$, (cf.~example \eqref{eqn:Genus0e1Example}).  For this subclass of integrals, the regularization factor $(1-x_3)^{-e_1}$ only contributes with the scalar $(1-x_3)^{-s_{23}}=x_{23}^{-s_{23}}$ and the corresponding entries of the regularized limit $\bC_1$ can be calculated as
\begin{align}
	\label{eqn:C1limitSel}
	&\lim_{x_3\rightarrow x_2} x_{23}^{-s_{23}} \SI{i_4,i_5,\ldots,i_\snol}(0,1,x_3)\nnl
	&\qquad=\int_{\mathcal{C}(x_3\to x_2)}\prod_{i=4}^L dx_i\, \prod_{0\leq x_j <x_l<x_3}x_{lj}^{s_{jl}}\prod_{0\leq x_n<x_3}x_{2n}^{s_{2n}+s_{3n}} \prod_{k=4}^L \frac{1}{x_{k i_k}}\nnl
	&\qquad=\SI{i_4,i_5,\ldots,i_\snol}(0,1,x_3=x_2)|_{s_{23}=0}^{\tilde{s}_{ij}=s_{ij}+\delta_{i2}s_{3j}}\,,
\end{align}
Thus the regularization $x_{23}^{-s_{23}}$ cancels the factor $x_{23}^{s_{23}}$ in the Selberg seed $\Sel$, which would otherwise render the integral vanishing. Moreover, the punctures $x_2$ and $x_3$ have merged, such that the associated Mandelstam variables, and hence, the momenta of the external states, are added to yield effective Mandelstam variables $\tilde{s}_{ij}=s_{ij}+\delta_{i2}s_{3j}$ for $i,j\in \{1,\dots,L\}\setminus \{3\}$, $i<j$.

The resulting differential form and integration domain in the integral \eqref{eqn:C1limitSel} are known from the type-$(L,3)$ Selberg integrals defined on the configuration space $\mathcal{F}_{\snol,3}$. Thus, in this limit, the forms in $\Selbld(x_3)$ can be expressed as linear combinations of the Parke--Taylor forms of $L$-point string amplitudes, which are discussed in the next subsection. In terms of the disk picture \ref{fig:disk0}, we are modifying the relative distances on the boundary by taking the limit $x_3\to x_2=1$. Upon identification of the points $x_2$ and $x_3$ we find the transition
\begin{equation}
\mathcal{F}_{\snol+1,4}\to \mathcal{F}_{\snol,3}
\end{equation}
with the $L$ insertion points $x_1=0,x_2=1,x_4,x_5,\ldots,x_L$ and $x_{L+1}=\infty$, which is the setup suitable for describing $L$-point amplitudes.\\[4pt]

\noindent\textbf{Boundary value $\bC^\El_0$:} For the limit $x_3\to 0$, we are facing
the following situation
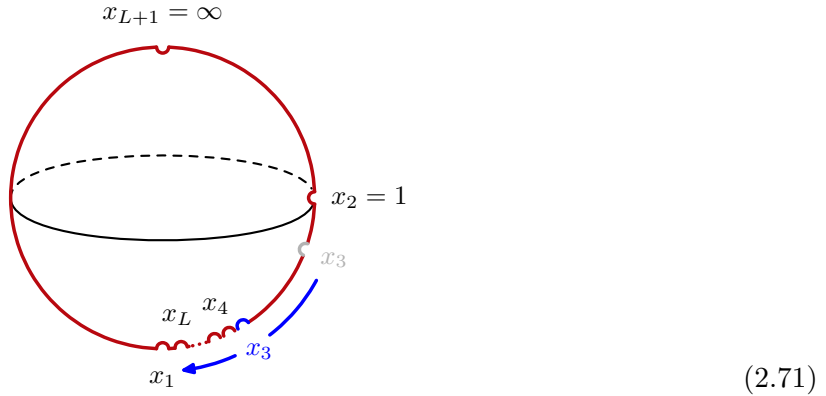
\begin{equation}
  \phantom{mmmm}\mpostuse{genuszerox3to0}
\end{equation}
This limit can be described in the basis
$\mathcal{B}_{2,2,\dots,2}$, since for this choice, the maximum eigenvalue of
$e_0$ is given by 
\begin{equation}
s_\mmax=s_{1,3,4,\dots ,L}\,.
\end{equation}
This can be seen by repeating the observation that led to \eqn{eqn:e1B222} for $e_1$: deriving \eqn{eqn:KZexampleBasis} for $\mathcal{B}_{2,2,\dots,2}$, assembles all the $s_{ij}$ with $i,j\neq 2,\snol+1$ in the matrix $e_0$. Therefore, the regularization factor $z_0^{-e_0}$ in $\bC_0$ can at most contribute with a factor $x_3^{-s_\mmax}$ to each integral.

The behavior of these entries for $x_3\to x_1=0$ may be determined using the change of variables $x_i=x_3 w_i$ for $0=x_1\leq x_i<x_2=1$, such that in particular $w_1=0$ and $w_3=1$. This yields for $i_k\neq 2$
\begin{align}
\label{eqn:C0entriesCalculation}
&\lim_{x_3\to 0}x_3^{-s_\mmax}\Sel[i_4,\dots,i_\snol](0,1,x_3)\nnl
&=\lim_{x_3\to 0}x_3^{-s_\mmax}\int_{\mathcal{C}(x_3)}\prod_{i=4}^\snol dx_i\, \prod_{0\leq x_j <x_l<x_3}\hspace{-0.4cm}x_{lj}^{s_{jl}}\prod_{0\leq x_m<x_3}\hspace{-0.2cm}x_{3m}^{s_{3m}}\prod_{0\leq x_n<x_2}\hspace{-0.2cm}x_{2n}^{s_{2n}} \prod_{k=4}^\snol \frac{1}{x_{k i_k}}\nnl
&=\lim_{x_3\to 0}\hspace{-0.5cm}
\int\limits_{\substack{\phantom{x}\\0=w_1<w_i<w_3=1}}
\hspace{-0.6cm}\prod_{i=4}^\snol dw_i\, \hspace{-0.2cm}\prod_{0\leq w_j <w_l<x_3}\hspace{-0.4cm}w_{lj}^{s_{jl}}\prod_{0\leq w_m<w_3}\hspace{-0.3cm}w_{3m}^{s_{3m}}\prod_{0\leq x_n<x_2}\hspace{-0.2cm}(1{-}x_3 w_n)^{s_{2n}} \prod_{k=4}^\snol \frac{1}{w_{k i_k}}\nnl
&=\int_{0=w_1<w_i<w_3=1 }\prod_{i=4}^\snol dw_i\, \prod_{0\leq w_j <w_l<x_3}w_{lj}^{s_{jl}}\prod_{0\leq w_m<w_3}w_{3m}^{s_{3m}} \prod_{k=4}^\snol \frac{1}{w_{k i_k}}\nnl
&=\SI{i_4,i_5,\ldots,i_\snol}(0,1,w_3=1)|_{s_{2j}=0}\,,
\end{align}
which is, as for the $x_3\to 1$ limit, an integral with integrand defined on $\mathcal{F}_{\snol,3}$.

Note that if we would not restrict to the basis $\mathcal{B}_{2,2,\dots,2}$ and there were $r$ indices $k_j\in\{4,5\dots,L\}$ such that $i_{k_j}=2$, then the change of variables would leave $r$ factors of $x_3$ in the quotient of the measure and the denominator
\begin{equation}
\prod_{k=4}^\snol \frac{dx_k}{x_{k i_k}}=x_3^r \prod_{k=4, k\not\in \{k_j\}}^\snol \frac{dw_k}{w_{k i_k}}\prod_{j=1}^r \frac{dw_{k_j}}{x_3 w_{k_j}-1}\,,
\end{equation}
which vanishes for $x_3\to 0$. Therefore, the entries of $\bC_0$ are linear
combinations of integrals
\begin{align}
\label{eqn:x3To0Selberg}
&\lim_{x_3\to 0}x_3^{-s_\mmax}\Sel[i_4,\dots,i_\snol](0,1,x_3)\nnl
&=\begin{cases}
\SI{i_4,i_5,\ldots,i_\snol}(0,1,w_3=1)|_{s_{2j}=0}&\text{if } \Sel[i_4,\dots,i_\snol](0,1,x_3)\in \mathcal{B}_{2,2,\dots,2}\,,\\
0&\text{otherwise}\,.
\end{cases}
\end{align}
\noindent\textbf{Mandelstam variables:}
According to \eqn{eqn:C1limitSel}, the Mandelstam variables $s_{3j}$ associated to the momentum of the auxiliary insertion point $x_3$ are redundant in $\bC_1$: they simply appear as a splitting of the effective momentum $\tilde{s}_{2j}=s_{2j}+s_{3j}$ associated to the insertion point at $x_2=1$ and thus, may be chosen to be set to zero. This choice is more subtle in the boundary value $\bC_0$ with the non-vanishing entries being calculated according to \eqn{eqn:C0entriesCalculation}: here, the Mandelstam variables $s_{3j}$ are not at all redundant, i.e.\ an artificial splitting of the momentum contribution, but encode the full momentum of the insertion point $w_3=1$. Thus, it may be expected that setting this momentum to zero effectively removes one external state, leaving an integrand defined on $\mathcal{F}_{\snol-1,3}$. This expectation is true for certain linear combinations of Selberg integrals, as argued in the next subsection. 

\noindent\textbf{Summary of subsection:}
The vector of type-$(2,L)$ Selberg integrals $\Selbld(x_3)$ with integrands on $\mathcal{F}_{L+1,4}$ encodes for $N=L$ the $\nol$- and $(\nol{-}1)$-point amplitudes, in the regularized limits $\bC_0$ and $\bC_1$, which can be related to each other using the Drinfeld associator $\Phi(e_0,e_1)$, with $e_0$ and $e_1$ determined by the KZ \eqn{eqn:KZexample}, as follows 
\begin{equation}\label{genus0diagram}
	\mpostuse{genuszerostructure}
\end{equation}
where the exact degeneracy to the amplitudes as $s_{3j}\to 0$ and the corresponding map $\Phi(e_0,e_1)|_{s_{3j}=0}$ will be explored in the next subsections.

\subsection{Open string amplitudes at genus zero}\label{subsec:Relate}
Open-string tree-level amplitudes arise as correlators between vertex-operators inserted at the boundary of the disk worldsheet. Usually one makes use of the conformal symmetry of the worldsheet in order to place the boundary of the disk at the real line.
Evaluating the correlators allows to frame the amplitude in the form
\cite{Broedel:2013tta}:
\begin{align}
A_\open(1,\snol,\snol-1,...,2,\snol+1;\ap) 
&= {\bf Z}^T  \, {\bf MK}  \,\, {\bf A}_\YM\,.
\end{align}
While the Yang--Mills tree-level amplitudes ${\bf A}_\YM$ can be obtained (for example) from BCFW recursion relations \cite{Britto:2004ap,Britto:2005fq}, the object ${\bf MK}$ is known as the momentum kernel and can be represented as a matrix of dimension $(\snol{-}2)!\times(\snol{-}2)!$. A recursive formula is known for any multiplicity \cite{Stieberger:2009hq,BjerrumBohr:2010hn}.  The vector ${\bf Z}$ consists of $(\snol{-}2)!$ so-called $Z$-integrals \cite{Broedel:2013tta} 
\begin{equation}
\label{eqn:Zint}
Z(q_1,q_2,\ldots,q_{\snol+1})= \int_{\mathcal{C}(x_2=1)} \prod_{i=3}^{\snol} dx_i\, \KN
\frac{x_{1,\snol+1}x_{2,\snol+1}x_{12}}{x_{q_1 q_2}x_{q_2 q_3}\cdots x_{q_{\snol} q_{\snol+1}}x_{q_{\snol+1} q_1}}\,,
\end{equation}
where the factor $x_{1,\snol+1}x_{2,\snol+1}x_{12}$ together with the fixing of the coordinates $(x_1,x_2,x_{\snol+1})=(0,1,\infty)$ (cf.~\eqn{eqn:fixedPunctures}) corresponds to dividing out the gauge volume $\CV_\text{CKG}$ of the conformal Killing group $\SL(2,\ZC)$. The quotient together with the integration measure is called Parke--Taylor form, while $\KN$ is called Koba--Nielsen factor\footnote{Here, for consistency with other articles, we have written the absolute value, despite all $x_{ji}$ are real and positive in our conventions. } and defined by
\begin{equation}
\label{eqn:KN}
\KN=\prod_{0=x_1\leq x_i<x_j\leq x_2=1} \hspace{-0.2cm}|x_{ij}|^{s_{ij}}=\prod_{0=x_1\leq x_i<x_j\leq x_2=1}\hspace{-0.2cm}\exp(s_{ij} \log |x_{ij}|)\,. 
\end{equation}
Note that we have defined the Selberg seed in \eqn{eqn:SelbergzeroSeed} in exactly the same way: it is constructed to equal the $(L{+}1)$-point Koba--Nielsen factor
\begin{equation}\label{eqn:SandKN}
\Sel=\KN\,.
\end{equation}
Since $\log x_{ij}$ is (almost) the genus-zero string propagator, the Koba--Nielsen factor can easily be identified as a generating functional of graphs connecting the vertex operators, where each edge connecting vertex operators at positions $x_i$ and $x_j$ is weighted by the corresponding Mandelstam variable~$s_{ij}$. 

Iterated integrals in $x_i$ over various derivatives of the Koba--Nielsen factor, in particular the $Z$-integrals defined in \eqn{eqn:Zint}, fall in the class of Selberg integrals \cite{Selberg44}. It is only those integrals, which need to be calculated in order to determine the full open-string tree-level amplitude at any multiplicity.  

Concretely, in \rcite{Broedel:2013aza} a vector of iterated integrals has been constructed, which is related via a basis transformation to the vector of genus-zero Selberg integrals $\Selbld(x_3)|_{\CB_{1,1,\dots,1}}$ in the basis $\CB_{1,1,\dots,1}$
\begin{align}
\hat{\bF}(x_3)&=\boldsymbol{B}  \Selbld(x_3)|_{\CB_{1,1,\dots,1}}\,.
\end{align}
The transformation matrix (of the corresponding bases of twisted forms) $\boldsymbol{B}$ has been calculated in \rcite{AKtbp}, its non-vanishing entries are polynomials over $\ZZ$ of degree $L{-}3$ in $s_{ij}$. According to the KZ \eqn{eqn:KZexampleBasis}, the function $\hat{\bF}(x_3)$ satisfies a KZ equation with matrices $\hat{\boldsymbol{e}}_0=\boldsymbol{B}\boldsymbol{e}_0\boldsymbol{B}^{-1}$ and $\hat{\boldsymbol{e}}_1=\boldsymbol{B}\boldsymbol{e}_1\boldsymbol{B}^{-1}$, where $\boldsymbol{e}_i$ are the matrices in the KZ equation satisfied by the Selberg integrals $\Selbld(x_3)|_{\CB_{1,1,\dots,1}}$. Moreover, the first $(L{-}3)!$-entries of the regularized limit as $x_3\to 1$ are linear combinations of the integrals in \eqn{eqn:C1limitSel} and contain the $L$-point, tree-level amplitudes with effective Mandelstam variables $\tilde{s}_{ij}=s_{ij}+\delta_{i2}s_{3j}$ for $i,j\in \{1,\dots,L\}\setminus \{3\}$, $i<j$, such that 
\begin{align}
\hat{\bC}_1 = \lim_{x_3\rightarrow 1} (1-x_3)^{-\hat{e}_1} \bhF(x_3)&=\begin{pmatrix}
{\bf Z}^T \,  {\bf MK}|_{	L-\text{point},\, \tilde{s}_{ij}=s_{ij}+\delta_{i2}s_{3j}}\\\vdots
\end{pmatrix}\,,
\end{align}
where the (labels of the) Mandelstam variables $\tilde s_{ij}$ correspond to the insertion points $x_1<x_{L}<x_{L-1}<\dots<x_4<x_2=1$. On the other hand, the only non-vanishing entries of the regularized boundary value
\begin{align}
\hat{\bC}_0 &= \lim_{x_3 \to 0} x^{-\hat{e}_0}\bhF(x_3)
\end{align}
degenerate in the soft limit $s_{3j}\to 0$ to the $(L{-}1)$-point, tree-level amplitudes
\begin{align}
\lim_{s_{3j}\to 0}\hat{\bC}_0 &= \begin{pmatrix}
{\bf Z}^T  \, {\bf MK}|_{	(L{-}1)-\text{point},\,	 s_{ij}}\\ 0\\\vdots\\0
\end{pmatrix} \,,
\end{align}
where the Mandelstam variables $s_{ij}$ are associated to the insertion points $w_1=0<w_L<w_{L-1}<\dots<w_4=1$. The limit $s_{3j}\to 0$ effectively leads to a merging of the punctures $w_4$ and $w_3$ in the integrals \eqref{eqn:x3To0Selberg}, since these integrals are linearly combined in $\hat{\bC}_0$ such that the integrands are total derivatives with respect to $w_4$ in the limit $s_{3j}\to 0$. Thus, taking the limit $s_{3j}\to 0$ of the associator equation 
\begin{align}
\hat{\bC}_1&=\boldsymbol{\Phi}(\hat{e}_0,\hat{e}_1)\hat{\bC}_0
\end{align}
leads to a recursion relating the $L$-point to the $(L{-}1)$-point, genus-zero amplitudes
\begin{align}
\begin{pmatrix}
{\bf Z}^T  \, {\bf MK}|_{	L-\text{point}, \,s_{ij}}\\\vdots
\end{pmatrix}&=\boldsymbol{\Phi}(\hat{e}_0,\hat{e}_1)|_{s_{3j}=0}  \begin{pmatrix}
{\bf Z}^T \,  {\bf MK}|_{	(L{-}1)-\text{point}, \,s_{ij}}\\ 0\\\vdots\\0
\end{pmatrix}\,.
\end{align}

\section{Genus one (one-loop level)}
\label{sec:genusone}
\setcounter{theorem}{0}
In this section, we develop and explore the genus-one version of the concepts from \secref{sec:genuszero} and link the resulting formalism to the evaluation of open-string configuration\hyp{}space integrals at one loop. The genus-one recursion is remarkably similar to the genus-zero recursion of \rcite{Broedel:2013aza} reviewed in \subsecref{subsec:Relate}. 

While the genus-zero recursion relates $N$-point configuration\hyp{}space integrals to $(N{-}1)$-point versions thereof and is thus a recursion in the number of external legs, the genus-one mechanism relates $N$-point one-loop configuration\hyp{}space integrals to $(N{+}2)$-point tree-level configuration\hyp{}space integrals, thus linking objects\footnote{Notation and limits depicted in figure \eqref{genus0genus1} will be introduced and explained in the course of this section.} occurring at different genera:
\begin{equation}\label{genus0genus1}
	\mpostuse{genuszerogenusone}\,.
\end{equation}
In the genus-zero recursion, the Drinfeld associator effectively adds an additional puncture to an $(N{-}1)$-point interaction resulting in an $N$-point tree-level interaction.  On the other hand, as shown below, the genus-one recursion amounts to two external states of the $(N{+}2)$-point tree-level interaction being glued together by the elliptic analogue of the Drinfeld associator, the elliptic Knizhnik-Zamolodchikov-Bernard (KZB) associator, to form a genus-one worldsheet of $N$ external string states.

In the current section, we follow the structure of the previous \secref{sec:genuszero}: in \subsecref{subsec:eMPLeMZV} to \subsecref{subsec:BoundaryValues} we introduce elliptic iterated integrals, a genus-one version of Selberg integrals, the elliptic KZB associator and the KZB equation for an auxiliary marked point.  In the subsequent \subsecref{ssec:openstringcalc}, the relation to open-string configuration\hyp{}space integrals is drawn, where we also discuss some practicalities.  In \secref{sec:examples}, the first orders in $\ap$ of the two-, three- and four-point one-loop configuration\hyp{}space integrals are calculated using the genus-one associator mechanism and are shown to match known results.

\subsection{Singularities, iterated integrals and elliptic multiple zeta values}
\label{subsec:eMPLeMZV}
In the following, we will consider the annulus formed by open-string worldsheets at one loop as embedded into a torus with $A$-cycle (red) and $B$-cycle (blue), where the ratio of the respective lengths, the modular parameter, is denoted by $\tau$.
\begin{equation}
 \mpostuse{fundamentaldomain}
\end{equation}
Suitable differentials on the torus are generated by the expansion \eqref{eqn:differentials} of the Eisenstein--Kronecker series $F(z,\eta,\tau)$ in $\eta$. This defines -- in distinction to the genus-zero scenario --  an infinite number of differentials $g^{(n)}(z,\tau)\dz$. The index $n$ labelling the functions $g^{(n)}$ is called its \textit{weight}. While $g^{(0)}=1$ is trivial, the function $g^{(1)}$ has poles at $z\in \ZZ\tau+ \ZZ$ and can be expanded in $q=\exp(2 \pi i \tau)$ as follows \cite{Broedel:2014vla}:
\begin{equation}
	g^{(1)}(z,\tau)= \pi \cot (\pi z)+4 \pi \sum\limits_{m=1}^\infty \sin (2 \pi m z) \sum\limits_{n=1}^\infty q^{mn}\,. 
\end{equation}
All $g^{(n)}$ with $n\geq 2$ are holomorphic in the fundamental elliptic domain \mbox{$z=s+\tau t$}, $s,t\in [0,1)$. The Eisenstein--Kronecker series $F(z,\eta,\tau)$ is one-periodic, but only quasi-periodic in $z$. Therefore, the integration kernels $g^{(n)}$ are also one-periodic
\begin{equation}
\label{eqn:1periodicityKernel}
g^{(n)}(z+1,\tau)=g^{(n)}(z,\tau)\,,
\end{equation}
but not $\tau$-periodic in $z$, thus, they can not be elliptic functions. Furthermore, they also inherit the following symmetry property from $F(z,\eta,\tau)$:
\begin{equation}
\label{eqn:symmetryKernel}
g^{(n)}(-z,\tau)=(-1)^ng^{(n)}(z,\tau)\,.
\end{equation}
Despite not being elliptic, the functions $g^{(n)}$ can be
considered to be genus-one generalizations of the integration kernels defining
the multiple polylogarithms \eqref{eqn:MPLdef}, which also lead to meromorphic,
but multi-valued functions. 

The integrals over the kernels $g^{(n)}$ lead to elliptic polylogarithms
\cite{Levin, BrownLev}: due to their periodicity in
\eqn{eqn:1periodicityKernel} they are single-valued functions on the annulus, but can be thought of as multi-valued functions on the torus. This is
equivalent to the behavior of the ordinary logarithm at genus zero: on each
Riemann sheet the logarithm is single-valued, while it is a multi-valued
function in the complex plane. 

Due to the poles of the integration kernel $g^{(1)}$, iterated integrals of $g^{(1)}$ need
to be regularized. Moreover, its poles will -- in certain limits -- act as the link
between the string propagators at Riemann surfaces of genus zero and genus one.
Corresponding to the differentials introduced in \eqn{eqn:differentials}, one
can define a class of iterated integrals $\Gt$ called elliptic multiple
polylogarithms:
\begin{equation}
\label{eqn:defGt}
\Gt(\begin{smallmatrix}n_1,\,n_2,\,\dots,\,n_k \\ a_1,\,a_2,\,\dots,\,a_k\end{smallmatrix}; z,\tau)=
\int_0^zdz'\, g^{(n_1)}(z'-a_1,\tau)\Gt(\begin{smallmatrix}n_2,\,\dots,\,n_k \\ a_2,\,\dots,\,a_k\end{smallmatrix}; z',\tau)\,,
\end{equation}
which due to their nature as iterated integrals obey shuffle relations
\begin{align}
	&\Gt(A_1,A_2,\ldots,A_j; z,\tau)\Gt(B_1,B_2,\ldots,B_k; z,\tau)\nnl
	&\qquad=\Gt\big((A_1,A_2,\ldots,A_j)\shuffle(B_1,B_2,\ldots,B_k); z,\tau\big)
\end{align}
in terms of combined letters
$A_i=\begin{smallmatrix}n_i\\a_i\end{smallmatrix}$.

The integral over $g^{(1)}$ will be of particular interest below:
$\Gtargzt{1}{0}$ requires regularization because of an endpoint divergence at
the lower integration boundary due to the pole at $z=0$. The standard
regularization procedure -- which we are going to use here -- is called
\textit{tangential basepoint regularization} and is discussed in detail for
example in \rcites{Deligne89,Brown:ICM14}. In short, we subtract the endpoint divergence by defining\footnote{The limit $\epsilon\to 0$ is taken within the unit interval. Unless stated otherwise, the same holds for any limits in this article}
\begin{align}\label{sec:eMPL:DefReg}
\Gt_\reg(\begin{smallmatrix}1\\0 \end{smallmatrix}; z,\tau)
&=\lim_{\epsilon\rightarrow 0}\int_{\epsilon}^z dz\, g^{(1)}(z,\tau) +\log(1-e^{2\pi i \epsilon})\nnl
&=\log(1-e^{2\pi i z})-\pi i z+4\pi \sum_{k,l>0}\frac{1}{2\pi k}\left(1-\cos(2\pi k z)\right)q^{kl}\,.
\end{align}
Considering $z\in(0,1)$, the following properties can be read off from the above $q$-expansion
\begin{align}\label{eqn:propertiesGamma1}
	\Gt_\reg(\begin{smallmatrix}1\\0 \end{smallmatrix}; z\pm 1,\tau)&=\Gt_\reg(\begin{smallmatrix}1\\0 \end{smallmatrix}; z,\tau)\mp \pi i\nnl
	\Gt_\reg(\begin{smallmatrix}1\\0 \end{smallmatrix}; -z,\tau)&=\Gt_\reg(\begin{smallmatrix}1\\0 \end{smallmatrix}; z,\tau)+\pi i\,,
\end{align}
where we place the branch cut of the logarithm such that $\log(-1)=\pi i$. This
implies in particular invariance under $z\to 1-z$ for $0<z<1$:
\begin{equation}\label{eqn:InvarianceGamma1}
\Gt_\reg(\begin{smallmatrix}1\\0 \end{smallmatrix}; z,\tau)=\Gt_\reg(\begin{smallmatrix}1\\0 \end{smallmatrix}; 1-z,\tau)\,.
\end{equation}
In addition, we find the following asymptotic behavior for $z\to 0$ 
\begin{equation}\label{eqn:asympGammaz0}
\Gt_\reg(\begin{smallmatrix}1\\0 \end{smallmatrix}; z,\tau)\sim \log(-2\pi i z)
\end{equation}
and $z\to 1$
\begin{equation}\label{eqn:asympGammaz1}
\Gt_\reg(\begin{smallmatrix}1\\0 \end{smallmatrix}; z,\tau)\sim \log(-2\pi i (1-z)) \,.
\end{equation}
The above regularization procedure is an algebra homomorphism, i.e.~compatible with the shuffle product. From now on, we will use the regularized iterated integrals exclusively and omit the subscript when noting $\Gt$. Furthermore, we are going to keep the dependence on $\tau$ implicit for all integration kernels $g^{(n)}$ and all iterated elliptic integrals $\tilde{\Gamma}$.

In the same way as products of terms of the form $1/x_{ij}$ can be related by partial fractioning \eqref{eqn:partialFractioning}, there is a genus-one analogue for the Kronecker series: the Fay identity. In terms of the functions $g^{(n)}(z)$ it can be phrased as
\begin{align}
	\label{eqn:Fay}
&g^{(n_1)}(t-x) g^{(n_2)}(t)\nnl
 &=  - (-1)^{n_1} g^{(n_1+n_2)}(x) + \sum_{j=0}^{n_2} \binom{ n_1 - 1 + j}{j} g^{(n_2-j)}(x) g^{(n_1+j)}(t-x) \notag  \\
 & \ \ \ \ \ + \sum_{j=0}^{n_1} \binom{n_2-1+j}{j} (-1)^{n_1+j} g^{(n_1-j)}(x) g^{(n_2+j)}(t) 
\end{align}
and derived from a similar property obeyed by the generating function $F(z,\eta,\tau)$. 

For compactness, we will use a notation similar to definition \eqref{eqn:MPLdef} in terms of words from an alphabet for the elliptic multiple polylogarithms $\Gt$ defined in \eqn{eqn:defGt} with $a_1=a_2=\dots=a_k=0$. Concretely, since there are infinitely many integration kernels $g^{(n)}$, the alphabet is infinite as well and denoted by $\{x^{(0)},x^{(1)},\dots\}$. For a word $w=x^{(n_1)}\dots x^{(n_k)}\in\{x^{(0)},x^{(1)},\dots\}^{\times}$, we denote the corresponding elliptic multiple polylogarithm by
\begin{equation}
\Gt_w(z)=\Gt(x^{(n_1)}\cdots x^{(n_k)}; z)=\Gtargz{n_1,...,n_k}{0,...,0}\,.
\end{equation}
For $w\neq  (x^{(1)})^n$, one finds
\begin{align}\label{ellKZ:zeroValue}
\lim_{z\rightarrow 0}\Gt_w(z)=0\,,
\end{align}
while the regularization \eqref{sec:eMPL:DefReg} implies logarithmic
divergences for words $w=(x^{(1)})^n$ in the limit $z\to 0$:
\begin{align}\label{ellKZ:zeroValueRegularized}
\Gt_{(x^{(1)})^n}(z)\sim \frac{1}{n!}\log(-2\pi i z)^n\,.
\end{align}
Due to the one-periodicity of $g^{(1)}$, this divergence also appears at the upper integration boundary for words $w=(x^{(1)})^n$ as $z\to 1$. The corresponding regularization procedure is particularly important for elliptic multiple zeta values to be discussed in the next paragraph. 

Considering the	limit $z\to 1$ leads to the genus-one analogues of MZVs defined in \eqn{eqn:mzv}. These so-called elliptic multiple zeta values (eMZVs) \cite{Enriquez:Emzv,Matthes:Thesis,Broedel:2014vla} are defined in terms of regularized iterated integrals $\Gt_w$ with $w=x^{(n_1)}\dots x^{(n_k)}\in X\setminus x^{(1)}X$, i.e.\ $n_1\neq 1$, at $z=1$:
\begin{equation}\label{sec:eMZV:def}
	\omega(n_k,\dots,n_1;\tau)=\omega(w^t;\tau)=\lim_{z\to 1}\Gt_{w}(z,\tau)=\lim_{z\to 1}\Gt(\begin{smallmatrix}n_1 &\dots &n_k\\ 0&\dots &0 \end{smallmatrix}; z,\tau)\,,
\end{equation}
where $w^t$ denotes the reversal of the word $w$. In order to extend this definition to all words $w\in X$, the singularity of $\Gt_{x^{(1)}w}(z,\tau)$ at $z=1$ has to be regularized. This can be done similarly as for the multiple polylogarithms in \eqn{eqn:regGenusZero}, and is elaborated on in detail in \appref{app:eMZVreg}. The main result is the following definition of the regularized eMZVs $\omega_\reg(w^t;\tau)$: for any word $w\in X\setminus x^{(1)}X$  or $w=(x^{(1)})^n$ they are defined by
\begin{equation}\label{eqn:regEMZV}
	w\mapsto \omega_\reg(w^t;\tau)= \begin{cases}\omega(w^t;\tau)& \text{if }w\in X\setminus (x^{(1)}X )\,,\\
	0& \text{if } w=(x^{(1)})^n,\, n\in\ZN\,.
	\end{cases}
\end{equation}
Again, the remaining cases $w\in x^{(1)}X$ and $w\neq (x^{(1)})^n$ can be related to the above situations using the shuffle algebra. As for the elliptic multiple polylogarithms, from now on unless stated otherwise, all elliptic multiple zeta values are assumed to be regularized and simply denoted by $\omega(w^t)$ omitting the subscript and the $\tau$-dependence in $\omega_\reg(w^t;\tau)$. 

In the same way as the shuffle algebra is preserved when regularizing iterated integrals $\Gt$ in \eqn{sec:eMPL:DefReg}, this is true for the corresponding MZVs: (regularized) eMZVs inherit the shuffle algebra, the properties implied by the Fay identity and some further properties from the elliptic multiple polylogarithms such as the reflection identity
\begin{equation}
\omega(n_k,\dots, n_1)=(-1)^{n_1+\dots+n_k}\omega(n_1,\dots, n_k)
\end{equation} 
due to property \eqref{eqn:symmetryKernel} of the integration kernels.  Furthermore, even elliptic zeta values are related to the (genus-zero) zeta values according to 
\begin{align}
\omega(2m;\tau)&=-2\zeta_{2m}\,.
\end{align}
Numerous other relations between eMZVs can be retrieved from
\cite{eMZVWebsite}.
%


\subsection{Genus-one Selberg integrals}\label{ssec:genSelberg}

In order to repeat the construction described for genus zero in \subsecref{subsec:SelbergIntegrals}, we will need to find a genus-one generalization of the Selberg seed function defined in \eqn{eqn:SelbergzeroSeed} which can be used to construct genus-one Selberg integrals. The genus-one Selberg seed should depend on the positions of insertion points inserted on the boundaries of an open-string worldsheet at one loop. Such worldsheets are quotients of a genus-one Riemann surface, where the corresponding involution is induced by complex conjugation. For simplicity, we restrict our discussion to oriented worldsheets where all insertion points are located on one boundary. This scenario corresponds to planar open-string interactions at one loop with the relevant geometry being the annulus with one punctured boundary. A generalization to the non-planar case, where points are allowed on both boundaries is not expected to pose any structural obstacles. Upon embedding the annulus into a torus, the relevant boundary is identified with the $A$-cycle and parametrised by the unit interval. In contrast to the genus-zero labelling \eqref{eqn:genus0ordering}, the positions of the insertion points are going to be denoted by and ordered according to
\begin{equation}
0=z_1<z_L<z_{L-1}<\dots <z_2<1=z_1\text{ mod } \ZZ\,,
\end{equation}
where we have used the symmetries of the torus to fix $z_1=0$. 
\begin{equation}
\mpostuse{genusonegeneral}
\end{equation}
Therefore -- in analogy to the genus-zero scenario -- we expect to find iterated integrals with integrands defined on the configuration space of $k$-punctured $A$-cycles $[0,1]\setminus\{z_1,\dots,z_k\}$ of tori with purely imaginary modular parameter $\tau$ and $k$ fixed punctures:
\begin{align}
&\mathcal{F}^{\tau}_{\snol,k}=\nnl
&\quad\{(z_{k+1},z_{k+2},\dots,z_{\snol})\in ([0,1]\setminus\{z_1,\dots,z_k\})^{\snol-k}|\forall i\neq j: z_i\neq z_j\}.
\end{align}
Remembering the basic properties of the genus-zero Selberg seed defined in \eqn{eqn:SelbergzeroSeed}, its generalization to genus one is straightforward. Defining 
\begin{equation}
\Gt_{ij}=\Gt(\begin{smallmatrix}1\\0 \end{smallmatrix}; z_{ij},\tau)=\Gt_{x^{(1)}}(z_{ij},\tau)\,,
\end{equation}
where
\begin{align}\label{eqn:zij}
z_{ij}&=z_i-z_j\,,
\end{align}
one can simply replace $\log x_{ji}=G_{e_0}(x_{ji})$ in the genus-zero Selberg seed by
$\Gt_{ji}=\Gt_{x^{(1)}}(z_{ji},\tau)$.

At this point, it is very natural to define a suitable genus-one analogue of the Selberg integrals \eqref{eqn:Selbergzero}:
\begin{definition}
	\label{def:SelbergInt}
	Let $L\geq 2$, $0=z_1<z_L<...<z_2<1$ and $\tau$ the modular parameter of the torus $\ZC/(\ZZ+\tau\ZZ)$. Let the empty genus-one Selberg integral (or genus-one Selberg seed) be
	\begin{equation}
	\label{eqn:SelbergSeed}
	\SelE=\SIE{}{}(z_1,\dots,z_L)=\prod_{0=z_1\leq z_i< z_j\leq z_2}\exp\left(s_{ij}\Gt_{ji}\right)\,.
	\end{equation}
	Genus-one Selberg integrals are then defined recursively by
	\begin{align}
	\label{eqn:Selberg}
	&\SIE{n_{k+1},\dots,n_L}{i_{k+1},\dots,i_L}(z_1,\dots,z_k)\nnl
	&\phantom{bbbbbb}=\int_0^{z_k}dz_{k+1}\, g^{(n_{k+1})}_{k+1,i_{k+1}}\SIE{n_{k+2},\dots,n_L}{i_{k+2},\dots,i_L}(z_1,\dots,z_{k+1})\,
	\end{align}
	where $1\leq i_p<p$ for $k+1\leq p\leq L$, $n_{k{+}1},\dots, n_L$ are non-negative integers and
	\begin{equation}
	g^{(n)}_{ij}=g^{(n)}_{i,j}=g^{(n)}(z_i-z_j,\tau)\,.
	\end{equation}
\end{definition}
For all genus-one Selberg integrals as well as for the genus-one Selberg seed, we will indicate the dependence on $\tau$ by the upper index and by using partial derivatives. 

The sum $n_{k+1}+\dots+n_L$ is called the \textit{weight} of the Selberg integral. This notation, where instead of the actual shifts $a_i$ from \eqn{eqn:defGt} the index of a position variable $z_i$ is used, will allow for rather compact equations when manipulating genus-one Selberg integrals.  Moreover, as for the genus-zero Selberg integrals, the shift $z_{i_{k+1}}$ in the integration kernel $g^{(n_{k+1})}_{k+1,i_{k+1}}$ can only be a variable which has not yet been integrated out, which leads to the genus-one analogue of the admissibility condition in \eqn{eqn:admissibility}:
\begin{equation}
1\leq i_p<p\qquad \forall p\in\{k+1,\dots,L\}\,,
\end{equation}
whereas the corresponding integrals at genus one are again called \textit{admissible}. 

Similar to the situation for genus-zero Selberg integrals, convergence is determined by the values of the complex parameters $s_{ij}$. Since the pole structure of genus-one Selberg integrals matches the corresponding structure at genus zero, the conditions discussed and referred to apply for genus-one Selberg integrals as well. We will assume the parameters $s_{ij}$ to be fixed accordingly throughout the remainder of this article. 

The expression for the Selberg seed \eqref{eqn:SelbergSeed} is already very close to the one-loop Koba--Nielsen factor $\KNE$ appearing in the one-loop string amplitudes below.  In particular, $G_{e_0}$ and $\Gt_{x^{(1)}}$ are the regularized integrals as defined in \eqns{eqn:regGenusZero}{sec:eMPL:DefReg}, respectively. A key observation for our construction is the relation between these two functions which is stated in \eqn{ellKZ:zeroValueRegularized}: the polylogarithm $G_{e_0}(-2\pi i z)$ describes the asymptotic behaviour of the elliptic polylogarithm $\Gt_{x^{(1)}}(z,\tau)$ as $z\to 0$.
\vspace*{6pt}

In order to be equipped for the next subsections, let us collect a couple of identities for genus-one Selberg integrals. Derivatives of the function $\Gt_{ij}$ can be redirected to another index via   
\begin{equation}
\frac{\partial}{\partial z_i}\Gt_{ij}=g^{(1)}(z_i-z_j)=-\frac{\partial}{\partial z_j}\Gt_{ij}\,.
\end{equation}
In the above language, the Fay identity \eqref{eqn:Fay} takes the form
\begin{align}\label{sec:G1:FayIdentity}
g^{(m)}_{kj}g^{(n)}_{ki}&=(-1)^{m+1} g^{(m+n)}_{ji}+\sum_{r=0}^n\binom{m+r-1}{m-1}g^{(n-r)}_{ji}g^{(m+r)}_{kj}\nnl
&\phantom{=}+\sum_{r=0}^m(-1)^{m-r}\binom{n+r-1}{n-1}g^{(m-r)}_{ji}g^{(n+r)}_{ki}\,.
\end{align}
The left-hand side of \eqn{sec:G1:FayIdentity} is admissible, when
w.l.o.g.~$i<j<k$: 
if this condition is met, the right-hand side is a $\ZZ$-linear combination of admissible products. 

The Fay identity is the reason why all integration kernels $g^{(n)}_{ij}$ are included in the definition of the genus-one Selberg integrals \eqref{eqn:Selberg} rather than only $g^{(1)}_{i j}$: application of the Fay identity introduces weights $n\neq 1$, such that a closed system with respect to integration by parts and the Fay identity requires all integration kernels $g^{(n)}_{ij}$.

When discussing a recursive solution for genus-one Selberg integrals below, various derivatives will have to be taken with respect to insertion points $z_i$, which is thoroughly discussed in \appref{app:KZBgeneral}. Here we would like to collect some key properties used in the calculations below. Taking the regularization prescription in \eqn{sec:eMPL:DefReg} into account, we find 
\begin{align}\label{sec:G1:BoundaryS}
	\SelE|_{z_i=z_j}&=0\qquad \text{ for }i\neq j\,,
\end{align}
which is the property analogous to \eqn{eqn:genus0Svanishing}. Taking a derivative of the one-loop Selberg seed with respect to a particular variable yields
\begin{align}
\label{eqn:derivativeSE}
\frac{\pd}{\pd z_i}\SelE&=\sum_{k\neq i}s_{ik}\,g^{(1)}_{ik}\SelE\,.
\end{align}

Considering the class of type-$(k,L)$ genus-one Selberg integrals for a fixed $L$ and a given number of integrations $L-k$, it is natural to ask for a basis.  There are two operations relating different genus-one Selberg integrals: one can integrate by parts and one can apply Fay identities. The question of a basis for this type of integrals is a very old one and amounts to determining a basis of the corresponding twisted de Rham cohomology, similar to the fibration basis in genus zero mentioned in the discussion above definition \eqref{eqn:basisGenusZero} of the bases for genus-zero Selberg integrals. One possible representation for a basis of the twisted de Rham cohomology on genus one was suggested in \cite{Mafra:2019xms}. 

Since a reduction to a basis is convenient, but not necessary in our construction, we do not try to rigorously provide a genus-one analogue of the fibration basis. Instead, we note certain observations for a class of genus-one Selberg integrals with fixed $L$ and a fixed number of integrations $L{-}k$: 
\begin{itemize}
	\item for an index $n_p=0$, the corresponding integration kernel
		$g^{(0)}_{p,i_p}=1$ is a constant, thus, we can always choose
		$i_p=1$ in this case.
	\item as for the genus-zero Selberg integrals, for an index $n_p=1$,
		integration by parts yields a linear equation for the integrals
		due to the partial derivative of the Selberg seed
		\eqref{eqn:derivativeSE}.  Hence, for each index $n_p=1$, we
		expect to be able to reduce the class of integrals from $1\leq
		i_p<p$ to $1\leq i_p\neq i'_p<p$ for any $1\leq i'_p<p$ by such
		an integration by parts identity and applications of the Fay
		identity (to recover admissible integrals).  However, no
		further such simplifications are expected for the indices
		$n_p>1$.
\end{itemize}

In \subsecref{ssec:KZBeq} below, we are going to consider a differential
equation for a vector of genus-one Selberg integrals of length $L{-}2$,
which are the relevant genus-one Selberg integrals containing the one-loop and
tree-level configuration\hyp{}space integrals:
\begin{align}
\label{eqn:relevantGenusOneSelberg}
\SIE{n_{3},\dots,n_L}{i_{3},\dots,i_L}(z_1=0,z_2)&=\int_0^{z_2}dz_{3}\, g^{(n_{3})}_{3,i_{3}}\SIE{n_{4},\dots,n_L}{i_{4},\dots,i_L}(z_1=0,z_2,z_{3})\nnl
&=\int_{\mathcal{C}(z_2)}\prod_{i=3}^L dz_i\, \Sel^\El \prod_{k=3}^L g^{(n_k)}_{k,i_k}\,,
\end{align}
where $1\leq i_k<k$ and the integration region is given by
(cf.~\eqn{eqn:integrationRegion}):
\begin{equation}
	\mathcal{C}(z_i)=\{0=z_1<z_L<z_{L-1}<\dots<z_i\}\,,
\end{equation}
such that the integral over this domain reads
\begin{equation}
\int_{\mathcal{C}(z_2)}\prod_{i=3}^L dz_i=\int_0^{z_2}dz_3\int_0^{z_3}dz_4\dots \int_0^{z_{L-1}}dz_L\,.
\end{equation}
The integrals defined in \eqn{eqn:relevantGenusOneSelberg} are the genus-one generalization of the Selberg integrals \eqref{eqn:relevantSelberg} relevant for the tree-level amplitude recursion. As for this genus-zero class, the differential equation satisfied by the vector of these genus-one Selberg integrals leads to an associator equation relating one-loop to tree-level configuration\hyp{}space integrals.

Using the considerations about a fibration basis above, we will at least reduce the class of iterated integrals defined in \eqn{eqn:relevantGenusOneSelberg} to a spanning set
\begin{align}
\label{eqn:basisGenusOne}
\mathcal{B}^\El_{i'_3,i'_4,\dots,i'_\snol}&=\Big\lbrace  \SIE{n_{3},\dots,n_L}{i_{3},\dots,i_L}(0,z_2)|n_k\geq 0\text{ and } 1\leq i_k<k\nnl
&\phantom{bbbbbbb}\text{ such that } i_k\neq i'_k \text{ if }n_k=1
\text{ and }i_k=1\text{ if }n_k=0 \Big\rbrace
\end{align}
similar to the genus-zero basis \eqref{eqn:basisGenusZero}. We also allow $i'_k=0$ if we only intend to reduce the kernels with $n_k=0$ and include all the kernels with $n_k=1$, which certainly does not yield a basis, but a spanning set reduced by the redundant labeling of $g^{(0)}_{k,i_k}=1$. In other words, the labels $i'_k$ in $\mathcal{B}^\El_{i'_3,i'_4,\dots,i'_\snol}$ denote that the integrals defined by the set $\mathcal{B}^\El_{i'_3,i'_4,\dots,i'_\snol}$ are the genus-one Selberg integrals from \eqn{eqn:relevantGenusOneSelberg}, where for $3\leq k\leq L$ any kernel of the form $g^{(1)}_{k, i'_k}$ is rewritten in terms of the kernels $g^{(1)}_{k, i_k}$ with $1\leq i_k<k$ and $i_k\neq i'_k$ using integration by parts and the Fay identity. Similarly, any kernel $g^{(0)}_{k,i_k}=1$ is simply denoted by $g^{(0)}_{k,1}=1$.

\subsection{Generating function for iterated integrals
\texorpdfstring{$\Gt$}{Gt} and the KZB associator}
\label{ssec:KZBassociator}

Before writing down a differential equation of KZB type for a vector of genus-one Selberg integrals in \subsecref{ssec:KZBeq} below, which is the genus-one generalization of the KZ equation~\eqref{eqn:KZexample}, let us consider its formal solution\footnote{As for the KZ equation, we are rather interested in relating a certain regularized boundary value to another regularized boundary value using an associator equation, than completely solving the equation. A rigorous discussion on solutions of the elliptic KZB equation can e.g.\ be found in \rcite{Felder:1995iv}} in terms of the so-called (elliptic) KZB associator, originally described\footnote{KZB equations are the higher-genus generalization of the KZ equation \cite{Bernard:1987df,Bernard:1988yv}. In this article, we exclusively consider the elliptic KZB equation and the $A$-cycle component of the elliptic KZB associator. Therefore, we simply refer to these genus-one objects as KZB equation and KZB associator, respectively, while the genus-zero analogues are called KZ equation and Drinfeld associator.} in \rcite{Enriquez:EllAss}. In fact, although usually represented in a language using a derivation algebra, we would like to point out that the equation as well as its formal solution is very naturally expressed in terms of the canonical iterated integrals $\Gt$ on the annulus.

By following exactly the same line of arguments as in \appref{app:KZassociator}, let us start from a generating function~\footnote{For this subsection, we explicitly denote the $\tau$-dependence of the functions in order to keep track of the analytic behavior of certain limits. For example, in the asymptotic behavior shown in \eqns{eqn:genusOneAsymptBehaviour0}{eqn:genusOneAsymptBehaviour1}, the right-hand side is $\tau$-independent.}
\begin{align}\label{ellKZ:LE}
\LL^\El(z)&=\sum_{w\in X} w \Gt_w(z,\tau)
\end{align}
of elliptic multiple polylogarithms $\Gt_w(z,\tau)$, which can be shown to satisfy the differential equation 
\begin{align}\label{ellKZ:ODE}
\frac{\partial}{\partial z} \LL^\El(z)&=\sum_{n\geq 0}  g^{(n)}(z,\tau)x^{(n)}\LL^\El(z)\,.
\end{align}
This differential equation is known as the Knizhnik--Zamolodchikov--Bernard equation (or KZB equation, for short) \cite{Bernard:1987df,Bernard:1988yv}. As for the genus-zero case, the asymptotic behavior around $z=0$ is determined by the asymptotics of the iterated integrals in \eqns{ellKZ:zeroValue}{ellKZ:zeroValueRegularized} which amounts to
\begin{equation}
\label{eqn:genusOneAsymptBehaviour0}
\LL^\El(z)\sim \exp\left(x^{(1)} \Gt(\begin{smallmatrix}1\\0 \end{smallmatrix}; z,\tau) \right)\sim(-2 \pi i z)^{x^{(1)}}\,.
\end{equation}
Due to the one-periodicity \eqref{eqn:1periodicityKernel} of the integration kernels $g^{(n)}$, the KZB equation is invariant under $z\mapsto z-1$ and, hence, there is another solution of the differential \eqn{ellKZ:ODE}, $\LL_1^\El(z)$, with the following asymptotics near $z=1$
\begin{equation}
\label{eqn:genusOneAsymptBehaviour1}
\LL_1^\El(z)\sim \exp\left(x^{(1)} \Gt(\begin{smallmatrix}1\\0 \end{smallmatrix};z,\tau) \right)\sim (-2\pi i (1-z))^{x^{(1)}}\,.
\end{equation}
As for the genus-zero case, the associator
\begin{align}
\Phi^\El&=\big(\LL_1^\El(z)\big)^{-1}\LL^\El(z)
\end{align}
is independent of $z$, which can be verified straightforwardly by taking the derivative of both sides of $L_1^\El\, \Phi^\El = L^\El$ and using the differential \eqn{ellKZ:ODE}. Thus, the elliptic associator $\Phi^\El$ can be expressed in the limit $z\rightarrow 1$, which yields the generating series of regularized eMZVs
\begin{align}\label{ellKZ:PhiE_eMZV}
\Phi^\El&= \lim_{z\rightarrow 1}\exp\left(-x^{(1)} \Gt(\begin{smallmatrix}1\\0 \end{smallmatrix};z,\tau) \right)\LL^\El(z)\nnl
&=\sum_{w\in X}w\, \omega(w^t;\tau)\,.
\end{align}
The last equation follows from definition \eqref{ellKZ:LE} and the cancellation of the divergent integrals due to the exponential prefactor in \eqn{ellKZ:PhiE_eMZV}. This is exactly the same mechanism which lead to the expression of the Drinfeld associator in terms of the regularized multiple zeta values in \eqn{KZ:PhiMZV} and effectively implements the appropriate regularization. Considering letters up to $x^{(2)}$ only, the first couple of terms of the KZB associator read
\begin{align*}
	\Phi^\El&= 1 + x^{(0)}\omega(0;\tau) + x^{(1)}\omega(1;\tau) + x^{(2)}\omega(2;\tau) + \nnl
		    &\quad+x^{(0)}x^{(0)}\omega(0,0;\tau)+x^{(0)}x^{(1)}\omega(1,0;\tau)+x^{(0)}x^{(2)}\omega(2,0;\tau)\nnl
		    &\quad+x^{(1)}x^{(0)}\omega(0,1;\tau)+x^{(1)}x^{(1)}\omega(1,1;\tau)+x^{(1)}x^{(2)}\omega(2,1;\tau)\nnl
		    &\quad+x^{(2)}x^{(0)}\omega(0,2;\tau)+x^{(2)}x^{(1)}\omega(1,2;\tau)+x^{(2)}x^{(2)}\omega(2,2;\tau)+\cdots\nnl
		    &=1 + x^{(0)} - 2 \zeta_2 x^{(2)} \nnl
		    &\quad+ \frac{1}{2} x^{(0)}x^{(0)} - (x^{(0)}x^{(1)}-x^{(1)}x^{(0)}) \omega(0, 1;\tau)  - \zeta_2 (x^{(0)}x^{(2)} + x^{(2)}x^{(0)}) \nnl
		    &\quad+ \big(x^{(1)}x^{(2)}-x^{(2)}x^{(1)}\big)\big(\omega(0,3;\tau)-2\zeta_2\omega(0,1;\tau)\big)
		    + 5 \zeta_4 x^{(2)}x^{(2)} + \cdots
\end{align*}
The elliptic associator $\Phi^\El$ provides an associator equation similar to \eqn{eqn:genusZeroAssociatorEq} at genus zero: it connects the regularized boundary values of an arbitrary solution $\FF^\El(z)$ of the KZB equation
\begin{align}
	\label{eqn:KZBFE}
\frac{\partial}{\partial z} \FF^\El(z)&=\sum_{n\geq 0}  g^{(n)}(z,\tau)x^{(n)}\FF^\El(z)\,,
\end{align}
which are regularized in order to compensate the asymptotic behavior shown in \eqns{eqn:genusOneAsymptBehaviour0}{eqn:genusOneAsymptBehaviour1}
\begin{equation}
\label{eqn:genus1BoundaryValues}
C_0^\El=\lim_{z\rightarrow 0}(-2 \pi i z)^{-x^{(1)}}\FF^\El(z),\,\,C_1^\El=\lim_{z\rightarrow 1}(-2 \pi i (1{-}z))^{-x^{(1)}}\FF^\El(z)\,.
\end{equation}
The calculation is similar to the genus-zero case (cf.~\eqn{DrinfeldC0C1}) and the result is the genus-one associator equation
\begin{align}
\label{eqn:genusOneAssociatorEq}
\Phi^\El C_0^\El&=\lim_{z\rightarrow 0}(\LL_1^\El(z))^{-1}\LL^\El(z) (-2 \pi i z)^{-x^{(1)}}\FF^\El(z)\nnl
&=\lim_{z\rightarrow 1}(\LL_1^\El(z))^{-1} \FF^\El(z)\nnl
&=C^\El_1\,.
\end{align}

\subsection{KZB equation for an auxiliary point}
\label{ssec:KZBeq}

The one-loop version of the recursive construction of open-string amplitudes will again facilitate a differential equation for an extra marked point: the point $z_2$, which is the variable parametrizing the integration domain of the integrals in \eqn{eqn:relevantGenusOneSelberg}. The relevant case for the calculation of the $(L{-}1)$-point genus-one configuration\hyp{}space integrals below is $k{=}2$.  The configuration\hyp{}space integrals to be calculated are contained in boundary values corresponding to limits of the variable $z_2$:
\begin{itemize}
	\item in the limit $z_2\to 1=z_1\text{ mod } \ZZ$, the integration domain closes and amounts to one complete boundary of the annulus: it leads to $(L{-}1)$-point genus-one configuration\hyp{}space integrals with integrands defined on $ \CF^{\tau}_{L-1,1}$.
	\item when taking $z_2\to 0=z_1$, genus-one Selberg integrals degenerate to tree-level string corrections, since the integration domain gets confined to a genus-zero domain and the resulting integrands are defined on $ \CF_{L+1,3}$.
\end{itemize}
The two associated boundary values can be related by the genus-one associator equation \eqref{eqn:genusOneAssociatorEq} providing the genus-one analogue of the amplitude recursion of \rcite{Broedel:2013aza}. After establishing the KZB equation in this subsection, the boundary values will be discussed in \subsecref{subsec:BoundaryValues} below. 

Consider a vector of Selberg integrals with fixed upper labels, but
lower labels stretching over all admissible values:
\begin{equation}
	\Selbld^\El_{(n_{k+1},\ldots,n_L)}=\colvec{\SIE{n_{k+1},\ldots,n_L}{1,\ldots,1}(z_1,...,z_{k})\\\vdots\\\SIE{n_{k+1},\ldots,n_L}{k,\ldots,L-1}(z_1,...,z_k)}\,.
\end{equation}
For the three-point example to be evaluated below, we have to consider integrals with $k=2, L=4$, such that we are going to work with vectors like
\begin{equation}
  \Selbld^\El_{(2,1)}=\colvec{
\SIE{2,1}{1,1}(z_1,z_2)\\
\SIE{2,1}{1,2}(z_1,z_2)\\
\SIE{2,1}{1,3}(z_1,z_2)\\
\SIE{2,1}{2,1}(z_1,z_2)\\
\SIE{2,1}{2,2}(z_1,z_2)\\
\SIE{2,1}{2,3}(z_1,z_2)}\,. 
\end{equation}
The entries are going to be ordered canonically. As agreed on in the discussion of the spanning set $\mathcal{B}^\El_{i_3',i_4',\dots,i_L'}$ defined in \eqn{eqn:basisGenusOne}, whenever there is an $n_k=0$, we write $i_k=1$ and we generally do not incorporate integration by parts identities to reduce the number of independent integrals, that is, we usually work with the set of integrals $\mathcal{B}^\El_{0,0,\dots,0}$. Accordingly, if none of the labels $n_{3},\ldots,n_L$ is zero, the vector $\Selbld^\El_{(n_3,\ldots,n_L)}$ has $(L{-}1)!$ components. 

In establishing the KZB equation for a vector of Selberg integrals, we are going to take derivatives of $\Selbld^\El_{(n_3,\ldots,n_L)}(z_1=0,z_2)$ with respect to the auxiliary point $z_2$. As will be pointed out below, taking a derivative of a Selberg integral of weight $w$ will lead to a combination of genus-one Selberg integrals of weights between zero and $w+1$. Accordingly, we collect all Selberg vectors of weight $w$ into a larger
vector $\SelbldEw(z_2)$: 
\begin{equation}
	\SelbldEw(z_2)=\colvec{\Selbld^\El_{(n_3,n_4,\dots,n_L)}(z_1=0,z_2)}_{ \sum_{k=3}^L n_k=w}\,.
\end{equation}
and combine all those vectors $\SelbldEw(z_2)$ into an infinitely large vector in order of increasing $w$: 
\begin{equation}
	\SelbldE(z_2)=\colvec{
		\Selbld^\El_0 \\
		\Selbld^\El_1 \\
		\Selbld^\El_2\\
		\vdots
	}\,.
\end{equation}
The vector $\SelbldE(z_2)$ is the genus-one analogue of the genus-zero Selberg vector $\Selbld(x_3)$ defined in \eqn{eqn:defSeL}, which satisfies the KZ \eqn{eqn:KZexample}. 
\begin{theorem}{(Elliptic KZB-system)}
	\label{thm:maintheorem}
	Let $\SelbldE(z_2)$ be the vector of genus-one Selberg integrals of type $(2,L)$ with
	auxiliary point $z_2$. The derivative with respect to the auxiliary point $z_2$ can be written in the form 
	\begin{align}
		\label{eqn:KZBz2}
		\frac{\pd}{\pd z_{2}} \SelbldE(z_2)
 	&=\sum_{n\geq 0}g^{(n)}_{21}x^{(n)} \SelbldE(z_2)\,,
	\end{align}
	which is a system of elliptic KZB-type. The non-vanishing entries of the matrices $x^{(n)}$ are $\ZZ$-linear combinations of the parameters $s_{ij}$.
\end{theorem}
\begin{proof}
	The proof is constructive, that is, the derivative of any entry of $\SelbldE(z_2)$ is explicitly brought in a form to fit \eqn{eqn:KZBz2} by a combinatorial algorithm. The algorithm consists of two parts: in the first part the expression is rewritten in a way such that the derivative acts on the Selberg seed exclusively, which can be evaluated straighforwardly. This is achieved using integration by parts. In the second part, Fay identities are used iteratively in order to rewrite the result as linear combination of admissible Selberg integrals. The coefficients are shown to consist of polynomials of degree one in the parameters $s_{ij}$ and a factor $g_{21}^{(n)}$.

Due to the length of describing the combinatorial algorithm, we refrain from providing the proof in the main text, rather we would like to ask the reader to consult \appref{app:KZBgeneral} instead. 
\end{proof}
\noindent\textbf{Example:} In order to illustrate the mechanism, let us consider the $z_2$-derivative of the Selberg vector $\Selbld^\El_{(0,1)}(z_1=0,z_2)$: 
\begin{small}
\begin{align}
\label{eqn:exampleGenusOneDerivative}
&\frac{\pd}{\pd z_2}\Selbld^\El_{(0,1)}\nnl
&\quad=g^{(0)}_{21}\!
\setlength{\arraycolsep}{1.4pt}
\begin{pmatrix}
-s_{24}&-s_{24}&0&0&0&0&-s_{23}&0&0&0&0\\
s_{14}&s_{14}\!+\!s_{34}&s_{34}&0&0&0&0&-s_{23}\!-\!s_{34}&s_{34}&0&s_{34}\\
0&-s_{24}&\!-s_{24}&0&0&0&0&s_{24}&-s_{23}\!-\!s_{24}&0&-s_{24}
\end{pmatrix}\!\!
\colvec{\Selbld^\El_{(0,2)} \\\Selbld^\El_{(1,1)} \\\Selbld^\El_{(2,0)}}\nnl
&\quad\phantom{=}+g^{(1)}_{21}\begin{pmatrix}
s_{12}\!+\!s_{24}&-s_{24}&0\\
-s_{14}&s_{12}\!+\!s_{14}&0\\
0&0&s_{12}
\end{pmatrix}
\Selbld^\El_{(0,1)}+g^{(2)}_{21}\begin{pmatrix}
-s_{24}\\s_{14}\\0
\end{pmatrix} \Selbld^\El_{(0,0)}\,.
\end{align}
\end{small}\\[-4pt]
An immediate observation is in place: considering the weight of the derivative to be one, taking the weight $n$ of each function $g_{21}^{(n)}$ into account and adding the weight of the genus-one Selberg integrals, the total weight is conserved in each term of the above equation.  

Starting from the above equation, one can collect all occurring vectors $\Selbld^\El_{(n_3,n_4)}$ of weight two into the weight-two-vector  
\begin{equation}
	\label{eq:SewExample}
	\Selbld^\El_2=\colvec{\Selbld^\El_{(0,2)} \\\Selbld^\El_{(1,1)} \\\Selbld^\El_{(2,0)}}\,,
\end{equation}
where the three subvectors are given by\\
\begin{minipage}[t]{0.5\textwidth}
\begin{align*}
	\Selbld^\El_{(0,2)}&=\colvec{
	\SIE{0,2}{1,1}(z_1,z_2)\\
	\SIE{0,2}{1,2}(z_1,z_2)\\
\SIE{0,2}{1,3}(z_1,z_2)} \nnl[13pt]
	\Selbld^\El_{(2,0)}&=\colvec{
	\SIE{2,0}{1,1}(z_1,z_2)\\
	\SIE{2,0}{2,1}(z_1,z_2)}
\end{align*}
\end{minipage}
\begin{minipage}[t]{0.5\textwidth}
\begin{equation}
	\Selbld^\El_{(1,1)}=\colvec{
	\SIE{1,1}{1,1}(z_1,z_2)\\
	\SIE{1,1}{1,2}(z_1,z_2)\\
	\SIE{1,1}{1,3}(z_1,z_2)\\
	\SIE{1,1}{2,1}(z_1,z_2)\\
	\SIE{1,1}{2,2}(z_1,z_2)\\
	\SIE{1,1}{2,3}(z_1,z_2)}\,.
\end{equation}
\end{minipage}
So the 11-component vector $\Selbld^\El_2$ captures the combinatorics from distributing weight two on the two slots $(n_3,n_4)$ as well as the combinatorics of the labels $i_k$ for each of those pairs. Neatly, the particular ordering does not play a role in the formalism to be described, however, we will follow the sorting convention in \eqn{eq:SewExample}.\\[6pt]
\textbf{Block structure of coefficient matrices $\boldsymbol{x^{(n)}}$:} Let us further investigate the structure of the differential system \eqref{eqn:KZBz2} and in particular the matrices $x^{(n)}$. As visible in example \eqref{eqn:exampleGenusOneDerivative}, taking a derivative of a Selberg integral will increase the weight by one. Using the algorithm described in \appref{app:KZBgeneral}, one can thus write the $z_2$-derivative on $\Selbld^\El_{w}$ as 
\begin{align}
\label{eqn:KZBz2Weight}
\frac{\pd}{\pd z_{2}} \Selbld^\El_{w}(z_2)
&=\sum_{n=0}^{w+1}g^{(n)}_{21}x_w^{(n)} \Selbld^\El_{w+1-n}(z_2)\,,
\end{align}
where the factor $x^{(n)}$ does not contribute to the weight, but $g_{21}^{(n)}$ does. From counting the weights, one can thus deduce that the matrices $x^{(n)}$ in \eqn{eqn:KZBz2} ought to be block-(off-)diagonal, where the size of the blocks corresponds the lengths of the Selberg vectors of weight $w$.  Schematically, one finds
\begin{small}
\begin{align}
\label{eqn:blockstructure}
\sbox1{$\setlength{\arraycolsep}{1.3pt}\begin{array}{c}\rowcolor{MidnightBlue!20}\phantom{0}\end{array}$}
\sbox2{$\setlength{\arraycolsep}{1.3pt}\begin{array}{cc}\rowcolor{MidnightBlue!20}\phantom{0}&\phantom{0}\\\rowcolor{MidnightBlue!20}\phantom{0}&\phantom{0}\end{array}$}
\sbox3{$\setlength{\arraycolsep}{1.3pt}\begin{array}{cccc}
		\rowcolor{MidnightBlue!20}\phantom{0}&\phantom{0}&\phantom{0}\\
		\rowcolor{MidnightBlue!20}\phantom{0}&\phantom{0}&\phantom{0}\\
		\rowcolor{MidnightBlue!20}\phantom{0}&\phantom{0}&\phantom{0}
\end{array}$}
\sbox4{$\setlength{\arraycolsep}{1.3pt}\begin{array}{cccc}
		\rowcolor{MidnightBlue!20}\phantom{0}&\phantom{0}&\phantom{0}&\phantom{0}\\
		\rowcolor{MidnightBlue!20}\phantom{0}&\phantom{0}&\phantom{0}&\phantom{0}\\
		\rowcolor{MidnightBlue!20}\phantom{0}&\phantom{0}&\phantom{0}&\phantom{0}\\
		\rowcolor{MidnightBlue!20}\phantom{0}&\phantom{0}&\phantom{0}&\phantom{0}
\end{array}$}
\sbox5{$\begin{array}{l}\phantom{\hspace{-12pt}}\end{array}$}
\sbox6{$\begin{array}{l}\phantom{\hspace{-12pt}}\\\phantom{\hspace{-12pt}}\end{array}$}
\sbox7{$\begin{array}{l}\phantom{\hspace{-12pt}}\\\phantom{\hspace{-12pt}}\\\phantom{\hspace{-12pt}}\end{array}$}
\sbox8{$\begin{array}{l}\phantom{\hspace{-12pt}}\\\phantom{\hspace{-12pt}}\\\phantom{\hspace{-12pt}}\\\phantom{\hspace{-12pt}}\end{array}$}
\arraycolsep=0pt
\frac{\pd }{\pd z_2} \SelbldE= g_{21}^{(0)} x^{(0)} \SelbldE + g_{21}^{(1)}
\underbrace{
\left(
\setlength{\arraycolsep}{0pt}
\begin{array}{c|c|c|c|c}
\usebox{1}&&&&\\
\hline
  &\usebox{2}&&&\\
\hline
  &&\usebox{3}&&\\
\hline
  &&&\usebox{4}&\\\hline
  &&&&\hspace{4.2pt}\ddots
\end{array}
\right)
}_{x^{(1)}}
\underbrace{
\left(
\begin{array}{c}
	\usebox{5}\Selbld^\El_0(z_2) \\\hline
	\usebox{6}\Selbld^\El_1(z_2) \\\hline
	\usebox{7}\Selbld^\El_2(z_2) \\\hline
	\usebox{8}\Selbld^\El_3(z_2) \\\hline
	\vdots
\end{array}
\right)}_{\SelbldE}
+g_{21}^{(2)} x^{(2)} \SelbldE + \cdots\,,
\end{align}
\end{small}\\
where only the blue blocks are non-vanishing. Given the blocks in the above equation, the other matrices will have the following structure:
\begin{align}\label{eqn:xnBlockForm}
  %
  \def\arraystretch{1.5}
  x^{(0)}&=\left(
  \setlength{\arraycolsep}{2pt}
  \begin{array}{c|c|c|c|c}
	  \phantom{x_0^{(0)}}&\cb x_0^{(0)}&&&\\
  \hline
  &&\cb x_1^{(0)}&&\\
  \hline
  &&&\cb x_2^{(0)}&\\
  \hline
  &&&&\cb x_3^{(0)}\\\hline
  &&&&\hspace{4.2pt}\ddots
  \end{array}
  \right)\,,
  \qquad
  x^{(1)}=\left(
  \setlength{\arraycolsep}{2pt}
  \begin{array}{c|c|c|c|c}
  \cb x_0^{(1)}&&&&\\
  \hline
    &\cb x_1^{(1)}&&&\\
  \hline
    &&\cb x_2^{(1)}&&\\
  \hline
    &&&\cb x_3^{(1)}&\\\hline
    &&&&\hspace{4.2pt}\ddots
  \end{array}
  \right)\,,\nnl
  x^{(2)}&=\left(
  \setlength{\arraycolsep}{2pt}
  \begin{array}{c|c|c|c|c}
  \phantom{\usebox{1}}&&&&\\\hline
  \cb x_1^{(2)}&&&&\\
  \hline
    &\cb x_2^{(2)}&&&\\
  \hline
    &&\cb x_3^{(2)}&&\\
  \hline
    &&&\cb x_3^{(2)}&\hspace{4.2pt}\ddots
  \end{array}
  \right)\,,\qquad \dots\,,
\end{align}
where the blocks of the individual matrices are labeled by $x^{(n)}_w$.\\[6pt]
\textbf{Truncation of the KZB system.} In practice, the infinitely long vector $\SelbldE(z_2)$ and the matrices $x^{(n)}$ of infinite dimension need to be truncated at a certain maximal total weight $\wmax$
\begin{equation}
\Selbld^\El_{\leq\wmax}(z_2)=
\colvec{
	\Selbld^\El_0 \\
	\Selbld^\El_1 \\
	\vdots\\
	\Selbld^\El_{\wmax}\\
}\,.
\end{equation}
Taking the $z_2$-derivative on the finite-length vector
$\Selbld^\El_{\leq\wmax}(z_2)$ leads to the differential equation
\begin{align}
\label{eqn:KZBz2MaxWeight}
\frac{\pd}{\pd z_{2}} \Selbld^\El_{\leq \wmax}(z_2)
&=\sum_{n=0}^{\wmax+1}g^{(n)}_{21}x^{(n)}_{\leq \wmax} \Selbld^\El_{\leq \wmax}(z_2) + r_{\wmax}\Selbld^\El_{\wmax+1}(z_2)\,,
\end{align}
where the remainder $r_{\wmax}$ prevents \eqn{eqn:KZBz2MaxWeight} to be a complete KZB equation.  However, as will be discussed below, this remainder may be disregarded when calculating one-loop configuration\hyp{}space integrals up to a particular order in $\ap$. 
 
The matrices $x^{(n)}_{\leq \wmax}$ for $0\leq n\leq \wmax{+}1$ correspond to the upper-left $(\wmax{+}1)\times(\wmax{+}1)$ block matrices of the matrices $x^{(n)}$. Explicitly:
\begin{align}\label{eqn:xnBlockFormwmax}
  %
  \def\arraystretch{1.5}
  x_{\leq\wmax}^{(0)}\!\!\!\!=\left(
  \setlength{\arraycolsep}{0.7pt}
  \begin{array}{c|c|c|c|c}
	  \phantom{x_0^{(0)}}&\cb x_0^{(0)}&&&\\
  \hline
  &&\cb x_1^{(0)}&&\\
  \hline
  &&&\cb \,\ddots\,\,\,&\\
  \hline
  &&&&\cb x_{\wmax-1}^{(0)}\\\hline
  &&&&\hspace{4.2pt}
  \end{array}
  \right),
  \,\,
  x_{\leq\wmax}^{(1)}\!\!\!\!=\left(
  \setlength{\arraycolsep}{0.7pt}
  \begin{array}{c|c|c|c|c}
  \cb x_0^{(1)}&&&&\\
  \hline
    &\cb x_1^{(1)}&&&\\
  \hline
    &&\cb x_2^{(1)}&&\\
  \hline
    &&&\cb \,\ddots\,\,\,&\\\hline
    &&&& \cb x_{\wmax}^{(1)}
  \end{array}
  \right),
\end{align}
\begin{align}
  x_{\leq\wmax}^{(2)}=\left(
  \setlength{\arraycolsep}{1pt}
  \begin{array}{c|c|c|c|c}
  \phantom{\usebox{1}}&&&&\\\hline
  \cb x_1^{(2)}&&&&\\
  \hline
    &\cb x_2^{(2)}&&&\\
  \hline
    &&\cb \,\ddots\,\,&&\\
  \hline
    &&&\cb x_\wmax^{(2)}&\phantom{x}
  \end{array}
  \right)\,,\ldots,
  x_{\leq\wmax}^{(\wmax+1)}=\left(
  \setlength{\arraycolsep}{1pt}
  \begin{array}{c|c|c|c|c}
  \phantom{x_1^{(2)}}&\phantom{x}&\phantom{x}&\phantom{x}&\phantom{x}\\\hline
  &&&&\\
  \hline
  &&&&\\
  \hline
  &&&&\\
  \hline
  \cb x_\wmax^{(\wmax+1)}&&&&
  \end{array}
  \right)\,.
\end{align}
Moreover, the remainder $r_{\wmax}$ is the $(\wmax{+}1){\times}1$
block submatrix of the first $\wmax+1$ blocks of the $(\wmax+2)$-column of the
matrix $x^{(0)}_{\leq \wmax+1}$:
\begin{align}\label{eqn:rnBlockFormwmax}
  %
  \arraycolsep=2pt
  \def\arraystretch{1.5}
  x_{\leq\wmax+1}^{(0)}=\left(
  \begin{array}{c|c|c|c|c|c}
	  \phantom{x_0^{(0)}}&\cb x_0^{(0)}&&&&\\
  \hline
			     &&\cb x_1^{(0)}&&&\\
  \hline
			     &&&\cb \,\ddots\,\,\,&&\\
  \hline
			     &&&&\cb x_{\wmax-1}^{(0)}&\\
  \hline
			     &&&&&\cb x_{\wmax}^{(0)}\\
  \hline
			     &&&&&\hspace{4.2pt}
  \end{array}
  \right),\quad
  r_\wmax = 
  \left(\begin{array}{c}
	  \\\hline
	  \\\hline
	  \vdots\\\hline
	  \\\hline\cb x_\wmax^{(0)}
  \end{array}
  \right)\,.
\end{align}
The value of $\wmax$ necessary for the calculation of the configuration\hyp{}space integrals up to a particular order in $\ap$ is going to be determined in \subsecref{ssec:openstringcalc}.

\subsection{Boundary values for the KZB equation}
\label{subsec:BoundaryValues}
Having derived a (modified) KZB equation for the genus-one Selberg integrals, we would like to apply the genus-one associator equation \eqref{eqn:genusOneAssociatorEq} in order to evaluate genus-one configuration\hyp{}space integrals from genus-zero configuration\hyp{}space integrals. This amounts to proving the following proposition. 
\begin{proposition}
	\label{prop:bdryvalues}
	The regularized boundary values
	\begin{align}
		\bC_1^\El&=\lim_{z_2\rightarrow 1}(-2 \pi i (1{-}z_2))^{-x^{(1)}}\SelbldE(z_2)\text{ and }
	\bC_0^\El&=\lim_{z_2\rightarrow 0}(-2 \pi i z_2)^{-x^{(1)}}\SelbldE(z_2)
	\end{align}
	are related by the A-cycle component $\Phi(x^{(0)},x^{(1)},x^{(2)},...)$ of the KZB associator via
	\begin{equation}\label{eqn:assocEq}
	\bC_1^\El=\Phi(x^{(0)},x^{(1)},x^{(2)},...)\bC_0^\El\,.
	\end{equation}
	The regularized boundary value $\bC_1^\El$ contains $(L{-}1)$-point configuration\hyp{}space integrals at genus one whereas $\bC_0^\El$ contains $(L{+}1)$-point configuration\hyp{}space integrals at genus zero.
\end{proposition}
\begin{proof}
While the statement in \eqn{eqn:assocEq} follows straightforwardly from the discussion in \subsecref{ssec:KZBassociator} and the form of the KZB equation in theorem~\ref{thm:maintheorem}, the boundary values $\bC_0^\El$ and $\bC_1^\El$ will be explicitly constructed and shown to contain the respective configuration\hyp{}space integrals below. Following definition \eqn{eqn:genus1BoundaryValues}, we will have to evaluate
\begin{align}
\label{eqn:genus1SelbergBoundaryValues}
\bC_0^\El&=\lim_{z_2\rightarrow 0}(-2 \pi i z_2)^{-x^{(1)}}\SelbldE(z_2)\nnl
	 &=\lim_{z_2\rightarrow 0}\colvec{(-2 \pi i z_2)^{-x_0^{(1)}}\Selbld^\El_0(z_2)\\ (-2 \pi i z_2)^{-x_1^{(1)}}\Selbld^\El_1(z_2)\\\vdots}=\colvec{\bC_{0,0}^\El\\ \bC_{0,1}^\El\\\vdots},\nnl 
\bC_1^\El&=\lim_{z_2\rightarrow 1}(-2 \pi i (1-z_2))^{-x^{(1)}}\SelbldE(z_2)\nnl
	 &=\lim_{z_2\rightarrow 1}\colvec{(-2 \pi i (1-z_2))^{-x_0^{(1)}}\Selbld^\El_0(z_2)\\ (-2 \pi i (1-z_2))^{-x_1^{(1)}}\Selbld^\El_1(z_2)\\\vdots}=\colvec{\bC_{1,0}^\El\\ \bC_{1,1}^\El\\\vdots},
\end{align}
where $\bC^\El_{0,w}$ and $\bC^\El_{1,w}$ denote the regularized limits of the subvectors $\Selbld^\El_w(z_2)$ of weight $w$ and the second equality in the above equations follows from the block-diagonal form of the matrix $x^{(1)}$. Switching to finite matrix size, we define 
\begin{align}
\label{eqn:genus1SelbergFiniteBoundaryValues}
\bC_{0,\leq \wmax}^\El&=\lim_{z_2\rightarrow 0}(-2 \pi i z_2)^{-x_{\leq \wmax}^{(1)}}\Selbld^\El_{\leq \wmax}(z_2)\,,\nnl
 \bC_{1,\leq \wmax}^\El&=\lim_{z_2\rightarrow 1}(-2 \pi i (1-z_2))^{-x_{\leq \wmax}^{(1)}}\Selbld^\El_{\leq \wmax}(z_2)\,.
\end{align}
\textbf{Boundary value $\bC^\El_1$:} Considering the limit $z_2\to 1$, we first determine the behavior of the component integrals $\SIE{n_{3},\dots,n_L}{i_{3},\dots,i_L}(z_2)$ and include the regularization factor $-2 \pi i (1-z_2)^{-x^{(1)}}$ afterwards.
\begin{equation}\label{genus1z2to1}
	\mpostuse{genusonez2to1}
\end{equation}
According to \eqn{eqn:asympGammaz1} and using \eqref{eqn:InvarianceGamma1}, the genus-one Selberg seed degenerates as
\begin{align}\label{eqn:C1StringAmplitudes}
	\lim_{z_2\to 1}(-2\pi i (1-z_2))^{-s_{12}}\SelE
&=
\hspace{-0.4cm}\prod_{0=z_1< z_i< z_j< z_2}\hspace{-0.4cm}\exp\left(s_{ij}\Gt_{ji}\right)\prod_{j> 2}\exp\left(\left(s_{1j}+s_{2j}\right)\Gt_{j1}\right)\nnl
&= \SelE\Big|_{(L{-}1)\text{-point}}^{\tilde{s}_{ij}=s_{ij}+\delta_{i1}s_{2j}}\,.
\end{align}
The term on the right-hand side of the above equation is the genus-one Selberg seed for $L{-}1$ insertion points on the boundary of the annulus. Since the insertion points $z_2$ and $z_1$ merge in the limit, effective Mandelstam variables $\tilde{s}_{ij}=s_{ij}+\delta_{i1}s_{2j}$ for $i,j\in \{1,\dots,L\}\setminus \{2\}$, $i<j$ have to be assigned to the insertion points, such that $\tilde{s}_{1j}$ associated to $z_1$ includes the contribution from $z_2$:
\begin{equation}
  \tilde{s}_{1j}=s_{1j}+s_{2j}\,.
\end{equation}
The behavior is the same as in the genus-zero case in \eqn{eqn:C1limitSel}: the momentum of the external state which corresponds to one of the fixed insertion points receives two contributions, one coming from the state at $z_1=0$ and the other from a state at the same position of the boundary of the annulus
$z_2=z_1\mod \ZZ$ due to the merged auxiliary insertion point $z_2$. 

Accordingly, the genus-one Selberg integral defined in
\eqn{eqn:relevantGenusOneSelberg} as a function on the configuration space of the $A$-cycle with
two positions fixed, degenerates at lowest order in $(1-z_2)$ to an integral with integrand defined on $\mathcal{F}^{\tau}_{L-1,1}$
\begin{align}\label{eqn:genus1SelbergLimit1}
&\lim_{z_2\to 1}(-2 \pi i (1-z_2))^{-s_{12}}\SIE{n_{3},\dots,n_L}{i_{3},\dots,i_L}(z_1=0,z_2)\nnl
&=\int_{\mathcal{C}(z_2\to 1)}\prod_{i=3}^L dz_i\, \SelE\Big|_{(L{-}1)\text{-point}}^{\tilde{s}_{1j}=s_{1j}+s_{2j}} \prod_{k=3}^L g^{(n_k)}_{k,i_k}\big|_{z_2\equiv z_1=0}\nnl
&=\SIE{n_{3},\dots,n_L}{i_{3},\dots,i_L}(0,z_2=1)\Big|^{\tilde{s}_{1j}=s_{1j}+s_{2j}}_{z_2\equiv z_1=0}\,.
\end{align}
Similar to \eqn{eqn:e1B222}, the relevant eigenvalues of the matrices $x^{(1)}_w$ appearing in the regularization factor $-2 \pi i (1-z_2)^{-x^{(1)}_w}$ in $\bC_1^\El$ from \eqn{eqn:genus1SelbergBoundaryValues} are $s_{12}$, such that the regularization factor contributes the factor $(-2 \pi i (1-z_2))^{-s_{12}}$ from \eqn{eqn:genus1SelbergLimit1} and the entries of $\bC_{1}^\El$ are given by the degenerate genus-one Selberg integrals in \eqn{eqn:genus1SelbergLimit1}. They are the $(L{-}1)$-point genus-one configuration\hyp{}space integrals \cite{Broedel:2014vla}.  This concludes the proof of the statement about the boundary value $\bC_{1}^\El$ in proposition~\ref{prop:bdryvalues}. In \rcite{Broedel:2020tmd} this and the analogous statement for the boundary value $\bC^\El_0$ below are proven using generating series of the open-string configuration-space integrals.\\[4pt]

\noindent\textbf{Boundary value $\bC^\El_0$:} The boundary value $\bC^\El_0$ is obtained by confining the region of integration to an infinitesimal interval as $z_2\to 0=z_1$. As for the genus-zero calculation in \eqn{eqn:C0entriesCalculation}, the main tool to investigate this degeneration and the corresponding behavior of genus-one Selberg integrals is a change of variables $z_i=z_2 x_i$, where $x_i$ are points in the unit interval on the real line whereas $z_i$ are located on the boundary of an annulus.
\begin{equation}
	\phantom{mm}\mpostuse{genusonez2to0}
\end{equation}
According to the discussion below definition \ref{def:SelbergInt} in \subsecref{ssec:genSelberg} and as a consequence of \eqn{eqn:asympGammaz0}, the genus-one Selberg seed $\SelE$ degenerates at lowest order in $z_2$ for $z_2\to 0$ up to a proportionality factor to the genus-zero Selberg seed $\Sel$ for the $L$ points $0=x_1<x_L<x_{L-1}<\dots<x_2=1$ on the unit interval, which is (cf.~\eqn{eqn:SandKN}) precisely the $(L{+}1)$-point genus-zero Koba--Nielsen factor defined in \eqn{eqn:KN}:
\begin{align}
\label{eqn:z20Limit}
\lim_{z_2\to 0}(-2 \pi i z_2)^{-s_{12\dots L}}\SelE
&=\prod_{0=x_1< x_i< x_j< x_2=1}x_{ji}^{s_{ij}}\nnl
&=\Sel|_{L\text{-point}}\nnl
&=\KN|_{(L+1)\text{-point}}\,.
\end{align}

The discussion of the eigenvalues of $x^{(1)}_w$ is similar to the genus-zero case and thoroughly addressed in \rcite{Broedel:2020tmd}. It turns out that the maximal and therefore dominant eigenvalue of $x^{(1)}_w$ is $s_{12\dots L}$, such that the regularization $(-2 \pi i z_2)^{-x^{(1)}_w}$ contributes the factor $(-2 \pi i z_2)^{-s_{12\dots L}}$ in \eqn{eqn:z20Limit}. Thus, the entries of $\bC_{0}^\El$ are given by
\begin{align}\label{eqn:genus1SelbergLimit0}
&\lim_{z_2\to 0}(-2 \pi i z_2)^{-s_{12\dots L}}\SIE{n_{3},\dots,n_L}{i_{3},\dots,i_L}(z_1=0,z_2)\nnl
&=\lim_{z_2\to 0}z_2^{L-2}\int_{\mathcal{C}(x_2=1)}\prod_{i=3}^L dx_i\, \Sel|_{L\text{-point}} \prod_{k=3}^L g^{(n_k)}(z_2 x_{k, i_k},\tau)\nnl
&=\begin{cases}
\int_{\mathcal{C}(x_2=1)}\prod_{i=3}^L dx_i\, \Sel|_{L\text{-point}} \prod_{k=3}^L \frac{1}{x_{k i_k}}&\text{if } n_1=n_2=\dots =n_k=1\,,\\
0&\text{otherwise}\,.
\end{cases}
\end{align}
The only non-vanishing entries are the ones for which all integration kernels have weight one, i.e.\ $n_k=1$, since only their pole can compensate for the $z_2^{L-2}$ factor from the measure.

A similar behavior was observed for the genus-zero boundary value which led to \eqn{eqn:x3To0Selberg}. Moreover, these simple poles ensure that the only non-vanishing integrals are exactly the degenerate genus-zero Selberg integrals $\SI{i_3,i_4,\ldots,i_\snol}(0,1,x_2=1)$ found in the genus-zero regularized boundary values $\bC_0$ and $\bC_1$ in \eqns{eqn:x3To0Selberg}{eqn:C1limitSel}, respectively. However, here we recover integrals with integrands defined on $\CF_{L+1,3}$ with the $L{+}1$ insertion points $0=x_1<x_L<x_{L+1}<\dots <x_2=1<x_{L+1}=\infty$ (cf.~\eqn{eqn:genus0ordering}). As discussed in \subsecref{subsec:Relate}, these integrals are related to the $(L{+}1)$-point genus-zero configuration\hyp{}space integrals by a basis transformation. This concludes the proof of proposition~\ref{prop:bdryvalues}. 
\end{proof}

\noindent\textbf{Mandelstam variables:} In contrast to both the genus-zero discussion and the limiting situation $\bC_1^\El$, in the boundary value $\bC_0^\El$ the Mandelstam variables $s_{2j}$ in \eqn{eqn:z20Limit} associated to the auxiliary insertion point $z_2$ are not redundant: the auxiliary genus-one momentum $k^\oneloop_2$ associated to $z_2$ encodes the genus-zero momentum $k^\tree_2$ associated to the tree-level insertion point $x_2$ 
\begin{equation}
k^\oneloop_2=k^\tree_2\,.
\end{equation}
In order to keep track of how this momentum contributes to the one-loop momenta, two distinct processes have to be considered: first, the topological change by the identification of $x_1$ with $x_{L+1}$ giving the genus-one insertion point $z_1$ depicted in figure \eqref{genus0genus1} and second, the merging of $z_2\to 1 =z_1\text{ mod }\ZZ$ shown in figure  \eqref{genus1z2to1}. In the first case, the momenta $k^\tree_1$ and $k^\tree_{L+1}$ associated to $x_1$ and $x_{L+1}$, respectively, yield the joint contribution to the one-loop momentum associated to $z_1$
\begin{equation}
k^\oneloop_1=k^\tree_1+k^\tree_{L+1}\,.
\end{equation}
The second limit is the merging of $z_2$ to $z_1$, which adds the momentum $k^\oneloop_2$ associated to $z_2$ to the momentum $k^\oneloop_1$ and we expect to find the effective momentum
\begin{equation}
\tilde{k}^\oneloop_1=k^\oneloop_1+k^\oneloop_2=k^\tree_1+k^\tree_{L+1}+k^\tree_{2}
\end{equation}
for the insertion point $z_1=z_2\text{ mod }\ZZ$ of the $(L{-}1)$-point one-loop interaction in the regularized boundary value $\bC_1^\El$, where we denote the one-loop momenta $k^\oneloop_i$ in the limit $z_2\to z_1=1\text{ mod }\ZZ$ by a tilde as depicted in figure \eqref{eqn:momentumrelation}. However, from our calculations of $\bC_1^\El$ in \eqn{eqn:C1StringAmplitudes} we see that the Mandelstam variables associated to $\tilde{k}^\oneloop_1$ are
\begin{equation}\label{eqn:splitMandelstam1}
\tilde{s}_{1j}=s_{1j}+s_{2j}\,.
\end{equation}
Therefore, the actual one-loop momentum associated to $z_1=z_2\text{ mod }\ZZ$ turns out to be
\begin{equation}\label{eqn:splitMom1}
\tilde{k}^\oneloop_1=k^\tree_1+k^\tree_{2}\,.
\end{equation}
This is in agreement with simultaneous momentum conservation in the tree-level and one-loop interaction if and only if
\begin{equation}\label{eqn:kL1Tree}
k^\tree_{L+1}=0\,,
\end{equation}
which can be interpreted as follows: reversing the first procedure discussed above (see figure \eqref{genus0genus1}), that is, going in the direction from genus one to genus zero, the momentum $k_1^\oneloop$ associated to $z_1$ is expected to split in a certain way and to contribute to the two tree-level momenta $k^\tree_1$ and $k^\tree_{L+1}$ accordingly. From \eqn{eqn:kL1Tree} follows that these two contributions are very unequal: while the momentum associated to $x_1$ obtains the full contribution $k^\tree_1=k_1^\oneloop$, the momentum associated to $x_{L+1}$ goes away empty-handed $k^\tree_{L+1}=0$. Note that the momenta associated the remaining tree-level insertion points $x_i$ for $i=3,4,\dots,L$ are exactly the one-loop momenta associated to the punctures $z_i$ for any $0<z_2\leq 1=z_1\text{ mod }\ZZ$:
\begin{equation}
\tilde{k}_i^\oneloop=k_i^\oneloop=k_i^\tree\quad \text{for }i=3,4,\dots,L\,.
\end{equation}
\begin{equation}
	\label{eqn:momentumrelation}
	\mpostuse[width=12.6cm]{momentumrelation}
\end{equation}

\noindent\textbf{Summary of subsection:} The regularized boundary value $\bC_0^\El$ is found to only have finitely many non-vanishing entries which are degenerate genus-zero Selberg integrals and hence linear combinations of $(N{+}2)$-point tree-level configuration\hyp{}space integrals, where $N=L-1$.  In turn, as will be discussed in detail in the next subsection, the entries of $\bC_1^\El$ given by \eqn{eqn:genus1SelbergLimit1} contain the $N$-point one-loop configuration\hyp{}space integrals. 

Therefore, the genus-one Selberg vector $\SelbldE(z_2)$ indeed interpolates between the genus-zero and genus-one configuration\hyp{}space integrals and the corresponding associator equation  
\begin{equation}\label{eqn:genus1AssociatorEquation}
 \bC_1^\El=\Phi^\El\, \bC_0^\El
\end{equation}
provides a link between genus-zero and genus-one integrals. Combining this associator equation with the genus-zero recursion for the genus-zero configuration\hyp{}space integrals from \rcite{Broedel:2013aza,AKtbp} based on the genus-zero associator equation \eqref{eqn:genusZeroAssociatorEqIntro} yields a recursion in genus and the number of external states to calculate the genus-one configuration\hyp{}space integrals.

The consideration about the contributions of the insertion points defining the genus-one Selberg integrals to the Mandelstam variables in the configuration\hyp{}space integrals appearing in the boundary values $\bC_0^\El$ and $\bC_1^\El$ leads to a geometric interpretation of the associator \eqn{eqn:genus1AssociatorEquation}: the $N$-point one-loop worldsheet is obtained from the $(N{+}2)$-point tree-level worldsheet by an effective gluing of the two legs corresponding to the insertion points $x_1=0$ and $x_{L+1}=\infty$ on the Riemann sphere. By momentum conservation the Mandelstam variables associated to the insertion point $z_1$ in the one-loop configuration\hyp{}space integrals of $\bC_1^\El$ are given by the sum $\tilde{s}_{1j}=s_{1j}+s_{2j}$.

\subsection{Open-string amplitudes at genus one}
\label{ssec:openstringcalc}

Let us finally employ the associator \eqn{eqn:genus1AssociatorEquation} to calculate the $\ap$-expansion of $N$-point one-loop con\-figu\-ra\-tion-space integrals \cite{Green:1982sw, Dolan:2007eh, Broedel:2014vla} in open string theory up to any desired order in $\ap$ from $(N{+}2)$-point tree-level configuration\hyp{}space integrals. 

While relating various entries of the regularized boundary values to known representations of configuaration-space integrals at genus zero and genus one, we will simultaneously single out the relevant parts of the matrix equation \eqref{eqn:genus1AssociatorEquation}. 

The main goal is the calculation of the $N$-point one-loop configuration\hyp{}space integral up to a desired maximal order in $\ap$ denoted by $o^\oneloop_\mmax$.
As observed in the previous subsection, integrals with integrands defined on $\CF^{\tau}_{N,1}$ for the $N$-point one-loop configuration\hyp{}space integrals arise in the $z_2\to1$ limit of genus-one Selberg integrals with $L=N{+}1$ marked points. Simultaneously, $(N{+}2)$-point tree-level configuration\hyp{}space integrals are encoded in the \mbox{$z_2\to 0$} limit of the same genus-one Selberg integrals.   

As pointed out at the end of \subsecref{ssec:KZBeq} above, for practical calculations the infinite genus-one Selberg vector has to be truncated to $\Selbld^\El_{\leq \wmax}(z_2)$. Given the target values $N$ and $o^\oneloop_\mmax$ for the calculation, let us determine $\wmax$ as well as various other parameters for the calculation. 

Each of the objects on the right-hand side in \eqn{eqn:genus1AssociatorEquation} has an expansion in the parameter $\ap$: since $x^{(n)}\propto \ap$ (cf.~\eqn{eqn:Mandelstam}, the expansion in word length of the elliptic KZB associator is exactly its $\ap$-expansion. The $\ap$-expansion of the tree-level integrals in $\bC_0^E$ can be obtained from the recursions in \rcites{Mafra:2016mcc,Broedel:2013aza}. Therefore, the maximal target $\ap$-order $o^\oneloop_\mmax$ of the one-loop configuration\hyp{}space integrals on the right-hand side is reached, when the KZB-associator is expanded up to $\ap$-order
\begin{equation}\label{eqn:lmax}
  l_\mmax=o^\oneloop_\mmax-o^\tree_\mmin
\end{equation}
where $o^\tree_\mmin$ denotes the leading (e.g.~minimal) order in the $\ap$-expansion of tree-level integrals in $\bC_0^\El$. This order turns out to be given by \cite{Mafra:2011nw}
\begin{equation}
o^\tree_\mmin=2-L=3-N\,.
\end{equation}

In order to determine $\wmax$, we need to think about the positions of the relevant information within the vectors $\bC^\El_0$ and $\bC_1^\El$: on the one hand, according to \eqn{eqn:genus1SelbergLimit0} the non-vanishing subvector of $\bC_0^\El$ which includes the tree-level configuration\hyp{}space integrals is contained in the weight 
\begin{equation}
	\label{eqn:w0}
w_0=L-2
\end{equation}
subvector $\bC_{0,w_0}^\El$ of $\bC_0^\El$. On the other hand, the one-loop configuration\hyp{}space integrals are contained in the weight
\begin{equation}
	\label{eqn:w1}
w_1=L-5-d
\end{equation}
subvector $\bC_{1,w_1}^\El$. The quantity $d$ denotes the number of additional factors of $g^{(n)}$ appearing in higher-point one-loop configuration\hyp{}space integrals and is given by $d=0$ for $L\leq 8$ and $d\geq 0$ otherwise \cite{Broedel:2014vla}.  For all calculations in this article, $d=0$ holds. The relevant part of the elliptic KZB associator is the submatrix $\Phi^\El_{w_1,w_0}$, which satisfies the equation
\begin{equation}\label{eqn:truncatedassociator}
\bC_{1,w_1}^\El=\Phi^\El_{w_1,w_0} \bC_{0,w_0}^\El\,.
\end{equation}
Since for all amplitude situations we find $w_1<w_0$, the submatrix $\Phi^\El_{w_1,w_0}$ is located above the diagonal of $\Phi^\El$. 

Here comes the block-(off-)diagonal form of the matrices $x^{(n)}$ depicted in \eqref{eqn:xnBlockForm} into play, which ensures that for a certain word length $l$, only finitely many words $w=x^{(n_1)}\dots x^{(n_l)}$ contribute non-trivially to $\Phi^\El_{w_1,w_0}$. The non-trivial contribution to $\Phi^\El_{w_1,w_0}$ at each word length $l$, which is the order $l$ in the $\ap$-expansion of the associator since $x^{(n)}\propto \ap$, is a finite sum $\sum_w  w \,\omega(w^t)$ of products $w=x_{\leq \wmax(l)}^{(n_1)}x_{\leq \wmax(l)}^{(n_2)}\dots x_{\leq \wmax(l)}^{(n_l)}$, where
	\begin{equation}\label{eqn:wmax}
	\wmax(l)=\max(l+w_1-w_0,w_0)
	\end{equation}
	and $(n_1,n_2,\dots,n_l)$ is a length-$l$, ordered partition of $\wmax(l)$, i.e.\
	$n_1+n_2+\dots+n_l=\wmax(l)$, which satisfies for each $r\in \{1,2,\dots,l\}$ the
	additional conditions
	\begin{equation}\label{eqn:cond}
	0\leq i-\sum_{s=1}^{r-1}(n_s-1)\leq \wmax(l)\,,\qquad 0\leq j+n_l-1\leq \wmax(l)\,.
	\end{equation}
Therefore, the $\ap$-expansion of $\Phi^\El_{w_1,w_0}$ up to some
	maximal order $l_\mmax$ in $\ap$ or maximal word length,
	respectively, can be calculated by finite-dimensional submatrices of $x^{(n)}$,
	which are the matrices $x^{(n)}_{\leq w_\text{max}}$ for the maximal weight $\wmax(l_\mmax)$:
	\begin{equation}
	\Phi^\El(x^{(n)})_{w_1,w_0}=\Phi^\El(x_{\leq \wmax(l_\mmax)}^{(n)})_{w_1,w_0}+\CO\Big((\ap)^{l_\mmax+1}\Big)\,.
	\end{equation}
In other words, the associator submatrix $\Phi^\El(x^{(n)})_{w_1,w_0}$ can be deduced from a truncated associator, which is determined by evaluating the matrix products of truncated representations of letters, taking only words up to length $l_\mmax$ and weight $\wmax$ into account. The truncated matrix representations $x^{(n)}_{\leq \wmax}$ of the letters can be obtained from the modified KZB \eqn{eqn:KZBz2MaxWeight}. Since word length $l_\mmax$ and maximal weight $\wmax(l_\mmax)$ are finite quantities, all sums consist of a finite number of terms and all matrices are of finite size. The process yields the finite-dimensional, truncated associator equation 
\begin{equation}\label{eqn:finiteAssociatorEqn}
\bC^\El_{1,\leq \wmax(l_\mmax)}+\CO\Big((\ap)^{o^\oneloop_\mmax+1}\Big)=\Phi^\El_{l_\mmax}(x^{(n)}_{\leq \wmax(l_\mmax)})\, \bC^\El_{0,\leq \wmax(l_\mmax)}\,,
\end{equation}
where $\Phi^\El_{l_\mmax}$ is the truncation of $\Phi^\El$ at the maximal word length $l_\mmax$. The finite subvectors
\begin{align}
\bC^\El_{0,\leq \wmax(l_\mmax)}&=\lim_{z_2\to 0}(-2 \pi i z_2)^{-x^{(1)}_{\leq \wmax(l_\mmax)}} \Selbld^\El_{\leq \wmax(l_\mmax)}(z_2)\,,\nnl
\bC^\El_{1,\leq \wmax(l_\mmax)}&=\lim_{z_2\to 1}(-2 \pi i (1-z_2))^{-x^{(1)}_{\leq \wmax(l_\mmax)}}\Selbld^\El_{\leq \wmax(l_\mmax)}(z_2)
\end{align}
of $\bC^\El_0$ and $\bC^\El_1$, respectively, contain the $(L{+}1)$-point tree-level string corrections at weight $w_0=L-2\leq\wmax(l_\mmax)$ and the $(L{-}1)$-point one-loop corrections at $w_1=L-5-d$. Denoting by $\Phi^\El_{l_\mmax}(x^{(n)}_{\leq \wmax(l_\mmax)})_{w_1,w_0}$ the weight-$(w_1,w_0)$ submatrix of the truncated KZB associator $\Phi^\El_{l_\mmax}\big(x^{(n)}_{\leq \wmax(l_\mmax)}\big)$, the relevant truncated vector equation which relates the string corrections to each other is 
\begin{equation}\label{eqn:finiteRelevantAssociatorEqn}
\bC^\El_{1,w_1}+\CO\Big((\ap)^{o^\oneloop_\mmax+1}\Big)=\Phi^\El_{l_\mmax}(x^{(n)}_{\leq \wmax(l_\mmax)})_{w_1,w_0}\, \bC^\El_{0,w_0}\,.
\end{equation}
where $\Phi^\El_{l_\mmax}(x^{(n)}_{\leq \wmax(l_\mmax)})_{w_1,w_0}$ is the weight-$(w_1,w_0)$ submatrix of the truncated elliptic KZB associator $\Phi^\El_{l_\mmax}(x^{(n)}_{\leq \wmax(l_\mmax)})$.

\subsection{Relating boundary values at genus zero and one}
\label{ssec:relation}

In this section, we briefly discuss how the regularized boundary value $C_0^\El$ of a function satisfying a KZB equation is related to a corresponding genus-zero limit $C_0$ of a solution of a KZ equation. This provides an explanation of why in the recursion described in the previous section, genus-zero configuration\hyp{}space integrals are obtained from genus-one Selberg integrals.

Before we focus on genus-one quantities, we determine the origin of the regularization used for the regularized genus-zero boundary value 
\begin{equation}
C_0=\lim_{x\to 0}x^{-e_0}F(x)
\end{equation}
of a solution $F(x)$ of the KZ equation 
\begin{equation}\label{eqn:KZ}
\frac{d}{dx}F(x)=\left(\frac{e_0}{x}+\frac{e_1}{x-1}\right)F(x)\,.
\end{equation}
For $0<x\ll1$, the KZ equation can be written as
\begin{equation}
\label{eqn:KZcloseTo0}
\frac{d}{d x}F(x)=\left(\frac{e_0}{x}-e_1+\CO(x)\right)F(x)
\end{equation}
up to linear order in $x$. Using this differential equation and the fact that $[e_0,e^{x e_1}]=\CO(x)$, the function $F(x)$ can be approximated by
\begin{equation}
\label{eqn:Fapproxi}
F(x)=e^{-x e_1} x^{e_0}f_0+\CO(x)
\end{equation}
for some constant $f_0$ in a neighborhood of zero. The regularization in $C_0$ ensures that this constant is exactly the regularized boundary value
\begin{equation}
C_0=f_0\,.
\end{equation}

The genus-one calculation can be carried out analogously, which naturally leads to a close relation to the constant $f_0$. For a function $F^\El(z)$ satisfying the KZB equation
\begin{align}\label{eqn:ellKZ}
\frac{\partial}{\partial z} F^\El(z)&=\sum_{n\geq 0}  g^{(n)}(z,\tau)x^{(n)}F^\El(z)\,,
\end{align}
letting $0<z\ll1$ leads to a similar situation as above: from the $q$-expansion of the integration kernels $g^{(n)}(z,\tau)$, we find that \cite{Broedel:2014vla}
\begin{equation}
g^{(n)}(z,\tau)=\begin{cases}
1&\text{if } n=0\,,\\
\frac{1}{z}+\CO(z)&\text{if }n=1\,,\\

-2\zeta_{2m}-2\frac{(-2 \pi i)^{2m}}{(2m-1)!}\sum\limits_{k,l>0}l^{2m-1}q^{kl}+\CO(z^2)&\text{if }n=2 m>0\,,\\
\CO(z)&\text{if }n=2m{+}1>1\,.
\end{cases}
\end{equation}
Therefore, we can assemble the even generators $x^{(2m)}$ and the corresponding order-zero prefactors into
\begin{equation}
x^{(e)}(\tau)=x^{(0)}-2\sum_{m>0}\left(\zeta_{2m}+\frac{(-2 \pi i)^{2m}}{(2m-1)!}\sum_{k,l>0}l^{2m-1}q^{kl}\right)x^{(2m)}
\end{equation}
in order to write the KZB \eqn{eqn:ellKZ} as
\begin{equation}
\frac{d}{d z}F^\El(z)=\left(\frac{x^{(1)}}{z}+x^{(e)}(\tau)+\CO(z)\right)F^\El(z)\,.
\end{equation}
This is a differential equation of the form \eqref{eqn:KZcloseTo0} of the KZ equation in the same regime. In other words, for small $z$, the operator
\begin{equation}
\nabla^{\text{KZB}}\big(x^{(n)}\big)=\sum_{n\geq 0}g^{(n)}(z) x^{(n)}
\end{equation}
on the right-hand side in the KZB equation \eqref{eqn:ellKZ} degenerates to the operator
\begin{equation}
\nabla^{\text{KZ}}(e_0,e_1)=\frac{e_0}{z}+\frac{e_1}{z-1}
\end{equation}
in the KZ equation \eqref{eqn:KZ} with $e_0=x^{(1)}$ and $e_1=x^{(e)}$:
\begin{equation}
\nabla^{\text{KZB}}\big(x^{(n)}\big)= \nabla^{\text{KZ}}\big(x^{(1)},x^{(e)}\big)+\CO(z)\,.
\end{equation}
Thus, as before for $F(x)$, the function $F^\El(z)$ can be approximated by 
\begin{equation}
\label{eqn:FEapproxi}
F^\El(z)=e^{z x^{(e)}(\tau)}z^{x^{(1)}} f^\El_0+\CO(z)\,,
\end{equation}
where $f^\El_0$ is some constant. Note that a similar degeneration to the genus-zero framework occurs for the generating series $L^\El(z)$ of elliptic multiple polylogarithms defined in \eqn{ellKZ:LE}: according to \eqn{eqn:genusOneAsymptBehaviour0}, for $e_0=x^{(1)}$ the series has at lowest order the same behavior as the generating series $L(z)$ of the multiple polylogarithms
\begin{equation}
L^\El(z)=(-2 \pi i z)^{x^{(1)}}(1+\CO(z))=(-2 \pi i )^{x^{(1)}}L(z)|_{e_0=x^{(1)}}(1+\CO(z))\,.
\end{equation}

We can conclude that the regularized boundary value
\begin{equation}
C_0^\El=\lim_{z\to 0}z^{-x^{(1)}}F^\El(z)=f_0^\El
\end{equation}
is indeed independent of $\tau$ and, upon comparing \eqn{eqn:Fapproxi} with \eqn{eqn:FEapproxi}, it is proportional (up to a constant matrix) to the corresponding genus-zero boundary value $C_0=f_0$ for a function $F(x)$ satisfying a KZ equation with $e_0=x^{(1)}$
\begin{equation}
C_0^\El=\lim_{z\to 0}z^{-x^{(1)}}F^\El(z)=f_0^\El\propto f_0=C_0\,.
\end{equation}
Note that in the case of matrix Lie algebras with $e_0\neq x^{(1)}$, but they have the same maximal eigenvalue, then the above argument modifies slightly but still applies analogously such that the elements of $C_0^\El$ turn out to be some linear combinations of the elements of $C_0$, which is exactly the situation observed in the recursion described in the previous section.

\section{Examples}\label{sec:examples}
\subsection{Example: two points}\label{ssec:2point}
As a first example, let us calculate the two-point one-loop string correction up to order $o^\oneloop_\mmax=2$ in $\ap$. While all essential steps are noted in this subsection, several lengthy details are outsourced to \appref{app:2point}.  The two-loop correction is non-trivial only, if the Mandelstam variables $s_{ij}$ are treated as independent parameters of the integrals, which do not satisfy any constraints like momentum conservation. The two-point configuration\hyp{}space integral reads \cite{Mafra:2019xms}
\begin{equation}\label{eqn:2loopCorrection}
S^\oneloop_{2\text{-point}}(\tilde{s}_{13})=\int_0^1 dz_3 \exp\left(\tilde{s}_{13} \Gt_{31}\right)=\sum_{n\geq 0}\tilde{s}_{13}^n\,\omega(\underbrace{1,\dots,1}_{n},0)\,,
\end{equation}
where $\tilde{s}_{13}$ is the Mandelstam variable associated to the loop momentum. Since the integral requires two vertex insertion points, the appropriate genus-one Selberg integral with an extra insertion point $z_2$ is of  length $L=3$ and the insertion points are ordered as 
\begin{equation}
  0=z_1<z_3<z_2<1=z_1\text{ mod }\ZZ
\end{equation}
on the boundary of the annulus. Indeed, in the limit $z_2\to 1$, the punctures $z_2$ and $z_1$ merge, leaving
the two punctures relevant for the one-loop string corrections. Thus, we
consider the iterated integrals
\begin{align}
	\label{eqn:twopointintegrals}
	\SIE{n_3}{i_3}(0,z_2)&=\int_0^{z_2} dz_3 \Sel^\El g^{(n_3)}_{3 i_3}\,,\quad 1\leq i_3 <3\,,\nnl
	\Sel^\El&=\exp\left(s_{13}\Gt_{31}+s_{12}\Gt_{21}+s_{23}\Gt_{23}\right)\,.
\end{align}
According to \eqn{eqn:w1}, the two-point one-loop correction can be found in the weight $w_1=0$ entry $\bC_{1,w_1}^\El$, while the tree-level correction resides at weight $w_0=1$ (cf.~\eqn{eqn:w0}). The $\ap$-expansion of the four-point tree-level correction turns out to start at order $o_\mmin^\tree=-1$, (cf.~\eqn{eqn:2pointC0}).  Therefore, consulting \eqns{eqn:lmax}{eqn:wmax}, it is sufficient to consider the truncated Selberg vector at maximal weight $\wmax=2$ to calculate the one-loop string corrections up to second order in $\ap$, i.e.~we only need to consider the vector
\begin{equation}
\Selbld^\El_{\leq 2}(z_2)=\begin{pmatrix}
\SIE{0}{1}(0,z_2)\\\SIE{1}{1}(0,z_2)\\ \SIE{2}{1}(0,z_2)\\ \SIE{2}{2}(0,z_2)
\end{pmatrix}\,
\end{equation}
where we use the reduced set of integrals $\mathcal{B}^\El_2$ obtained from the relations 
\begin{align}
	\SIE{0}{1}(0,z_2)&=\SIE{0}{2}(0,z_2)\,,\nnl s_{13}\SIE{1}{1}(0,z_2)&=-s_{23}\SIE{1}{2}(0,z_2)
\end{align}
to exclude the integrals $\SIE{0}{2}(0,z_2)$ and $\SIE{1}{2}(0,z_2)$ from our analysis.

Before we can explicitly check that the regularized boundary values indeed reproduce the tree-level and one-loop string corrections and apply the associator \eqn{eqn:finiteRelevantAssociatorEqn}, we have to determine the matrices $x^{(0)}_{\leq 2}$, $x^{(1)}_{\leq 2}$ and $x^{(2)}_{\leq 2}$ appearing in the modified KZB equation satisfied by $\Selbld^\El_{\leq 2}(z_2)$. Following the general algorithm in \appref{app:KZBgeneral} and performing the corresponding calculations shown in \appref{app:2point}, the partial differential equation can indeed be written in the form \eqref{eqn:KZBz2MaxWeight}:
\begin{equation}\label{sec:ex2:deq}
\frac{\partial}{\partial z_2}\Selbld^\El_{\leq 2}(z_2)=\left(g^{(0)}_{21}x^{(0)}_{\leq 2}+g^{(1)}_{21}x^{(1)}_{\leq 2}
+g^{(2)}_{21}x^{(2)}_{\leq 2}\right)
\Selbld^\El_{\leq 2}(z_2)+r_2 \Selbld^\El_{3}(z_2)\,,
\end{equation}
where $\Selbld^\El_{3}(z_2)=\left( \SIE{3}{1}(0,z_2), \SIE{3}{2}(0,z_2)\right)^T$ and the matrices are given by
\begin{equation}\label{eqn:2ptExamplex01}
x^{(0)}_{\leq 2}=\begin{pmatrix}
0&s_{13}&0&0\\
0&0&-s_{23}&-s_{23}\\
0&0&0&0\\
0&0&0&0
\end{pmatrix},\,\, x^{(1)}_{\leq 2}=\begin{pmatrix}
s_{12}&0&0&0\\
0&s_{123}&0&0\\
0&0&s_{12}+s_{23}&-s_{23}\\
0&0&-s_{13}&s_{12}+s_{13}
\end{pmatrix}
\end{equation}
and
\begin{align}\label{eqn:2ptExamplex2}
x^{(2)}_{\leq 2}&=\begin{pmatrix}
0&0&0&0\\
-s_{23}&0&0&0\\
0&s_{13}&0&0
\\
0&s_{13}&0&0
\end{pmatrix}\,,\qquad r_2 =\begin{pmatrix}
0&0\\
0&0\\
-2 s_{23}&- s_{23}
\\
-s_{13}&2 s_{13}
\end{pmatrix}\,. 
\end{align}
Now, we can evaluate the relevant entries of the regularized boundary values $\bC_{0,w_0=1}^\El$ and $\bC_{1,w_1=0}^\El$ explicitly: the latter involves the weight $w_1=0$ eigenvalue $x^{(1)}_{0}=s_{12}$ of $x^{(1)}_{\leq 2}$ in the regularization factor $(-2 \pi i (1-z_2))^{-x^{(1)}_{\leq 2}}$, which leads to the boundary value
\begin{equation}\label{sec.ex2:CE1Expansion}
\bC_{1,0}^\El=\lim_{z_2\rightarrow 1}(-2 \pi i (1-z_2))^{-s_{12}}\SIE{0}{1}(0,z_2)=S^\oneloop_{2\text{-point}}(\tilde{s}_{13})\,,
\end{equation}
given by the one-loop string correction $S^\oneloop_{2\text{-point}}(\tilde{s}_{13})$ with effective Mandelstam variable 
\begin{equation}
\tilde{s}_{13}=s_{13}+s_{23}\,,
\end{equation}
which is in agreement with our general considerations in
\eqn{eqn:C1StringAmplitudes}. On the other hand, the relevant eigenvalue of
$x^{(1)}_{\leq 2}$ for the boundary value $\bC_{0,1}^\El$ is
$x^{(1)}_1=s_{123}$, such that
\begin{align}\label{eqn:2pointC0}
\bC_{0,1}^\El&=\lim_{z_2\rightarrow 0} (-2 \pi i z_2)^{-s_{123}}\SIE{1}{1}(0,z_2)=\frac{1}{s_{13}}\frac{\Gamma(1+s_{13})\Gamma(1+s_{23})}{\Gamma(1+s_{13}+s_{23})}
\end{align}
yields indeed the well-known Veneziano amplitude for the four-point amplitude of open strings at tree-level. Since each Mandelstam variable comes with a factor of $\ap$, we find the leading order to be $o_\mmin^\tree=-1$.

Since according to \eqn{eqn:lmax}, the maximal order in $\ap$ or, equivalently, the maximal word length in the KZB associator is $l_\mmax=3$, the truncated associator \eqn{eqn:finiteAssociatorEqn} reads
\begin{equation}\label{eqn:assEqn2Point}
\begin{pmatrix}
S^\oneloop_{2\text{-point}}(\tilde{s}_{13})\\
\ast\\
\ast\\
\ast
\end{pmatrix}+\CO\Big((\ap)^{3}\Big)=\Phi_3^\El(x^{(n)}_{\leq 2})\, \begin{pmatrix}0\\
\frac{1}{s_{13}}\frac{\Gamma(1+s_{13})\Gamma(1+s_{23})}{\Gamma(1+s_{13}+s_{23})}\\
0\\
0
\end{pmatrix}\,.
\end{equation}
From the matrices given in \eqns{eqn:2ptExamplex01}{eqn:2ptExamplex2} and the truncation $\Phi^\El_{3}$ of the associator $\Phi^\El$ given by the generating series of eMZVs in \eqn{ellKZ:PhiE_eMZV}, we find that the only words contributing to the relevant $(w_1,w_0)=(0,1)$-submatrix $\Phi_3^\El(x^{(n)}_{\leq 2})_{0,1}$ are at
\begin{itemize}
	\item word length $1$: $x^{(0)}_{\leq 2}$
	\item word length $2$: the commutator
		\begin{align*}
		[x^{(1)}_{\leq 2},x^{(0)}_{\leq 2}]&=\begin{pmatrix}
		0&-s_{13}(s_{13}+s_{23})&0&0\\0&0&-2s_{13}s_{23}&-2s^2_{23}\\
		0&0&0&0\\
		0&0&0&0
		\end{pmatrix}
		\end{align*}
	\item word length $3$: the nested commutator
		\begin{align*}
		[x^{(1)}_{\leq 2},[x^{(1)}_{\leq 2},x^{(0)}_{\leq 2}]]&=
		\setlength{\arraycolsep}{1.2pt}
		\begin{pmatrix}
			0&s_{13}(s_{13}{+}s_{23})^2&0&0\\0&0&-2s_{13}s_{23}(s_{13}{+}s_{23})&2s_{23}^2(s_{13}{+}s_{23})\\
		0&0&0&0
		\\
		0&0&0&0
		\end{pmatrix}
		\end{align*}
		and the products
		\begin{align*}
		x^{(0)}_{\leq 2}x^{(0)}_{\leq 2}x^{(2)}_{\leq 2}&=
		\setlength{\arraycolsep}{2pt}
		\begin{pmatrix}
		0&-2s_{13}^2
		s_{23}&0&0\\
		0&0&0&0\\
		0&0&0&0\\
		0&0&0&0\end{pmatrix}\nnl
		x^{(0)}_{\leq 2}x^{(2)}_{\leq 2}x^{(0)}_{\leq 2}&=
		\setlength{\arraycolsep}{2pt}
		\begin{pmatrix}
		0&-s_{13}^2
		s_{23}&0&0\\
		0&0&2s_{13}
		s_{23}^2&2s_{13}
		s_{23}^2\\
		0&0&0&0\\
		0&0&0&0\end{pmatrix}\,.
		\end{align*}
\end{itemize}
The above list of contributions can be easily obtained from our general
analysis in \eqns{eqn:wmax}{eqn:cond}.

Evaluating all matrix products, the relevant $(w_1,w_0)$-submatrix of the
truncated KZB associator is explicitly given by the entry
\begin{align}
\Phi_3^\El\Big(x^{(n)}_{\leq 2}\Big)_{0,1}&=s_{13}\big( \omega(0)+(s_{13}+s_{23})\omega(1,0)+(s_{13}+s_{23})^2\omega(1,1,0)\nnl
&\phantom{=}-s_{13}s_{23}(\omega(0,2,0)+2\omega(2,0,0))\big)\,.
\end{align}
The $\ap$-expansion of the Veneziano amplitude can be obtained from the
identity  
\begin{align}
\frac{\Gamma(1+s_{13})\Gamma(1+s_{23})}{\Gamma(1+s_{13}+s_{23})}&=\exp\left(\sum_{n\geq 2}(-1)^n \frac{\zeta_n}{n}(s_{13}^n+s_{23}^n-(s_{13}+s_{23})^n)\right)\nnl
&=1-\zeta_2 s_{13}s_{23}+ \CO\Big((\ap)^3\Big)\,.
\end{align}
Using these two $\ap$-expansions, the right-hand side of the relevant part 
of the truncated associator \eqn{eqn:assEqn2Point} is given by 
\begin{align}\label{eqn:assEqn2PointRelevant}
&S^\oneloop_{2\text{-point}}(\tilde{s}_{13})+\CO\Big((\ap)^{3}\Big)\nnl
&\qquad\qquad=\Phi_3^\El\Big(x^{(n)}_{\leq 2}\Big)_{0,1} \frac{1}{s_{13}}\frac{\Gamma(1+s_{13})\Gamma(1+s_{23})}{\Gamma(1+s_{13}+s_{23})}\nnl
&\qquad\qquad=1+(s_{13}+s_{23})\omega(1,0)+(s_{13}+s_{23})^2\omega(1,1,0)+\CO\Big((\ap)^3\Big)\,,
\end{align}
where we have used the identity $
\omega(0,2,0)=-\zeta_2-2\omega(2,0,0)$ for the regularized eMZVs \cite{eMZVWebsite}. This reproduces indeed the two-point
one-loop string correction $S^\oneloop_{2\text{-point}}(\tilde{s}_{13})$ given in
\eqn{eqn:2loopCorrection} with the effective Mandelstam variable
$\tilde{s}_{13}=s_{13}+s_{23}$ up to second order in $\ap$. Simultaneously, this result
approves the validity of the (relevant part) of the truncated associator
\eqn{eqn:assEqn2PointRelevant}.

We have performed the calculation up to order $o_\mmax^\oneloop=4$ in
$\ap$.  In order to compare our result with the literature, in particular
with \rcite{Mafra:2019xms}, we translate our result into iterated integrals of
Eisenstein series\footnote{The conversion from the $\omega$-form of eMZVs to
their representation in terms of iterated integrals of Eisenstein series
$\gamma_0$ is thoroughly explained in \rcite{Broedel:2015hia}.} $\gamma_0$ and
use the one-loop open Green's function 
\begin{equation}
\label{eqn:CG}
\CG_{ij}=\Gt(\begin{smallmatrix}1\\0 \end{smallmatrix}; |z_{ij}|,\tau)+ \omega(0,1) 
\end{equation}
%
in the
definition \eqref{eqn:SelbergSeed} of the Selberg seed $\SelE$ and
in the one-loop string corrections
$S^\oneloop_{N\text{-point}}(\tilde{s}_{ij})$ rather than just $\Gt_{ij}$. The
additional term $\omega(0,1)$ vanishes in the sum $\sum_{i<j}
s_{ij}\left(\Gt(\begin{smallmatrix}1\\0 \end{smallmatrix}; |z_{ij}|,\tau)+\omega(0,1)\right)$
 if momentum conservation is imposed
and is, thus, physically irrelevant. Using these two adjustments, we find that
the relevant part of the right-hand side of the associator
\eqn{eqn:assEqn2Point} up to order $(\ap)^4$ is given by
\begin{align}\label{eqn:2pointEisenstein}
	S^\oneloop_{2\text{-point}}(\tilde{s}_{13})\Big|_{\CG_{ij}}\!\!&=1+\tilde{s}_{13}^2 \left(\frac{1}{4}  \zeta_2-3  \gamma_0(4,0)\right)\nnl
&\phantom{=}+ \tilde{s}_{13}^3\left(10  \gamma_0(6,0,0)-24 \zeta_2 \gamma_0(4,0,0)-\frac{1}{4}\zeta_3\right)\nnl
&\phantom{=}+\tilde{s}_{13} ^4 \Big(9  \gamma_0(4,0,4,0)-18  \gamma (4,4,0,0)-126  \gamma_0(8,0,0,0)\nnl
&\phantom{=bbbbb}-\frac{3}{4} \zeta_2 \gamma_0(4,0)-144  \zeta_4 \gamma_0(4,0,0,0)\nnl
&\phantom{=bbbbb}+240  \zeta_2 \gamma_0(6,0,0,0)+\frac{19}{64} 
\zeta_4\Big)+\CO\Big((\ap)^5\Big)\,.
\end{align}
Note that \eqns{eqn:assEqn2PointRelevant}{eqn:2pointEisenstein} show nicely on
a simple example, how using the associator \eqn{eqn:genus1AssociatorEquation}
relating the $(L{+}1)$-point tree-level to $(L{-}1)$-point one-loop string
corrections may geometrically be interpreted in terms of a gluing mechanism of
worldsheets as discussed at the end of \subsecref{subsec:BoundaryValues}:
starting with the four-point Veneziano amplitude, gluing together the external
legs of the string worldsheet which correspond to the two external states
labelled by the positions $x_1=0$ and $x_4=\infty$ on the Riemann
sphere yields a two-point genus-one worldsheet with punctures $z_1=z_2\text{ mod }\ZZ$ and $z_3$. The effective momentum propagating between $z_1=z_2\text{ mod }\ZZ$ and $z_3$ yields the Mandelstam variable
$\tilde{s}_{13}=s_{13}+s_{23}$ of the two-point one-loop
interaction.

\subsection{Example: three points}
The calculation for three points proceeds in analogy to the two-point example
without structural difficulties and complications. Naturally, the
dimensionality of the relevant matrices and vectors is larger, such that we do
not write them down explicitly but rather provide the results of the
computation.

The recursive algorithm requires one extra point on top of the three insertion
points present in three-point one-loop string correction integrals.
Correspondingly, we are going to consider the class of genus-one Selberg integrals
with $L=4$. The relevant integral is of the form 
\begin{equation}\label{eqn:3loopCorrection}
S^\oneloop_{3\text{-point}}(\tilde{s}_{ij})=\int_0^1 dz_3\int_0^{z_3}dz_4 \exp\left(\tilde{s}_{13} \Gt_{31}+\tilde{s}_{14} \Gt_{41}+\tilde{s}_{34} \Gt_{34}\right)\,,
\end{equation}
The above integral resides in the weight $w_1=0$ subvector of $\bC_1^\El$. We
are going to perform the calculation up to order $o^\oneloop_\mmax=3$ in
$\ap$. Since the corresponding five-point tree-level integrals start at
order $o^\tree_\mmin=-2$ and appear at weight $w_0=2$ in $\bC^\El_0$, the
required maximal weight for the truncation of the genus-one Selberg vector is
$\wmax=3$ according to \eqn{eqn:wmax}. The relevant finite-dimensional matrices
$x_{\leq 3}^{(n)}$ for $n=0,1,2,3$ are obtained from the algorithm in
\appref{app:KZBgeneral}, which leads to the modified KZB equation
\begin{align}
\frac{\pd}{\pd z_{2}} \Selbld^\El_{\leq 3}(z_2)
&=\sum_{n=0}^{4}g^{(n)}_{21}x^{(n)}_{\leq 3} \Selbld^\El_{\leq 3}(z_2) + r_{3}\Selbld^\El_{4}(z_2)\,.
\end{align}
Regularized boundary values can be calculated from the $x^{(1)}_{w_0=2}$ and
$x^{(1)}_{w_1=0}$ submatrices of $x^{(1)}_{\leq 3}$, which results in the
expected subvectors
\begin{equation}
\bC_{0,2}^\El=\lim_{z_2\to 0}(-2 \pi i z_2)^{-x^{(1)}_{2}} \Selbld^\El_2(z_2)=\begin{pmatrix}
0\\0\\0\\\SI{1,1}(0,1,x_2=1)\\\vdots\\\SI{2,3}(0,1,x_2=1)\\0\\0
\end{pmatrix}
\end{equation}
containing the five-point, genus-zero Selberg integrals for $z_2\to 0$ at weight $w_0=2$ and the three-point one-loop string correction for $z_2\to 1$ at weight $w_1=0$:
\begin{equation}
\bC_{1,0}^\El=\lim_{z_2\to 1}(-2 \pi i (1-z_2))^{-x^{(1)}_{0}} \Selbld^\El_0(z_2)=\begin{pmatrix}
S^\oneloop_{3\text{-point}}(\tilde{s}_{ij})
\end{pmatrix}
\end{equation}
with the effective Mandelstam variables
\begin{equation}
\tilde{s}_{1j}=s_{1j}+s_{2j}\,,\qquad \tilde{s}_{ij}=s_{ij}
\end{equation}
for $i,j\in
\{3,4\}$. The truncation of the KZB associator at $l_\mmax=5$ (cf.\
\eqn{eqn:lmax}), is required in order to use the finite associator
\eqn{eqn:finiteRelevantAssociatorEqn} 
\begin{equation}\label{eqn:associatorEq3Point}
\bC^\El_{1,0}+\CO\Big((\ap)^{4}\Big)=\Phi^\El_{5}\Big(x^{(n)}_{\leq 3}\Big)_{0,2}\, \bC^\El_{0,2}\,.
\end{equation}
The words contributing to the weight-$(0,2)$ submatrix $\Phi^\El_{5}(x^{(n)}_{\leq 3})_{0,2}$ of this truncation are determined with the mechanism described in \subsecref{ssec:openstringcalc}. The resulting $\ap$-expansion of the right-hand side of \eqn{eqn:associatorEq3Point} up to order $o^\oneloop_\mmax=3$ reads in terms of iterated integrals of Eisenstein series and the redefinition $\Gt_{ij}\mapsto \Gt(\begin{smallmatrix}1\\0 \end{smallmatrix}; |z_{ij}|,\tau) +\omega(0,1) = \CG_{ij}$
in the Selberg seed as follows:
\begin{align}
&S^\oneloop_{3\text{-point}}(\tilde{s}_{ij})\Big|_{\CG_{ij}}\nnl
&=\frac{1}{2}+\frac{1}{8} \left(\tilde{s}_{13}^2+\tilde{s}_{14}^2+\tilde{s}_{34}^2\right) \big(\zeta_2-12 \gamma_0(4,0)\big)\nnl
&\phantom{=}+\frac{1}{8}\Big(-\tilde{s}_{13} \tilde{s}_{34} \tilde{s}_{14} \big(-240 \gamma_0(6,0,0)+144
\zeta_2 \gamma_0(4,0,0)+\zeta_3\big)
\nnl
&\phantom{=+\frac{1}{8}\Big(}-\left(\tilde{s}_{13}^3+\tilde{s}_{14}^3+\tilde{s}_{34}^3\right) (-40
\gamma_0(6,0,0)+96 \zeta_2 \gamma_0(4,0,0)+\zeta_3)\Big)\nnl
&\phantom{=}+\CO\Big((\ap)^4\Big)\,,
\end{align}
which agrees with the known $\ap$-expansion of the three-point string correction.

\subsection{Example: four points}
\label{ssec:fourpoint}
If momentum conservation is imposed at the one-loop level, the first
non-trivial example is the four-point one-loop string correction. It is given
by the integral \cite{Green:1982sw} 
\begin{align}
S^\oneloop_{4\text{-point}}(\tilde{s}_{ij})&=\int_0^1 dz_3\int_{0}^{z_3} dz_4 \int_0^{z_4} dz_5 \prod_{0\leq z_i<z_j\leq z_3}\exp\left( \tilde{s}_{ij}\Gt_{ji}\right)\,,
\end{align}
where $i,j\in\{1,3,4,5\}$. The calculation of the $\ap$-expansion is
exactly the same as for the previous integrals: the one-loop integral is found
in the weight $w_1=0$ subvector of $\bC_1^\El$ and the six-point tree-level
integrals at the weight $w_0=3$ with $o^\tree_\mmin=-3$. Hence,
in order to obtain the expansion up to order
$o^\oneloop_\mmax=2$, the KZB associator can be truncated at the
maximal word length $l_{\mmax}=5$ and \eqn{eqn:wmax} requires the maximal
weight $\wmax=w_0=3$.  The matrices $x_{\leq 3}^{(n)}$ for $n=0,1,2,3$ are
obtained by forming the modified KZB \eqn{eqn:KZBz2MaxWeight}
\begin{align}
\frac{\pd}{\pd z_{2}} \Selbld^\El_{\leq 3}(z_2)
&=\sum_{n=0}^{4}g^{(n)}_{21}x^{(n)}_{\leq 3} \Selbld^\El_{\leq 3}(z_2) + r_{3}\Selbld^\El_{4}(z_2)\,.
\end{align}
As before, the subvectors of the regularized boundary values which contain the
six-point, three-level Selberg integrals for $z_2\to 0$ at weight $w_0=3$ and
the four-point one-loop string correction for $z_2\to 1$ at weight $w_1=0$ can
be calculated using the appropriate submatrices of $x^{(1)}_{\leq 3}$
and read 
\begin{equation}
\bC_{0,3}^\El=\lim_{z_2\to 0}(-2 \pi i z_2)^{-x^{(1)}_{3}} \Selbld^\El_3(z_2)=\begin{pmatrix}
0\\\vdots \\ 0\\\SI{1,1,1}(0,1,x_2=1)\\\vdots\\\SI{2,3,4}(0,1,x_2=1)\\0\\\vdots\\0
\end{pmatrix}
\end{equation}
and
\begin{equation}
\bC_{1,0}^\El=\lim_{z_2\to 1}(-2 \pi i (1-z_2))^{-x^{(1)}_{0}} \Selbld^\El_0(z_2)=\begin{pmatrix}
S^\oneloop_{4\text{-point}}(\tilde{s}_{ij})
\end{pmatrix}\,,
\end{equation}
respectively, with the effective Mandelstam variables
\begin{equation}
\tilde{s}_{1j}=s_{1j}+s_{2j}\,,\qquad \tilde{s}_{ij}=s_{ij}
\end{equation} 
for $i,j\in \{3,4,5\}$.  The truncated elliptic KZB associator at the maximal
length $l_\mmax=5$, with the contributing words calculated as usually,
leads to the finite associator \eqn{eqn:finiteRelevantAssociatorEqn} 
\begin{equation}\label{eqn:associatorEq4Point}
\bC^\El_{1,0}+\CO\Big((\ap)^{3}\Big)=\Phi^\El_{5}\Big(x^{(n)}_{\leq 3}\Big)_{0,3}\, \bC^\El_{0,3}\,.
\end{equation}
Expressed in terms of iterated integrals of Eisenstein series and using the redefinition $\Gt_{ij}\mapsto \Gt(\begin{smallmatrix}1\\0 \end{smallmatrix}; |z_{ij}|,\tau) +\omega(0,1) = \CG_{ij}$
\begin{align}
S^\oneloop_{4\text{-point}}(\tilde{s}_{ij})\Big|_{\CG_{ij}}&=\frac{1}{6}-\frac{\zeta (3) }{4 \pi
	^2}\left(\tilde{s}_{1,2}-2
\tilde{s}_{1,3}+\tilde{s}_{1,4}+\tilde{s}_{2,3}-2 \tilde{s}_{2,4}+\tilde{s}_{3,4}\right)\nnl
&\phantom{=}-6 \gamma_0(4,0,0) \left(\tilde{s}_{1,2}-2 \tilde{s}_{1,3}+\tilde{s}_{1,4}+\tilde{s}_{2,3}-2
\tilde{s}_{2,4}+\tilde{s}_{3,4}\right)\nnl
&\phantom{=}+\CO\Big((\ap)^2\Big)\,,
\end{align}
which agrees up to order $(\ap)^2$ with the $\ap$-expansion of the four-point configuration\hyp{}space integral.

\section{Summary and Outlook}
\label{sec:conclusion}

In this article, we have generalized the recursive formalism for the evaluation
of genus-zero Selberg integrals by Aomoto and Terasoma to genus one. After
establishing and discussing the genus-one formalism, we have put it to work to
evaluate several one-loop open-string scattering amplitudes.  

The original construction at genus zero is based on relating two boundary values of a Knizh\-nik-Za\-molod\-chi\-kov equation by the Drinfeld associator. The boundary values arise as two different limits of Selberg integrals and can be shown to contain integrals constituting the $N$-point and $(N{-}1)$-point open-string tree-level amplitudes respectively. Accordingly, the method allows to determine all tree-level string corrections at arbitrary order in $\ap$ recursively using a suitable representation of the Drinfeld associator. 

Our genus-one formalism is based on canonical generalizations of the above construction: at the heart there is now the elliptic Knizh\-nik-Za\-molod\-chi\-kov-Bernard equation, whose boundary values are related by the genus-one analogue of the Drinfeld associator, the elliptic KZB associator.  The boundary values arise as limits of genus-one Selberg integrals and can be shown to contain the one-loop $N$-point and the tree-level $(N{+}2)$-point open-string configuration\hyp{}space integrals. Thus all one-loop open-string corrections can be calculated using the elliptic associator equation \eqref{eqn:genus1AssociatorEquation} to any desired order in $\ap$. Our results obtained match the known expressions at multiplicity two, three and four.  

The original recursion at genus zero as well as our recursion at genus one have clear geometrical interpretations in terms of degenerations of the worldsheets: the extra marked point serves as variable in the KZ and KZB equations and thereby simultaneously parametrizes the degeneration of the worldsheets in the limits, which in turn define the boundary values. The class of iterated integrals leading to the Selberg integrals as well as the respective integration domains are very naturally defined in terms of the de Rham cohomology of the configuration spaces in question: at genus zero, the twisted forms appearing in the Selberg integrals give rise to a basis of the twisted de Rham cohomology of the configuration space of punctured Riemann spheres with fixed points on the real line. Similarly, the forms in the genus-one Selberg integrals form a closed system with respect to integration by parts, the Fay identity and taking derivatives.
\medskip

\noindent The following points deserve further investigation:
\begin{itemize}
  \item Very likely, recursions with an extra marked point can not only be
	  constructed for corrections to open-string amplitudes as done in this
	  article. Rather, it seems the formalism is extendable to a wide range
	  of string- and quantum field theories. An application or translation
	  to the calculation of scattering amplitudes in $\CN=4$
	  super-Yang--Mills theory in the multi-Regge limit might be a first
	  testing ground: several recursive structures as well as numerous
	  formal similarities are already visible in
	  \rcites{DelDuca:2016lad,DelDuca:2019tur}. Another environment for
	  amplitude recurrences, similar to our current construction, is
	  discussed and applied in
	  \rcites{Puhlfuerst:2015gta,Puhlfuerst:2015zqw}. It would be very
	  interesting to understand the relation between the two approaches. 
  \item Considering the step from genus zero to genus one, all generalizations
	  have been completely canonical. We do not see any structural
	  obstructions for establishing a similar recursion for higher genera.
	  Given the algebraic complexity of the genus-one construction already,
	  combinatorics will not only cause large matrix sizes, but also
	  originate from considering three geometric parameters in the period
	  matrix at genus two.   
  \item Our construction makes use of several genus-zero tools developed in the
	  context of \cite{Mizera:2019gea}, the most prominent example being
	  the matching of dimensions of the respective matrices, which
	  correspond to a basis of Selberg vectors w.r.t.~partial fraction and
	  integration by parts: the respective dimensions are exactly as
	  predicted by twisted de Rham theory.  
  \item A substantial part in establishing our genus-one recursion was devoted
	  to finding a useful and feasible way to single out a basis for
	  Selberg vectors. For higher orders in $\ap$ as well as for higher
	  multiplicity, a formulation of genus-one Selberg integrals in terms
	  of weighted graphs and Fay identities using weighted adjacency
	  matrices analogous to the genus-zero description in \cite{AKtbp}
	  might be the correct computational framework. 
  \item Most importantly, a formalism for calculating one-loop open-string
	  scattering amplitudes from a differential equation has been put
	  forward in \rcites{Mafra:2019ddf,Mafra:2019xms}. The constructions
	  are formally rather similar: both rely on an elliptic KZB equation.
	  While we are using an extra insertion point as differentiation
	  variable, Mafra and Schlotterer employ the modular parameter $\tau$
	  for this purpose. Our formulation employs iterated integrals for the
	  insertion points and the $\omega$-representations of eMZVs, while in
	  \rcites{Mafra:2019ddf,Mafra:2019xms} iterated $\tau$-integrals,
	  Eisenstein series and the $\gamma_0$-representation of eMZVs is
	  employed. There is little doubt that the formalisms can be shown to
	  finally be equivalent: at an algebraic level, several steps have been
	  undertaken in \rcite{Broedel:2020tmd}.  
  \item Our genus-one recursion is tailored to the calculation of
	  \textit{planar} open-string corrections, where vertex insertions are
	  allowed on only one of the boundaries of the annulus. An extension to
	  non-planar open-string amplitudes is expected to be straightforward:
	  in particular one ought to use doubly-periodic integration kernels
	  instead of the functions $g^{(n)}$. A construction
	  for non-planar one-loop string corrections already exists in
	  \rcites{Mafra:2019ddf,Mafra:2019xms}.
\end{itemize}


%


\appendix

\section{Generating function for polylogarithms and the Drinfeld associator}
\label{app:KZassociator}

Let us introduce the general strategy to relate two regularized boundary values of a KZ equation such as
\eqref{eqn:KZexample} by considering a representation of some Lie algebra
generators $e_0$ and $e_1$, as well as a function $\FF(x)$ with $x\in(0,1)$ and
values in the vector space the representations $e_0$ and $e_1$ act upon and which
satisfies the KZ equation 
\begin{equation}
\label{sec:genus0:KZ}
\frac{d}{dx} \FF(x)=\left(\frac{e_0}{x}+\frac{e_1}{x-1}\right)\FF(x)\,.
\end{equation}
Given this situation, one is often interested in calculating the limit of
$\FF(x)$ for $x\to 1$ while knowing the boundary value as $x\to 0$. As will be
reviewed in this section, there is an operator, the Drinfeld associator
$\Phi(e_0,e_1)$ \cite{Drinfeld:1989st,Drinfeld2}, which parallel transports the
(regularized) boundary value of $\FF(x)$ at $x\to 0$ to its (regularized) value
at $x\to 1$. It turns out that the Drinfeld associator is the generating series
of the regularized MZVs, which was originally shown in \rcite{Le} and which is
reviewed in this paragraph following the lines of \rcite{Brown:2013qva}. 

In order to construct the Drinfeld associator, we first investigate the following generating function of multiple polylogarithms
\begin{align}
\LL(x)&=\sum_{w\in\{e_0,e_1\}^{\times}} w\, G_w(x)\,.
\end{align}
The iterative definition \eqref{eqn:GPolylog} and the corresponding regularization prescription of the multiple polylogarithms implies that the series $\LL(x)$
satisfies the KZ equation
\begin{align}\label{sec:genus0:KZforL}
\frac{d }{d x }\LL(x)&=\left(\frac{e_0}{x}+\frac{e_1}{x-1}\right)\LL(x)\,.
\end{align} 
with the asymptotic behavior as $x\to 0$
\begin{equation}\label{KZ:asymptL}
\LL(x)\sim x^{e_0}\,.
\end{equation}
By the symmetry $x\mapsto 1-x$ of the KZ equation, there is another solution $\LL_1$ of \eqref{sec:genus0:KZforL} with the asymptotic behavior
\begin{equation}\label{KZ:asymptL1}
\LL_1(x)\sim (1-x)^{e_1}
\end{equation} 
as $x\to 1$. Now, let $\FF(x)$ be an arbitrary solution of the KZ
equation \eqref{sec:genus0:KZforL}. For this solution, regularized boundary values are defined via
\begin{align}
\label{eqn:C0C1def}
C_0 = \lim_{x \rightarrow 0} x^{-e_0} \FF(x) \ , \ \ \ C_1 = \lim_{x\rightarrow 1} (1-x)^{-e_1} \FF(x)\,.
\end{align}
For two functions $\FF_0(x)$ and $\FF_1(x)$ satisfying the KZ equation
\eqref{sec:genus0:KZ} the product $(\FF_1)^{-1}\FF_0$ is independent of $x$,
and by the asymptotics \eqref{KZ:asymptL}, \eqref{KZ:asymptL1} of $\LL(x)$ and
$\LL_1(x)$, respectively, the calculation 
\begin{equation}
\label{DrinfeldC0C1}
(\LL_1(x))^{-1}\LL(x)C_0=\lim_{x\rightarrow 0}(\LL_1(x))^{-1}\FF(x)=\lim_{x\rightarrow 1}(\LL_1(x))^{-1}\FF(x)=C_1
\end{equation}
shows that the product 
\begin{align}\label{KZ:PhiL}
\Phi(e_0,e_1)&=(\LL_1(x))^{-1}\LL(x)
\end{align}
maps the regularized boundary value $C_0$ to the regularized boundary value $C_1$
\begin{align}
\label{eqn:genusZeroAssociatorEq}
C_1&=\Phi(e_0,e_1)\, C_0\,.
\end{align}
The operator $\Phi(e_0,e_1)$ is the Drinfeld associator which is defined in
terms of the generating series of multiple polylogarithms $\LL(x)$ and the
corresponding solution $\LL_1(x)$. In order to write it as a generating series
of MZVs, its definition \eqref{KZ:PhiL} can be evaluated in the limit
$x\rightarrow 1$, since $\Phi(e_0,e_1)$ is independent of $x$: it is a product
of a function satisfying the KZ equation and an inverse of such a function.
This leads to the relation of the Drinfeld associator to the MZVs discovered
in \rcite{Le},
\begin{align}\label{KZ:PhiMZV}
\Phi(e_0,e_1)&=\lim_{x\rightarrow 1}(1-x)^{-e_1}\LL(x)\nnl
&=\sum_{w\in\{e_0,e_1\}^{\times}} w\zeta_w\nnl
& = 1 - \zeta_2 [e_0,e_1] -\zeta_3 [e_0+e_1,[e_0,e_1]] \notag \\
&\phantom{=}+\zeta_4 ([e_1,[e_1,[e_1,e_0]]]+\tfrac{1}{4}[e_1,[e_0,[e_1,e_0]]] \notag \\
& \phantom{=}-[e_0,[e_0,[e_0,e_1]]]+ \tfrac{5}{4} [e_0,e_1]^2)+  \ldots\;\; \,,
\end{align}
i.e.~the Drinfeld associator is a generating series for the (regularized) MZVs $\zeta_w$
defined in and below \eqn{eqn:mzv}. The limit $x\rightarrow 1$ is chosen to correspond to
taking the tangential base point in negative direction at $1$, such that the
contributions from $(1-x)^{-e_1}$ lead to the discussed regularization $\zeta_{e_1}=0$ of the
divergent terms in $\LL(x)$ by cancelling the positive integer powers of
$\log(1-x)$ in the divergent multiple polylogarithms $G_w(x)$.

\section{Regularization of elliptic multiple zeta values}
\label{app:eMZVreg}
In this section, we give a brief description how eMZVs may be regularized analogously to the regularization of the (genus-zero) MZVs.

The reversal of the ordering in the definition \eqref{sec:eMZV:def} and the
regularization of the iterated integrals $\Gt$ implies that only the eMZVs
\begin{equation}
\omega(n_k,\dots,n_1;\tau)=\omega(w^t;\tau)=\lim_{z\to 1}\Gt_{w}(z,\tau)=\lim_{z\to 1}\Gt(\begin{smallmatrix}n_1 &\dots &n_k\\ 0&\dots &0 \end{smallmatrix}; z,\tau)\,,
\end{equation}
labelled by the word $w=x^{(n_1)}\dots x^{(n_k)}\in X$ with $n_1=1$ inherit the
end point divergence at the upper integration boundary due to the $1/(z{-}1)$
asymptotics of $g^{(1)}(z,\tau)$ in the limit $z\to 1$. For example the
definition \eqref{sec:eMPL:DefReg} and the asymptotic behavior
\eqref{eqn:asympGammaz1} imply that if we would allow for $n_1=1$ in the
definition of the eMZVs, then
\begin{align}
\omega(1;\tau)&=\lim_{z\to 1}\Gt(\begin{smallmatrix}1\\ 0 \end{smallmatrix}; z,\tau)=\lim_{z \to 1}\log(-2 \pi i(1-z))\,\\ \omega(\underbrace{1,\dots,1}_{n};\tau)&= \frac{1}{n!}\omega(1;\tau)^n
\end{align}
are divergent and the $q$-expansion of $g^{(1)}$ implies
\begin{align}
\omega(0,1;\tau)&=\lim_{z\rightarrow 1}\Gt(\begin{smallmatrix}1 &0\\ 0&0 \end{smallmatrix}; z,\tau)\\
&=\lim_{z\to 1}\int_0^zdz'\, g^{(1)}(z',\tau) z'\nnl
&=\lim_{z\to 1}\log(-2 \pi i (1-z))-\frac{i\pi}{2} -2\sum_{k,l>0}\frac{q^{kl}}{k}\,,
\end{align}
such that
\begin{align}
\omega(1,0;\tau)&=\lim_{z\rightarrow 1}\Gt(\begin{smallmatrix}0 &1\\ 0&0 \end{smallmatrix}; z,\tau)\nnl
&=\lim_{z\rightarrow 1}\left(\Gt(\begin{smallmatrix}0\\ 0\end{smallmatrix}; z,\tau)\Gt(\begin{smallmatrix}1\\ 0 \end{smallmatrix}; z,\tau)-\Gt(\begin{smallmatrix}1&0\\ 0&0 \end{smallmatrix}; z,\tau)\right)\nnl
&=\omega(1;\tau)-\omega(0,1;\tau)\nnl
&=\frac{i\pi}{2} +2\sum_{k,l>0}\frac{q^{kl}}{k}
\end{align}
is free of any logarithmic divergence. Using the shuffle algebra, any (divergent) elliptic multiple zeta value can be expanded in powers of $\omega(1;\tau)$, such that the regularized eMZVs $\omega_\reg$ can be defined as being the convergent coefficient (of $1$) in this expansion. For example from above, we find at depth one
\begin{align}
\omega_\reg(1;\tau)&=0\,,
\end{align}
at depth two
\begin{equation}\label{sec:eMZV:w01}
\omega(0,1;\tau)=-\omega(1,0;\tau)+\omega(0)\omega(1;\tau)
\end{equation}
such that
\begin{equation}
\omega_\reg(0,1;\tau)=-\omega(1,0;\tau)=-\omega_\reg(1,0;\tau)
\end{equation}
and further examples of divergent eMZVs are at depth three and weight one
\begin{align}\label{sec:eMZV:w001}
\omega(0,0,1;\tau)&=-\omega(0,1,0;\tau)-\omega(1,0,0;\tau)+\omega(0,0;\tau)\omega(1;\tau)\nnl
&=-\omega(1,0,0;\tau)+\omega(0,0;\tau)\omega(1;\tau)
\end{align}
and at weight 2
\begin{align}\label{sec:eMZV:w101}
\omega(1,0,1;\tau)&=-2\omega(1,1,0;\tau)+\omega(1,0;\tau)\omega(1;\tau)\,,
\end{align}
as well as
\begin{align}\label{sec:eMZV:w011}
\omega(0,1,1;\tau)&=-\omega(1,1,0;\tau)-\omega(1,0,1;\tau)+\omega(0;\tau)\omega(1,1;\tau)\nnl
&=\omega(1,1,0;\tau)-\omega(1,0;\tau)\omega(1;\tau)+\omega(0;\tau)\omega(1,1;\tau)\,,
\end{align}
such that
\begin{align}\label{sec:eMZV:wReg}
\omega_\reg(0,0,1;\tau)&=-\omega_\reg(1,0,0;\tau)\,, \nnl\omega_\reg(1,0,1;\tau)&=-2\omega_\reg(1,1,0;\tau)\,, \nnl \omega_\reg(0,1,1;\tau)&=\omega_\reg(1,1,0;\tau)\,.
\end{align}
As for the regularized elliptic multiple polylogarithms, we generally omit the subscript in $\omega_\reg$ and always refer to the regularized versions when we write an elliptic multiple zeta value $\omega$.

\section{Proof of theorem \ref{thm:maintheorem}.}
\label{app:KZBgeneral}
\begin{proof}
It is to be shown, that the $z_2$-derivative 
of a genus-one $(k{=}2)$-Selberg integral (cf.~\eqn{eqn:relevantGenusOneSelberg}) is expressible as linear combination of admissible Selberg integrals with coefficients composed from $g_{21}^{(n)}$ and $\ZZ$-linear combinations of Mandelstam variables in order to recover the (matrix) KZB equation \eqref{eqn:KZBz2} for the Selberg vector $\SelbldE(z_2)$.

In order to prove the statement, we provide a constructive algorithm involving two steps: the first one is based on
integration by parts such that any partial derivative in the integrand of
$\frac{\partial}{\partial z_2}\SIE{n_{3},\dots,n_L}{i_{3},\dots,i_L}(0,z_2)$
only acts on the Selberg seed $\SelE=\prod_{0\leq z_i<z_j\leq z_2}\exp\left(
s_{ij}\Gt_{ji}\right)$. 
The second step is an iterative application of the Fay
identity to recover admissible products $\prod_{k=3}^L g^{(n_k)}_{k,i_k}$ in
the integrand, such that the integral can be written as a linear combination of
genus-one Selberg integrals.\\[4pt]
\noindent\textbf{Step 1}:
In order to conveniently describe the evaluation of 
\begin{equation*}
\frac{\partial}{\partial z_2}\SIE{n_{3},\dots,n_N}{i_{3},\dots,i_N}(0,z_2)  
\end{equation*}
let us start with a couple of definitions, which are reminiscent to the graphical notation in \rcite{AKtbp}. A product of the form 
\begin{equation}
\prod_{i=1}^{r-1} g^{(n_{k_{i+1}})}_{k_{i+1},k_i}\,,\qquad \text{where}\quad k_{i+1}>k_{i}\,,
\end{equation}
is called a $g$-\textit{chain} from $k_1$ to $k_r$ with \textit{weights}
$(n_{k_2},n_{k_3},\dots,n_{k_r})$. Furthermore, a $g$-\textit{chain with a
branch} at $k_j$ is a product of the form
\begin{equation}
\bigg(\prod_{i=1}^{j-1} g^{(n_{k_{i+1}})}_{k_{i+1},k_i}\bigg)g^{(n_{l_1})}_{l_{1},k_j}\prod_{i=1}^{s-1} g^{(n_{l_{i+1}})}_{l_{i+1},l_i}g^{(n_{m_1})}_{m_{1},k_j}\prod_{i=1}^{t-1} g^{(n_{m_{i+1}})}_{m_{i+1},m_i}\,,
\end{equation} 
with the $g$-\textit{subchains} from $k_1$ to $k_j$, from $k_j$ to $l_s$ and
from $k_j$ to $m_t$. If there exists a $g$-chain in the product $\prod_{k=3}^L
g^{(n_k)}_{k,i_k}$ from $k_1$ to $k_s$, $k_s$ is said to be $g$-\textit{chain
connected} to $k_1$. In order to formulate the first step in the algorithm, we
define for $1\leq k\leq L$ the set of all the integers which are $g$-chain
connected to $k$
\begin{equation}
U^{\vec{n},\vec{i}}_{k}=\{k\leq k'\leq L| k'\text{ is $g$-chain connected to } k \text{ in }\prod_{k=3}^L g^{(n_k)}_{k,i_k}\}\,,
\end{equation}
which, as indicated by the superscripts $\vec{n}=(n_3,\dots, n_L)$ and
$\vec{i}=(i_3,\dots,i_L)$, depends on the product $\prod_{k=3}^L
g^{(n_k)}_{k,i_k}$ and is the genus-one analogue of the set defined in
\eqn{eqn:U3}. Similarly, we define the set of all the integers to which $k$ is
$g$-connected
\begin{equation}
D^{\vec{n},\vec{i}}_{k}=\{3\leq k'\leq k| k\text{ is $g$-chain connected to } k' \text{ in }\prod_{k=3}^L g^{(n_k)}_{k,i_k}\}\,.
\end{equation}
Thus, the set $U^{\vec{n},\vec{i}}_{k}$ goes up the $g$-chain with possible
branches beginning at $k$ and the set $D^{\vec{n},\vec{i}}_{k}$ goes down the
$g$-chain beginning at $k$.  

Using the above notions, the derivative of
$\SIE{n_{3},\dots,n_L}{i_{3},\dots,i_L}(0,z_2)$ with respect to $z_2$ can  be
expressed as 
\begin{align}\label{app:algo:derivativeSafterIbP}
\frac{\partial}{\partial z_2}\SIE{n_{3},\dots,n_L}{i_{3},\dots,i_L}(0,z_2)&=\int_{\mathcal{C}(z_2)}\prod_{i=3}^L dz_i\, \left(\sum_{l\in U^{\vec{n},\vec{i}}_2}\frac{\partial}{\partial z_l}\SelE\right)\prod_{l=3}^L g^{(n_k)}_{k,i_k}\nnl
&=\int_{\mathcal{C}(z_2)}\prod_{i=3}^L dz_i\, \SelE\left(\sum_{l\in U^{\vec{n},\vec{i}}_2}\sum_{j\in U^{\vec{n},\vec{i}}_1} s_{lj} g_{lj}^{(1)}\right)\prod_{k=3}^L g^{(n_k)}_{k,i_k}\,.
\end{align}
where in the first line the derivative has been rewritten to act on the Selberg seed only and in the second line those derivatives have been performed explicitly using \eqref{eqn:derivativeSE}.

The validity of the manipulations can be seen as follows: the admissibility condition $1\leq i_k<k$ implies that the product of differential forms in the integrand of the Selberg integral is a product of $g$-chains starting at $1$ and $g$-chains starting at $2$
\begin{align}\label{app:algo:splitProd}
\prod_{k=3}^L g^{(n_k)}_{k,i_k}&=\prod_{k\in U^{\vec{n},\vec{i}}_1, k\geq 3} g^{(n_k)}_{k,i_k}\prod_{k\in U^{\vec{n},\vec{i}}_2, k\geq 3} g^{(n_k)}_{k,i_k}\,.
\end{align}
The $z_2$-derivative of the integrand of $\SIE{n_{3},\dots,n_L}{i_{3},\dots,i_L}(0,z_2)$  acts on $\SelE$ and the $g$-chains starting at 2 only:
\begin{align}\label{app:algo:derivIntegrans}
&\frac{\partial}{\partial z_2}\left(\SelE\prod_{k=3}^L g^{(n_k)}_{k,i_k}\right)\nnl
&=\left(\frac{\partial}{\partial z_2}\SelE\right)\prod_{k=3}^L g^{(n_k)}_{k,i_k}+\SelE\hspace{-0.3cm}\prod_{k\in U^{\vec{n},\vec{i}}_1, k\geq 3} \hspace{-0.3cm} g^{(n_k)}_{k,i_k}\left(\frac{\partial}{\partial z_2}\prod_{k\in U^{\vec{n},\vec{i}}_2, k\geq 3} g^{(n_k)}_{k,i_k}\right)\,.
\end{align}
Moreover, the first product in the last term of the above equation can be split into a product of all the (disjoint) $g$-chains (possibly with branches) starting at $2$ and ending at some $k\in U^{\vec{n},\vec{i}}_2$ (or several such terminal values in case of branches).
If we consider one such $g$-chain without a branch $g^{n_k}_{k,k_r}\prod_{i=1}^{r-1} g^{(n_{k_{i+1}})}_{k_{i+1},k_i}g^{n_{k_1}}_{k_1,2}$ for $k>k_{i+1}>k_i>2$, the partial derivative with respect to $z_2$ acts as follows
\begin{align}
&\SelE\left(\frac{\partial}{\partial z_2}g^{n_k}_{k,k_r}\prod_{i=1}^{r-1} g^{(n_{k_{i+1}})}_{k_{i+1},k_i}g^{n_{k_1}}_{k_1,2}\right)\nnl
&=\SelE g^{n_{k}}_{k,k_r}\prod_{i=1}^{r-1} g^{(n_{k_{i+1}})}_{k_{i+1},k_i}\frac{\partial}{\partial z_2}g^{n_{k_1}}_{k_1,2}\nnl
&=\SelE g^{n_{k}}_{k,k_r}\prod_{i=1}^{r-1} g^{(n_{k_{i+1}})}_{k_{i+1},k_i}\Big(-\frac{\partial}{\partial z_{k_1}}g^{n_{k_1}}_{k_1,2}\Big)\nnl
&=\left(\frac{\partial}{\partial z_{k_1}}\SelE\right)g^{n_{k}}_{k,k_r}\prod_{i=1}^{r-1} g^{(n_{k_{i+1}})}_{k_{i+1},k_i}g^{n_{k_1}}_{k_1,2}+\SelE g^{n_{k}}_{k,k_r}\prod_{i=1}^{r-1} g^{(n_{k_{i+1}})}_{k_{i+1},k_i}\Big(\frac{\partial}{\partial z_{k_1}}g^{n_{k_2}}_{k_2,k_1}\Big)g^{n_{k}}_{k_1,2}\nnl
&=\left(\frac{\partial}{\partial z_{k_1}}\SelE\right)g^{n_{k}}_{k,k_r}\prod_{i=1}^{r-1} g^{(n_{k_{i+1}})}_{k_{i+1},k_i}g^{n_{k_1}}_{k_1,2}+\SelE g^{n_{k}}_{k,k_r}\prod_{i=1}^{r-1} g^{(n_{k_{i+1}})}_{k_{i+1},k_i}\Big(-\frac{\partial}{\partial z_{k_2}}g^{n_{k_2}}_{k_2,k_1}\Big)g^{n_{k}}_{k_1,2}
\end{align}
where we have used integration by parts for the second-to-last equation and omitted the boundary terms, since they vanish in the iterated integral $\SIE{n_{3},\dots,n_L}{i_{3},\dots,i_L}(0,z_2)$. The above manipulation can iteratively be repeated until any partial derivative only acts on the factor $\SelE$, such that due to the product rule of the derivative we obtain 
\begin{align}
\SelE\left(\frac{\partial}{\partial z_2}g^{n_k}_{k,k_r}\prod_{i=1}^{r-1} g^{(n_{k_{i+1}})}_{k_{i+1},k_i}g^{n_{k_1}}_{k_1,2}\right)&=\left(\left(\sum_{i=1}^r \frac{\partial}{\partial z_{k_i}}+\frac{\partial}{\partial z_k}\right) \SelE\right)\prod_{i=1}^{r-1} g^{(n_{k_{i+1}})}_{k_{i+1},k_i}g^{n_{k_1}}_{k_1,2}\,.
\end{align}
The product rule ensures that the same holds for the $g$-chains with branches
as well. Therefore, we can continue with the calculation
\eqref{app:algo:derivIntegrans} and use the above procedure such that all the
partial derivatives only act on the Selberg seed. The calculation is the
following
\begin{align}
&\frac{\partial}{\partial z_2}\left(\SelE\prod_{k=3}^L g^{(n_k)}_{k,i_k}\right)\nnl
&=\left(\frac{\partial}{\partial z_2}\SelE\right)\prod_{k=3}^L g^{(n_k)}_{k,i_k}+\SelE\prod_{k\in U^{\vec{n},\vec{i}}_1, k\geq 3} g^{(n_k)}_{k,i_k}\left(\frac{\partial}{\partial z_2}\prod_{k\in U^{\vec{n},\vec{i}}_2, k\geq 3} g^{(n_k)}_{k,i_k}\right)\nnl
&=\left(\frac{\partial}{\partial z_2}\SelE\right)\prod_{k=3}^L g^{(n_k)}_{k,i_k}+\left(\sum_{l\in U^{\vec{n},\vec{i}}_2, l\geq 3}\frac{\partial}{\partial z_l}\SelE\right)\prod_{k\in U^{\vec{n},\vec{i}}_1, k\geq 3} g^{(n_k)}_{k,i_k}\prod_{k\in U^{\vec{n},\vec{i}}_2, k\geq 3} g^{(n_k)}_{k,i_k}\nnl
&=\left(\sum_{l\in U^{\vec{n},\vec{i}}_2}\frac{\partial}{\partial z_l}\SelE\right)\prod_{k=3}^L g^{(n_k)}_{k,i_k}\nnl
&=\SelE\left(\sum_{l\in U^{\vec{n},\vec{i}}_2}\sum_{j=1,j\neq l}^L s_{lj}g_{lj}^{(1)}\right)\prod_{k=3}^L g^{(n_k)}_{k,i_k}\nnl
&=\SelE\left(\sum_{l\in U^{\vec{n},\vec{i}}_2}\left(\sum_{j\in U^{\vec{n},\vec{i}}_2\setminus \{l\}} s_{lj}g_{lj}^{(1)}+\sum_{j\in U^{\vec{n},\vec{i}}_1} s_{lj}g_{lj}^{(1)}\right)\right)\prod_{k=3}^L g^{(n_k)}_{k,i_k}\nnl
&=\SelE\left(\sum_{l\in U^{\vec{n},\vec{i}}_2}\sum_{j\in U^{\vec{n},\vec{i}}_1} s_{lj}g_{lj}^{(1)}\right)\prod_{k=3}^L g^{(n_k)}_{k,i_k}\,,
\end{align}
where we have used the antisymmetry $g_{lj}^{(1)}=-g_{jl}^{(1)}$ for the last
equality. This completes the proof of \eqn{app:algo:derivativeSafterIbP}. 

\noindent\textbf{Step 2:} 
The integrals in \eqn{app:algo:derivativeSafterIbP} do not yet have
the desired form, i.e.\ a factor of $g_{21}^{(n)}$ times a product of the form
$g^{(n_k)}_{k,i_k}$ with $1\leq i_k<k$ for all $k\in \{3,\dots, L\}$. This form
can be obtained in a second step using the Fay identity
\eqref{sec:G1:FayIdentity}. Due to the decomposition in
\eqn{app:algo:splitProd}, any term in \eqn{app:algo:derivativeSafterIbP} can be
split into a product of a $g$-chain from $1$ to $j$ labeled by
$D^{\vec{n},\vec{i}}_j=\{j_1<j_2<\dots<j_s<j\}$ and a $g$-chain from $2$ to $l$
labeled by $D^{\vec{n},\vec{i}}_l=\{l_1<l_2<\dots<l_r<l\}$ and the remaining
factors:
\begin{align}
&s_{lj} g_{lj}^{(1)}\prod_{k=3}^L g^{(n_k)}_{k,i_k}\nnl
&\qquad=s_{kj} g_{kj}^{(1)}g_{j,j_s}^{(n_{j})}\prod_{i=1}^{s-1}g_{j_{i+1}j_i}^{(n_{j_{i+1}})}g_{j_1,1}^{(n_{j_1})}g_{l,l_r}^{(n_{l})}\prod_{i=1}^{r-1}g_{l_{i+1}l_i}^{(n_{l_{i+1}})}g_{l_1,2}^{(n_{l_1})}\hspace{-0.3cm}\prod_{k=3, k\not\in D^{\vec{n},\vec{i}}_l \cup D^{\vec{n},\vec{i}}_j}^L\hspace{-0.3cm} g^{(n_k)}_{k,i_k}\,.
\end{align}
The factor $g_{lj}^{(1)}$ connects the two $g$-chains starting at 1 and 2, such
that applying the Fay identity iteratively, the product
\begin{align}\label{app:algo:nonAdmisisbleProduct}
g_{lj}^{(1)}g_{j,j_s}^{(n_{j})}\prod_{i=1}^{s-1}g_{j_{i+1}j_i}^{(n_{j_{i+1}})}g_{j_1,1}^{(n_{j_1})}g_{l,l_r}^{(n_{l})}\prod_{i=1}^{r-1}g_{l_{i+1}l_i}^{(n_{l_{i+1}})}g_{l_1,2}^{(n_{l_1})}
\end{align}
can be written as a factor $g_{21}^{(n)}$ times a linear combination of
admissible factors. The complete procedure is the following: 
\begin{itemize}
	\item First, assume (without loss of generality, rename the labels
		otherwise) that $l<j$, such that the subscript $j$ in
		$g_{lj}^{n_l}$ can be lowered to $j_s$ using the Fay identity
		as follows:
	\begin{align}
		g_{lj}^{n_l}g_{j,j_s}^{n_j}&=(-1)^{n_l}g_{jl}^{n_l}g_{j,j_s}^{n_j}=(-1)^{n_l}g_{l,j_1}\begin{pmatrix}
		g_{j, j_s}\\ g_{j l} \end{pmatrix}_{n_l,n_j}\,, \end{align}
	where the product on the right-hand side is defined to be the sum
	obtained by the Fay identity \eqref{sec:G1:FayIdentity}. It is a $\ZZ$-linear
	combination of $g^{(n_l+n_j-i)}_{l, j_s}g^{(i)}_{j, j_s}$ and
	$g^{(n_l+n_j-i)}_{l, j_s}g^{(i)}_{j l }$ for $0\leq i\leq n_k+n_j$ with integer coefficients. Importantly, it is a linear combination of
	admissible factors and the index $j$ in $g^{(n_l)}_{l j}$ has been
	lowered to $j_s$.
	\item If $l<j_s$, we repeat this step with the products
		$g^{(n_l+n_j-i)}_{l, j_s} g^{(n_{j_s})}_{j_s,j_{s-1}}$.
		Similarly for lower indices $j_t$, unless we arrive at
		$g_{j_1,1}^{(n_{j_1})}$, where another application of the Fay
		identity leaves us with a linear combination of $g^{(n)}_{l,1}$
		and admissible factors times the product
		$g_{l,l_r}^{(n_{l})}\prod_{i=1}^{r-1}g_{l_{i+1},l_i}^{(n_{l_{i+1}})}g_{l_1,2}^{(n_{l_1})}$.
		The same procedure can be applied to
		$g^{(n)}_{l,1}g_{l,l_r}^{(n_{l})}\prod_{i=1}^{r-1}g_{l_{i+1},l_i}^{(n_{l_{i+1}})}g_{l_1,2}^{(n_{l_1})}$
		such that we are left with a linear combination of admissible
		factors times a factor $g_{21}^{(n)}$ and some integer coefficients. However, if we arrive at some $j_t$ such that
		$l>j_t$, we have to apply the Fay identity earlier to the
		product
		$g_{l,l_r}^{(n_{l})}\prod_{i=1}^{r-1}g_{l_{i+1},l_i}^{(n_{l_{i+1}})}g_{l_1,2}^{(n_{l_1})}$
		in order to recover admissible factors. 
	\item Thus, if we arrive at some $j_t$ with $l> j_t$, we apply the
		above procedure to the product $g_{l
		,l_r}^{(n_l)}\prod_{i=1}^{r-1}g_{l_{i+1},l_i}^{(n_{l_{i+1}})}g_{l_1,2}^{(n_{l_1})}$
		beginning with the factor
		\begin{equation}
		g_{l, j_t}^{(n)}g_{l ,l_r}^{(n_l)}= g_{j_t,l_r}\begin{pmatrix}
		g_{l,l_r}\\g_{l,j_t}
		\end{pmatrix}_{n,n_l}\,.
		\end{equation}
		As above, this process can be applied to lower $l_i$ unless we
		arrive either at $g_{l_1,2}^{(n_{l_1})}$ or at $ l_i<j_t$. In
		the latter case, we again proceed with the application of the
		Fay identity with respect to the $j_t$ index as in the previous
		step. In the former case, we arrive at a linear combination of
		$g^{(m)}_{j_t,2}$ and we are left with applying the procedure
		to the $j_t$ index unless we hit $j_1$. 
	\item The above procedure terminates once we could rewrite the product in
		\eqn{app:algo:nonAdmisisbleProduct} as a linear combination of
		$g^{(n)}_{21}$ times solely admissible factors and some integer coefficients. 
\end{itemize}

After applying step 1 and step 2 to any $z_2$-derivative of an admissible genus-one Selberg integral, one will obtain the desired form, that is, one component of the matrix equation \eqref{eqn:KZBz2}. 

Writing the weights of the genus-one Selberg vector as a sequence of the form
\mbox{$\vec{w}=(w_3,\dots,w_L)\in\ZN^{L-2}$}, such that the total weight is given by
$w=|\vec{w}|=w_3+\dots w_L$, and the admissible labelings $\vec{i}=(i_3,\dots,
i_L)\in\ZN^{L-2}$ with $1\leq i_k<k$, this algorithm converts the derivative of
the genus-one Selberg integral
$\SIE{w_{3},\dots,w_L}{i_{3},\dots,i_L}(0,z_2)=\SIE{\vec{w}}{\vec{i}}(0,z_2)$
given in \eqn{app:algo:derivativeSafterIbP} to a form similar to the KZB
equation
\begin{align}\label{app:algo:KZBSingleIntegral}
\frac{\partial}{\partial z_2}\SIE{\vec{w}}{\vec{i}}(0,z_2)&=\sum_{n=0}^{w+1}g^{(n)}_{21}\sum_{\begin{smallmatrix}
	\vec{m}\in \ZN^{L-2}:\\ m=w+1-n
	\end{smallmatrix}}\sum_{\vec{j}\text{ adm}} x^{\vec{w},\vec{i}}_{\vec{m},\vec{j}}\SIE{\vec{m}}{\vec{j}}(0,z_2)\,,
\end{align}
where $m=|\vec{m}|$ and the sum over $\vec{j}\in \ZN^{L-2}$ runs over the
admissible labelings, i.e.\ the vectors $\vec{j}$ such that $1\leq
(\vec{j})_i=j_i<2+i$.
Each coefficient
$x^{\vec{w},\vec{i}}_{\vec{m},\vec{j}}\in \ZZ[s_{ij}]$ either vanishes or is a
$\ZZ$-linear combination of the Mandelstam variables, determined by the above algorithm. Note that all the terms
$g^{(n)}_{21}\SIE{\vec{m}}{\vec{j}}(0,z_2)$ are of total weight $w+1=n+m$,
since $m=w+1-n$. This is a consequence of the above algorithm: the partial
derivatives in the last line of \eqn{app:algo:derivativeSafterIbP} only act on
the Selberg seed $\SelE$, which effectively multiplies $\SelE$ with
some $g_{lj}^{(1)}$. Hence, the integrand $\SelE\prod_{k=3}^L g^{(n_k)}_{k,
i_k}$ is multiplied with $g_{lj}^{(1)} $ which increases the total weight by
one. The application of the Fay identity in the second step of the algorithm
preserves this weight, which leads to the differential
\eqn{app:algo:KZBSingleIntegral}. This completes the proof.
\end{proof}
%
\noindent\textbf{Example for Step 1}:
As an example for step 1, let us consider $L=6$ and the following product $p(z)$ with a
branch at $k=3$
\begin{equation}
p(z)=\SelE g^{(n_6)}_{62}g^{(n_5)}_{53}g^{(n_4)}_{43}g^{(n_3)}_{32}\,.
\end{equation}
Upon discarding boundary terms, the partial derivative of $p(z)$ with respect to $z_2$ is
\begin{align}\label{app:algo:pExample}
\frac{\partial}{\partial z_2}p(z)&=\frac{\partial}{\partial z_2}\left(\SelE g^{(n_6)}_{62}g^{(n_5)}_{53}g^{(n_4)}_{43}g^{(n_3)}_{32}\right)\nnl
&=\left(\frac{\partial}{\partial z_2}\SelE\right) g^{(n_6)}_{62}g^{(n_5)}_{53}g^{(n_4)}_{43}g^{(n_3)}_{32}+\SelE \left(\frac{\partial}{\partial z_2}g^{(n_6)}_{62}\right)g^{(n_5)}_{53}g^{(n_4)}_{43}g^{(n_3)}_{32}\nnl
&\phantom{=}+\SelE g^{(n_6)}_{62}g^{(n_5)}_{53}g^{(n_4)}_{43}\left(\frac{\partial}{\partial z_2}g^{(n_3)}_{32}\right)\nnl
&=\left(\frac{\partial}{\partial z_2}\SelE\right) g^{(n_6)}_{62}g^{(n_5)}_{53}g^{(n_4)}_{43}g^{(n_3)}_{32}+\SelE \left(-\frac{\partial}{\partial z_6}g^{(n_6)}_{62}\right)g^{(n_5)}_{53}g^{(n_4)}_{43}g^{(n_3)}_{32}\nnl
&\phantom{=}+\SelE g^{(n_6)}_{62}g^{(n_5)}_{53}g^{(n_4)}_{43}\left(-\frac{\partial}{\partial z_3}g^{(n_3)}_{32}\right)\nnl
&=\left(\left(\frac{\partial}{\partial z_2}+\frac{\partial}{\partial z_6}+\frac{\partial}{\partial z_3}\right)\SelE\right) g^{(n_6)}_{62}g^{(n_5)}_{53}g^{(n_4)}_{43}g^{(n_3)}_{32}\nnl
&\phantom{=}+\SelE g^{(n_6)}_{62}\left(\frac{\partial}{\partial z_3}g^{(n_5)}_{53}\right)g^{(n_4)}_{43}g^{(n_3)}_{32}+\SelE g^{(n_6)}_{62}g^{(n_5)}_{53}\left(\frac{\partial}{\partial z_3}g^{(n_4)}_{43}\right)g^{(n_3)}_{32}\nnl
&=\left(\left(\frac{\partial}{\partial z_2}+\frac{\partial}{\partial z_6}+\frac{\partial}{\partial z_3}\right)\SelE\right) g^{(n_6)}_{62}g^{(n_5)}_{53}g^{(n_4)}_{43}g^{(n_3)}_{32}\nnl
&\phantom{=}+\SelE g^{(n_6)}_{62}\!\left(-\frac{\partial}{\partial z_5}g^{(n_5)}_{53}\!\right)g^{(n_4)}_{43}g^{(n_3)}_{32}+\SelE g^{(n_6)}_{62}g^{(n_5)}_{53}\!\left(-\frac{\partial}{\partial z_4}g^{(n_4)}_{43}\!\right)g^{(n_3)}_{32}\nnl
&=\left(\left(\frac{\partial}{\partial z_2}+\frac{\partial}{\partial z_6}+\frac{\partial}{\partial z_3}+\frac{\partial}{\partial z_5}+\frac{\partial}{\partial z_4}\right)\SelE\right) g^{(n_6)}_{62}g^{(n_5)}_{53}g^{(n_4)}_{43}g^{(n_3)}_{32}\nnl
&=\SelE\left(\sum_{k=2}^L s_{k1}g_{k1}^{(1)}\right) g^{(n_6)}_{62}g^{(n_5)}_{53}g^{(n_4)}_{43}g^{(n_3)}_{32}\,,
\end{align}
which is the result expected from \eqn{app:algo:derivativeSafterIbP}
since $U^{\vec{n},(2,3,3,2)}_{2}=\{2,3,4,5,6\}$.

\section{Explicit calculation of the one-loop two-point configuration\hyp{}space integrals}
\label{app:2point}
In this appendix, detailed calculations for the two-point example in \subsecref{ssec:2point} are provided. The configuration\hyp{}space contribution to the two-point amplitude is described by the $L=3$ Selberg integrals in \eqn{eqn:twopointintegrals}. The two-point one-loop amplitude with Mandelstam variable $s=s_{13}+s_{23}$ is reproduced for $n=0$, $i_3=1$ as the first entry of the boundary value 
\begin{align}
\bC_1^\El&=\lim_{z_2\rightarrow 1} (-2 \pi i (1-z_2))\begin{pmatrix}
\SIE{0}{1}(0,z_2)\\\SIE{1}{1}(0,z_2)\\ \SIE{2}{1}(0,z_2)\\ \SIE{2}{2}(0,z_2)\\ \vdots
\end{pmatrix}\,.
\end{align}
In order to evaluate the first entry of $\bC_1^\El$ we can use the
block-diagonal form of $x^{(1)}$ with the first block being
$x^{(1)}_{0}=s_{12}$ as shown below. Thus, the relevant entry of the
regularization factor for $z_2\to 1$ is $(-2 \pi i (1-z_2))^{-x^{(1)}_1}\sim
e^{-s_{12}\Gt_{21}}$ and the integral is given by 
\begin{align}\label{app.ex2:CE1Expansion}
&\lim_{z_2\rightarrow 1}(-2 \pi i (1-z_2))^{-s_{12}}\SIE{0}{1}(0,z_2)\nnl
&=\lim_{z_2\rightarrow 1}e^{-s_{12}\Gt_{21}}\int_0^{z_2} dz_3 \exp\left(s_{13}\Gt_{31}+s_{12}\Gt_{21}+s_{23}\Gt_{23}\right)\nnl
&=\int_0^{1} dz_3 \exp\left((s_{13}+s_{23})\Gt_{31}\right)\nnl
&=\sum_{n\geq 0}\frac{\left(s_{13}+s_{23}\right)^n}{n!}\int_0^{1} dz_3 \Gt_{31}^n\nnl
&=\sum_{n\geq 0}\frac{\left(s_{13}+s_{23}\right)^n}{n!}\int_0^{1} dz_3\, n!\Gt(\underbrace{\begin{smallmatrix}1&\dots&1\\ 0&\dots&0 \end{smallmatrix}}_{n}; z_3,\tau)\nnl
&=\sum_{n\geq 0}\left(s_{13}+s_{23}\right)^n \omega(\underbrace{1,\dots,1}_{n},0)\,.
\end{align}
The regularization of the above boundary value corresponds to the first
eigenvalue $s_{12}$ of $x^{(1)}$, which can be determined by bringing the
derivative of $\SIE{n_3}{i_3}(0,z_2)$ in KZB form 
\begin{align}\label{sec:ex2:deqS01}
\frac{\partial}{\partial z_2}\SIE{0}{1}(0,z_2)&=\int_0^{z_2}dz_3 s_{21}g^{(1)}_{21} S+\int_0^{z_2}dz_3 s_{23}g^{(1)}_{23} S\nnl
&=s_{21}g^{(1)}_{21}\SIE{0}{1}(0,z_2)+\int_0^{z_2}dz_3 s_{31}g^{(1)}_{31} S\nnl
&=s_{21}g^{(1)}_{21}\SIE{0}{1}(0,z_2)+s_{31}g^{(0)}_{31}\SIE{1}{1}(0,z_2)\,,
\end{align}
such that the first columns of the matrices $x^{(0)}$ and $x^{(1)}$ are given by
\begin{equation}
x^{(0)}=\begin{pmatrix}
0&s_{31}&0&0&\dots\\
\vdots& & & &
\end{pmatrix}\,, \qquad x^{(1)}=\begin{pmatrix}
s_{21}&0&0&0&\dots\\
\vdots & & & &
\end{pmatrix}\,.
\end{equation}
Note that we have used the integration by parts identity
\begin{align}\label{sec:ex:IbP2point}
s_{23}\SIE{1}{2}(0,z_2)+s_{13}\SIE{1}{1}(0,z_2)&=0\,.
\end{align}
The boundary value for $z_2\rightarrow 0$ is more subtle. In this limit, the
one-loop propagator degenerates to the tree level propagator and, in
particular, loses its $\tau$-dependence at the lowest order in $z_2$
\begin{equation}
\Gt_\reg(\begin{smallmatrix}1\\0 \end{smallmatrix}; z_2,\tau)=\log(-2 \pi i z_2)+\CO\Big(z_2^2\Big)\,,\qquad g^{(1)}(z_2,\tau)=\frac{1}{z_2}+\CO(z_2)
\end{equation}
such that, using the change of variables $z_i=z_2w_i$,
the lowest order in $z_2$ for $n_3=1$, $i_3=1$ is given by 
\begin{align}\label{sec:ex2:z2To0}
&\SIE{1}{1}(0,z_2)\nnl
&=\int_0^{z_2}dz_3\, \exp\left(s_{13}\Gt_{31}+s_{12}\Gt_{21}+s_{23}\Gt_{23}\right)g^{(1)}_{31}\nnl
&=\int_0^{1}dw_3\,z_2 (-2 \pi i z_2w_{3})^{s_{13}}(-2 \pi i z_2)^{s_{12}}(-2 \pi i z_2(1-w_{3}))^{s_{23}}\frac{1}{z_2 w_3}(1+\CO(z_2))\nnl
&=(-2 \pi i z_2)^{s_{123}}\int_0^{1}dw_3\, w_{3}^{s_{13}}(1-w_{3})^{s_{23}}\frac{1}{ w_3}(1+\CO(z_2))
\nnl
&=(-2 \pi i z_2)^{s_{123}}\left(\frac{1}{s_{13}}\frac{\Gamma(1+s_{13})\Gamma(1+s_{23})}{\Gamma(1+s_{13}+s_{23})}\right)(1+\CO(z_2))\,.
\end{align}
Therefore, at the lowest order in $z_2$, the integral $\SIE{1}{1}(0,z_2)$
degenerates to the four-point tree-level amplitude with Mandelstam variables
$s_{13}$ and $s_{23}$. Now, let us check that the regularization by the
factor $(-2 \pi i z_2)^{-x^{(1)}}$ projects out that lowest-order coefficient of
$z_2$. In order to obtain the appropriate eigenvalue of $x^{(1)}$, the
differential equation satisfied by $\SIE{1}{1}(0,z_2)$ has to be brought in KZB
form and the coefficient of $\SIE{1}{1}(0,z_2)$ itself has to be determined
\begin{align}
	&\frac{\partial}{\partial z_2}\SIE{1}{1}(0,z_2)\nnl
	&\quad=\int_0^{z_2}dz_3\, \exp\left(s_{13}\Gt_{31}+s_{12}\Gt_{21}+s_{23}\Gt_{23}\right)g^{(1)}_{31}\left(s_{12}g^{(1)}_{21}{+}s_{23}g^{(1)}_{23}\right)\nnl
&\quad=s_{12}g^{(1)}_{21}\SIE{1}{1}(0,z_2)-s_{23}\int_0^{z_2}dz_3\, \Sel^\El g^{(1)}_{31}g^{(1)}_{32}.
\end{align}
In order to bring the second integral into the appropriate form, the Fay identity
\begin{align}
g^{(1)}_{31}g^{(1)}_{32}&=g_{21}^{(2)}+g_{31}^{(2)}+g^{(2)}_{32}+g_{21}^{(1)}g_{32}^{(1)}-g_{21}^{(1)}g_{31}^{(1)}
\end{align}
has to be used, followed by an application of \eqn{sec:ex:IbP2point}
\begin{align}\label{sec:ex2:deqS11}
&\frac{\partial}{\partial z_2}\SIE{1}{1}(0,z_2)\nnl
&\quad=-s_{23}g^{(2)}_{21}\SIE{0}{1}(0,z_2)-s_{23}g^{(0)}_{21}\SIE{2}{1}(0,z_2)-s_{23}g^{(0)}_{21}\SIE{2}{2}(0,z_2)\nnl
&\quad\phantom{=}+s_{12}g^{(1)}_{21}\SIE{1}{1}(0,z_2)-s_{23}g^{(1)}_{21}\SIE{1}{2}(0,z_2)+s_{23}g^{(1)}_{21}\SIE{1}{1}(0,z_2)\nnl
&\quad=-s_{23}g^{(2)}_{21}\SIE{0}{1}(0,z_2)-s_{23}g^{(0)}_{21}\SIE{2}{1}(0,z_2)-s_{23}g^{(0)}_{21}\SIE{2}{2}(0,z_2)\nnl
&\quad\phantom{=}+\left(s_{12}+s_{13}+s_{23}\right)g^{(1)}_{21}\SIE{1}{1}(0,z_2)\,.
\end{align}
Therefore, we find that the appropriate eigenvalue of $x^{(1)}$ is indeed
$s_{123}=s_{12}+s_{13}+s_{23}$, such that according to \eqn{sec:ex2:z2To0} the
second, i.e.\ the weight-one, entry of $\bC_0^\El$ is given by the four-point
tree-level amplitude
\begin{equation}
\bC_0^\El=\lim_{z_2\rightarrow 0} e^{-x^{(1)}\Gt_{21}}\begin{pmatrix}
\SIE{0}{1}(0,z_2)\\\SIE{1}{1}(0,z_2)\\ \SIE{2}{1}(0,z_2)\\\SIE{2}{2}(0,z_2)\\ \vdots
\end{pmatrix}=\begin{pmatrix}
\ast\\\frac{1}{s_{13}}\frac{\Gamma(1+s_{13})\Gamma(1+s_{23})}{\Gamma(1+s_{13}+s_{23})}\\\ast \\ \ast\\ \vdots
\end{pmatrix}\,.
\end{equation}
As discussed in \subsecref{subsec:BoundaryValues}, since the eigenvalue of
$x^{(1)}$ can not be bigger than $s_{123}$ and we can only compensate the
Jacobian $z_2$ in \eqn{sec:ex2:z2To0} from the change of variables $z_3=z_2
w_3$ by the singular asymptotic behavior of
$g^{(1)}(z_{3},\tau)\rightarrow\frac{1}{z_2 w_{3}}$ for $z_2\rightarrow 0$, if
there would be another integration kernel $g^{(n_3)}(z_{3 i_3},\tau)$ with
$n_3\neq 1$ which is regular close to the origin, there would not be such a
compensation. Thus, all other entries of the boundary value $\bC_0^\El$ which
do not correspond to a singular integration kernel $g^{(1)}(z_{3 i_3},\tau)$
vanish and we obtain
\begin{align}\label{sec.ex2:CE0}
\bC_0^\El&=\begin{pmatrix}
0\\\frac{1}{s_{13}}\frac{\Gamma(1+s_{13})\Gamma(1+s_{23})}{\Gamma(1+s_{13}+s_{23})}\\0\\ 0\\ \vdots
\end{pmatrix}\,.
\end{align}

In order to check the consistency of the first entry of the vector equation 
\begin{align}\label{sec:ex2:AssociatorEq}
\bC_1^\El&=\Phi^\El\, \bC_0^\El
\end{align}
up to order $(\ap)^2$, we also need to calculate the derivative of $\Selbld^\El_2(z_2)$, which includes the following two derivatives: the first one is
\begin{align}
\frac{\partial}{\partial z_2} \SIE{2}{1}(0,z_2)&=\int_0^{z_2}dz_3\, \Sel^\El g^{(2)}_{31}\left(s_{21}g^{(1)}_{21}+s_{23}g^{(1)}_{23}\right)\nnl
&=s_{12}g^{(1)}_{21}\SIE{2}{1}(0,z_2)-s_{23}\int_0^{z_2}dz_3\, \Sel^\El g^{(2)}_{31}g^{(1)}_{32}\,,
\end{align}
where we can apply again the Fay identity
\begin{align}
g^{(1)}_{32}g^{(2)}_{31}
&=-(-1)^2 g^{(3)}_{12}+\sum_{r=0}^2\binom{r}{0}g^{(2-r)}_{21}g^{(1+r)}_{k2}+\sum_{r=0}^1\binom{r+1}{1}g^{(1-r)}_{12}g^{(2+r)}_{k1}\nnl
&=g^{(3)}_{21}+g_{21}^{(2)}g^{(1)}_{32}+g_{21}^{(1)}g^{(2)}_{32}+g_{21}^{(0)}g^{(3)}_{32}-g_{21}^{(1)}g_{31}^{(2)}+2g_{12}^{(0)}g_{31}^{(3)}\,.
\end{align}
Therefore, we find
\begin{align}\label{sec:ex2:deqS21}
&\frac{\partial}{\partial z_2} \SIE{2}{1}\nnl
&\quad=s_{12}g^{(1)}_{21}\SIE{2}{1}-s_{23}\big(g^{(3)}_{21}\SIE{0}{1}+g_{21}^{(2)}\SIE{1}{2}+g_{21}^{(1)}\SIE{2}{2}\nnl
&\quad
\phantom{=}+g_{21}^{(0)}\SIE{3}{2}-g_{21}^{(1)}\SIE{2}{1}+2g_{12}^{(0)}\SIE{3}{1}\big)\nnl
&\quad=g^{(0)}_{21}\left(-2 s_{23}\SIE{3}{1}-s_{23}\SIE{3}{2}\right)+g^{(1)}_{21}\left( (s_{12}+s_{23})\SIE{2}{1}-s_{23}\SIE{2}{2}\right)\nnl
&\quad\phantom{=}+g^{(2)}_{21}\left( -s_{32}\SIE{1}{2}\right)+g^{(3)}_{21}\left( -s_{32}\SIE{0}{1}\right)\nnl
&\quad=g^{(0)}_{21}\left(-2 s_{23}\SIE{3}{1}-s_{23}\SIE{3}{2}\right)+g^{(1)}_{21}\left( (s_{12}+s_{23})\SIE{2}{1}-s_{23}\SIE{2}{2}\right)\nnl
&\quad\phantom{=}+g^{(2)}_{21}\left( s_{13}\SIE{1}{1}\right)+g^{(3)}_{21}\left( -s_{32}\SIE{0}{1}\right)
\end{align}
and similarly
\begin{align}
&\frac{\partial}{\partial z_2} \SIE{2}{2}(0,z_2)\nnl
&\quad=\int_0^{z_2}dz_3\, \Sel^\El g^{(2)}_{32}\left(s_{21}g^{(1)}_{21}+s_{23}g^{(1)}_{23}\right)+\int_0^{z_2}dz_3\, \Sel^\El \frac{\partial}{\partial z_2}g^{(2)}_{32}\nnl
&\quad=s_{21}g^{(1)}_{21}\SIE{2}{2}-s_{23}\int_0^{z_2}dz_3\, \Sel^\El g^{(2)}_{32}
g^{(1)}_{32}-\int_0^{z_2}dz_3\, \Sel^\El \frac{\partial}{\partial z_3}g^{(2)}_{32}\nnl
&\quad=s_{21}g^{(1)}_{21}\SIE{2}{2}-s_{23}\int_0^{z_2}dz_3\, \Sel^\El g^{(2)}_{32}
g^{(1)}_{32}\nnl
&\qquad\phantom{=}+\int_0^{z_2}dz_3\, \Sel^\El (s_{31}g^{(1)}_{31}+s_{32}g^{(1)}_{32})g^{(2)}_{32}\nnl
&\quad=s_{21}g^{(1)}_{21}\SIE{2}{2}+s_{13}\int_0^{z_2}dz_3\, \Sel^\El g^{(2)}_{32}
g^{(1)}_{31}\,,
\end{align}
where we can again use
\begin{align}
g^{(1)}_{31}g^{(2)}_{32}
&=g^{(3)}_{21}+g_{12}^{(2)}g^{(1)}_{31}+g_{12}^{(1)}g^{(2)}_{31}+g_{12}^{(0)}g^{(3)}_{31}-g_{12}^{(1)}g_{32}^{(2)}+2g_{21}^{(0)}g_{32}^{(3)}\nnl
&=-g^{(3)}_{21}+g_{21}^{(2)}g^{(1)}_{31}-g_{21}^{(1)}g^{(2)}_{31}+g_{21}^{(0)}g^{(3)}_{31}+g_{21}^{(1)}g_{32}^{(2)}+2g_{21}^{(0)}g_{32}^{(3)}\,,
\end{align}
such that
\begin{align}\label{sec:ex2:deqS22}
&\frac{\partial}{\partial z_2} \SIE{2}{2}(0,z_2)\nnl
&\quad=g^{(0)}_{21}\left(s_{13}\SIE{3}{1}+2 s_{13}\SIE{3}{2}\right)+g^{(1)}_{21}\left( -s_{13}\SIE{2}{1}+(s_{12}+s_{13})\SIE{2}{2}\right)\nnl
&\quad\phantom{=}+g^{(2)}_{21}\left( s_{13}\SIE{1}{1}\right)+g^{(3)}_{21}\left( -s_{13}\SIE{0}{1}\right)\,.
\end{align}

The relevant $4\times 4$-submatrices $x^{(n)}_{\leq 2}$ of $x^{(n)}$ for
$n\in\{0,1,2\}$ appearing in the differential \eqn{sec:ex2:deq} of
$\Selbld^\El_{\leq 2}(z_2)$, i.e.\
\begin{equation}
\frac{\partial}{\partial z_2}\Selbld^\El_{\leq 2}(z_2)=\left(g^{(0)}_{21}x^{(0)}_{\leq 2}+g^{(1)}_{21}x^{(1)}_{\leq 2}
+g^{(2)}_{21}x^{(2)}_{\leq 2}\right)
\Selbld^\El_{\leq 2}(z_2)+r_2 \Selbld^\El_{3}(z_2)\,,
\end{equation}
can now be read off from the differential equations \eqref{sec:ex2:deqS01},
\eqref{sec:ex2:deqS11}, \eqref{sec:ex2:deqS21} and \eqref{sec:ex2:deqS22},
which gives the matrices in \eqns{eqn:2ptExamplex01}{eqn:2ptExamplex2}.
%


%
%

\bibliographystyle{alpha}
\bibliography{AmplitudeRecursions}

\end{document}

%% file: AmplitudeRecursionsmp.tex
\begin{mpostdef}
pair vpos[];
pair epos[];
pair ext[];
pair exta[];
path paths[];
picture pic;
picture pic[];
picture savepic;
path cpath;
path hpath;
xu:=1cm;
yu:=1cm;
\end{mpostdef}


\begin{mpostdef}
def pensize(expr s)=withpen pencircle scaled s enddef;
def fillshape(expr p,ci,tb,cb)=
  fill p withcolor ci;
  draw p pensize(tb) withcolor cb;
enddef;
def filldot(expr z,s,ci)=
  fillshape(fullcircle scaled s shifted z, ci, 0.5pt, 0.0white);
enddef;
def filltrig(expr z,s,r,ci)=
  fillshape(((dir 60)--(dir 180)--(dir 300)--cycle) scaled s rotated r shifted z, ci, 0.5pt, 0.0white);
enddef;
def fillsqr(expr z,s,r,ci)=
  fillshape(((dir 45)--(dir 135)--(dir 225)--(dir 315)--cycle) scaled s rotated r shifted z, ci, 0.5pt, 0.0white);
enddef;
def drawcross(expr z,s,r,t,c)=
  draw ((-0.5,-0.5)--(+0.5,+0.5)) scaled s rotated r shifted z pensize(t) withcolor c;
  draw ((+0.5,-0.5)--(-0.5,+0.5)) scaled s rotated r shifted z pensize(t) withcolor c;
enddef;

def midarrow (expr p, t) =
  fill arrowhead subpath(0,arctime(arclength(subpath (0,t) of p)+0.5ahlength) of p) of p;
enddef;

def drawprop expr p=draw p pensize(1.5pt) enddef;
def drawline expr p=draw p pensize(0.5pt) enddef;
def drawpos expr p=drawcross(p, 5pt, 0, 1.0pt, 0.0white) enddef;
def drawvertex expr p=filldot(p, 5pt, 0.5white+0.5red) enddef;
def drawblackdotscale expr p=filldot(p, 0.4*xu, black) enddef;
def drawreddotscale expr p=filldot(p, 0.4*xu, red) enddef;
def drawwhitedotscale expr p=filldot(p, 0.4*xu, white) enddef;
def drawblackdot expr p=filldot(p, 8pt, black) enddef;
def drawwhitedot expr p=filldot(p, 8pt, white) enddef;
def drawlinearrow expr p=draw p pensize(0.5pt); midarrow (p,0.5); enddef;
def drawproparrow expr p=draw p pensize(1.5pt); midarrow (p,0.5); enddef;
def drawproparrowdouble expr p=draw p pensize(1.5pt); midarrow (p,0.65); midarrow (reverse p,0.65); enddef;
def drawwhitedotmr (expr p, sz)=filldot(p, sz*xu, white) enddef;
def drawblackdotmr (expr p, sz)=filldot(p, sz*xu, black) enddef;
\end{mpostdef}
%

\begin{mpostdef}
u:=xu;									
a=2u;									
l=8u;									
u=12.5v;								
del=23.44;                             	
pair w$;								

let vector=color;                       
let xc=redpart; let yc=greenpart; let zc=bluepart;  
def dot(expr u,v)=
    (xc(u)*xc(v)+yc(u)*yc(v)+zc(u)*zc(v))
enddef;
def g(expr phi,lambda)=                 
    (cosd(phi)*cosd(lambda),cosd(phi)*sind(lambda),sind(phi))
enddef;
def G(expr omega,incl,Omega)=           
    (cosd(omega)*cosd(Omega)-sind(omega)*cosd(incl)*sind(Omega),
     cosd(omega)*sind(Omega)+sind(omega)*cosd(incl)*cosd(Omega),sind(omega)*sind(incl))
enddef;
def ellipse(expr ra,rb,an)=
    (fullcircle xscaled 2ra yscaled 2rb rotated an)
enddef;
def halfell(expr ra,rb,an)=
    (halfcircle xscaled 2ra yscaled 2rb rotated an)
enddef;

def circle(expr u,psi,a,f)=
    numeric s, c, e, c[]; path p[]; pair q[];

    c0=dot(u,g(phi,lambda));
    c1=dot(u,g(0,lambda+90));
    c2=dot(u,g(phi+90,lambda));

    e  = 1+-+c0;                        
    s  = sind(psi); c  = cosd(psi);     %
    q1 = (a,0); q2 = (0,a);             
    w$:=q0=c1*q1 + c2*q2;               
    p0 = origin--q0;                    
    p9 = ellipse(c*a,abs(c0)*c*a,angle(c2,-c1)) shifted (s*q0);

    if (psi=0):                         
        p1 = halfell(a,c0*a,angle(-c2,c1)); 
        p2 = halfell(a,c0*a,angle(c2,-c1)); 
    elseif (abs(c0)>c):                 
        if (s*c0>0):                    
            p1=p9;
            p2=origin;
        else:                           
            p1=origin;
            p2=p9;
        fi
    else:                               
        c5=e+-+s; c6=e*e;
        q3=((c1*s+c2*c5)*a/c6,(c2*s-c1*c5)*a/c6);
        q4=((c1*s-c2*c5)*a/c6,(c2*s+c1*c5)*a/c6);
        c3=xpart(p9 intersectiontimes ((s*q0)--1.1q3));
        c4=xpart(p9 intersectiontimes ((s*q0)--1.1q4));
        if (psi>0):
            p3 = subpath(c4,8) of p9 & subpath(0,c3) of p9;
            p4 = subpath(c3,c4) of p9;
        else:
            p3 = subpath(c4,c3) of p9;
            p4 = subpath(c3,8) of p9 & subpath(0,c4) of p9;
        fi

        if (c0>0):
            p1=p3; p2=p4;
        else:                           
            p1=p4; p2=p3;
        fi
    fi

    draw p1;
    if (f>1):
        draw p2 dashed evenly;
    fi
    if (f=1) or (f=3):
        drawarrow p0 dashed evenly;
    fi

enddef;
\end{mpostdef}
%


\begin{mpostdef}
color ecolor; ecolor:=(.72,.03,.06);
color bcolor; bcolor:=(.07,.05,.66);
color gcolor; 
gcolor:=(.0,.44,.0);
gcolor:=blue;
color faint; faint:=(.7,0.7,0.7);
vsize := 0.16xu;
rad:= 2xu;
cpath:= for i=0 upto 35: 
  dir (i*10) scaled rad ..
endfor cycle;
def vert(expr pos)=
  fill fullcircle scaled vsize shifted point pos of cpath withcolor white;
  draw halfcircle scaled vsize rotated (90+10*pos) shifted point pos of cpath withcolor ecolor pensize(1.3pt);
enddef;
def vertthick(expr pos)=
  fill fullcircle scaled vsize shifted point pos of cpath withcolor white;
  draw halfcircle scaled vsize rotated (90+10*pos) shifted point pos of cpath withcolor ecolor pensize(2.0pt);
enddef;
def vertcol(expr pos,col)=
  fill fullcircle scaled vsize shifted point pos of cpath withcolor white;
  draw halfcircle scaled vsize rotated (90+10*pos) shifted point pos of cpath withcolor col pensize(1.3pt);
enddef;
def drawbdry(expr startpos,endpos)=
  draw subpath(startpos,endpos) of cpath withcolor ecolor pensize(1.3pt);
enddef;
def drawbdryblack(expr startpos,endpos)=
  draw subpath(startpos,endpos) of cpath withcolor black pensize(1.3pt);
enddef;
def drawbdrythick(expr startpos,endpos)=
  draw subpath(startpos,endpos) of cpath withcolor ecolor pensize(2.0pt);
enddef;
def drawbdrythickblack(expr startpos,endpos)=
  draw subpath(startpos,endpos) of cpath withcolor black pensize(2.0pt);
enddef;
def drawbdrydashed(expr startpos,endpos)=
  draw subpath(startpos,endpos) of cpath withcolor ecolor dashed withdots scaled 0.5 pensize(1.3pt);
enddef;
def drawbdrydashedthick(expr startpos,endpos)=
  draw subpath(startpos,endpos) of cpath withcolor ecolor dashed withdots scaled 0.5 pensize(2.0pt);
enddef;
def vlabel(expr pos, vscal, tt)=
vert(pos);
label(tt,point pos of cpath scaled vscal);
enddef;

def vlabelcol(expr pos, vscal, tcol, vcol, tt)=
vertcol(pos, vcol);
label(tt,point pos of cpath scaled vscal) withcolor tcol;
enddef;

def ppath(expr p,spos,epos)=
subpath(arctime(spos*arclength(p))of p,arctime(epos*arclength(p)) of p) of p
enddef;
\end{mpostdef}

\begin{mpostfig}[label=genuszerogeneral]
	hpath:= (-2xu,0xu) ..controls (-2xu,0.75xu) and (2xu,0.75xu) .. ( 2xu,0xu);
	draw ppath(hpath,0.0,0.97) dashed evenly pensize (0.8pt) withcolor black;
        draw (-2xu,0xu) ..controls (-2xu,-0.75xu) and (2xu,-0.75xu) .. ( 2xu,0xu) pensize (0.8pt) withcolor black;
	drawbdry(31,36);
	drawbdry(0,29);
	drawbdrydashed(22,31);
	vlabel(27,0.78,btex $x_1=0$ etex);
	vlabel(0,1.36,btex $x_2=1$ etex);
	vlabel(9,1.13,btex $x_{L+1}=\infty$ etex);
	vlabel(34,1.2,btex $x_3$ etex);
	vlabel(32.5,1.2,btex $x_4$ etex);
	vlabel(31,1.2,btex $x_5$ etex);
	vlabel(28.5,1.2,btex $x_L$ etex);
	hpath:= subpath (29,31) of cpath scaled 1.15;
	draw ppath(hpath,0.07,0.75) dashed withdots scaled 0.7 pensize(1.3pt);

\end{mpostfig}

\begin{mpostfig}[label=genuszerox3to1]
	hpath:= (-2xu,0xu) ..controls (-2xu,0.75xu) and (2xu,0.75xu) .. ( 2xu,0xu);
	draw ppath(hpath,0.0,0.97) dashed evenly pensize (0.8pt) withcolor black;
        draw (-2xu,0xu) ..controls (-2xu,-0.75xu) and (2xu,-0.75xu) .. ( 2xu,0xu) pensize (0.8pt) withcolor black;

	drawbdry(31,36);
	drawbdry(0,29);
	drawbdrydashed(22,31);
	vlabel(27,0.78,btex $x_1=0$ etex);
	vlabel(0,1.36,btex $x_2=1$ etex);
	vlabel(9,1.13,btex $x_{L+1}=\infty$ etex);
	vlabelcol( 34, 1.2, gcolor, gcolor, btex $x_3$ etex);
	vlabel(32.5,1.2,btex $x_4$ etex);
	vlabel(31,1.2,btex $x_5$ etex);
	vlabel(28.5,1.2,btex $x_L$ etex);
	hpath:= (point 34 of cpath .. point 0 of cpath .. point 2 of cpath) scaled 1.15;
	drawarrow ppath(hpath,0.12,0.38) withcolor gcolor pensize(1.3pt);
	hpath:= subpath (28,31) of cpath scaled 1.15;
	draw ppath(hpath,0.3,0.8) dashed withdots scaled 0.7 pensize(1.3pt);
\end{mpostfig}

\begin{mpostfig}[label=genuszerox3to0]
	hpath:= (-2xu,0xu) ..controls (-2xu,0.75xu) and (2xu,0.75xu) .. ( 2xu,0xu);
	draw ppath(hpath,0.0,0.97) dashed evenly pensize (0.8pt) withcolor black;
        draw (-2xu,0xu) ..controls (-2xu,-0.75xu) and (2xu,-0.75xu) .. ( 2xu,0xu) pensize (0.8pt) withcolor black;

	drawbdry(30,36);
	drawbdry(0,28);
	drawbdrydashed(28,29);
	vlabel(27,1.2,btex $x_1$ etex);
	vlabel(0,1.36,btex $x_2=1$ etex);
	vlabel(9,1.13,btex $x_{L+1}=\infty$ etex);
	vlabelcol( 34, 1.2, faint, faint, btex $x_3$ etex);
	vlabelcol( 30.2, 1.2, gcolor, gcolor, btex $x_3$ etex);
	vlabel(29.6,0.8,btex $x_4$ etex);
	vlabel(29,0.6,btex ~ etex);
	vlabel(27.7,0.8,btex $x_L$ etex);
	hpath:= subpath(34,27) of cpath scaled 1.15;
	draw ppath(hpath,0.12,0.46) withcolor gcolor pensize(1.3pt);
	drawarrow ppath(hpath,0.65,0.9) withcolor gcolor pensize(1.3pt);
\end{mpostfig}

\begin{mpostfig}[label=genuszerostructure]
	label.rt(btex $\mathbf{C}_1$ on ${\cal F}_{N,3} = {\cal M}_{0,N}$ etex,(3xu,1xu));
	label.rt(btex \footnotesize{$s_{3j}\to 0$} etex,(6.5xu,1.2xu));
	label(btex $N$-point amplitudes etex,(10.2xu,1.2xu));
	label(btex on ${\cal M}_{0,N}$ etex,(10.2xu,0.85xu));
	drawarrow (6.4xu,1xu)--(8.1xu,1xu);
	label.rt(btex $\mathbf{C}_0$ on ${\cal F}_{N,3} = {\cal M}_{0,N}$ etex,(3xu,-1xu));
	label.rt(btex \footnotesize{$s_{3j}\to 0$} etex,(6.5xu,-0.8xu));
	label(btex $(N{-}1)$-point amplitudes etex,(10.2xu,-0.85xu));
	label(btex on ${\cal M}_{0,N-1}$ etex,(10.2xu,-1.2xu));
	drawarrow (6.4xu,-1xu)--(8.1xu,-1xu);
	label.rt(btex $\mathrm{\mathbf{S}}(x_3)$ on ${\cal F}_{N+1,4}$ etex, (0xu,0xu));
	label(btex \tiny$x_3\!\to\!x_2\!=\!1$ etex,(0xu,0xu)) rotated 16 shifted (3.2xu,0.4xu);
	drawarrow (2xu,0.2xu)--(4xu,0.8xu);
	label(btex \tiny$x_3\!\to\!x_1\!=\!0$ etex,(0xu,0xu)) rotated -16 shifted (3.2xu,-0.4xu);
	drawarrow (2xu,-0.2xu)--(4xu,-0.8xu);
	label.rt(btex $\Phi(e_0,e_1)$ etex,(4.8xu,0xu));
	drawarrow (4.8xu,-0.75xu)--(4.8xu,0.8xu);
	label.rt(btex $\Phi(e_0,e_1)|_{s_{3j}=0}$ etex,(10.2xu,0xu));
	drawarrow (10.2xu,-0.55xu)--(10.2xu,0.6xu);
\end{mpostfig}


\begin{mpostfig}[label=genuszerogenusone]
	hpath:= (-2xu,0xu) ..controls (-2xu,0.75xu) and (2xu,0.75xu) .. ( 2xu,0xu);
	draw ppath(hpath,0.0,0.97) dashed evenly pensize (0.8pt) withcolor bcolor;
        draw (-2xu,0xu) ..controls (-2xu,-0.75xu) and (2xu,-0.75xu) .. ( 2xu,0xu) pensize (0.8pt) withcolor bcolor;
	drawbdry(31,36);
	drawbdry(0,29);
	drawbdrydashed(22,31);
        drawbdryblack(9,27);
	vlabel(27,1.2,btex $x_1$ etex);
	vlabel(0,1.2,btex $x_2$ etex);
	vlabel(9,1.2,btex $x_{L+1}$ etex);
	vlabel(34,1.2,btex $x_3$ etex);
	vlabel(32.5,1.2,btex $x_4$ etex);
	vlabel(31,1.2,btex $x_5$ etex);
	vlabel(28.5,1.2,btex $x_L$ etex);
	hpath:= subpath (29,31) of cpath scaled 1.15;
	draw ppath(hpath,0.07,0.75) dashed withdots scaled 0.7 pensize(1.3pt);
  	pic:=currentpicture;
  	currentpicture:= nullpicture;
        
	hpath:= (-2xu,0xu) ..controls (-2xu,0.75xu) and (2xu,0.75xu) .. ( 2xu,0xu);
	draw ppath(hpath,0.0,0.97) dashed evenly pensize (1.6pt) withcolor bcolor;
        draw (-2xu,0xu) ..controls (-2xu,-0.75xu) and (2xu,-0.75xu) .. ( 2xu,0xu) pensize (1.6pt) withcolor bcolor;
	drawbdrythick(31,36);
	drawbdrythick(0,9);
	drawbdrythick(27,29);
	drawbdrydashedthick(22,31);
	drawbdrythickblack(9,27);
	vertthick(27);
	vertthick( 0);
	vertthick( 9);
	vertthick(34);
	vertthick(32.5);
	vertthick(31);
	vertthick(28.7);
	draw (0xu,2xu){dir 100}..(-3xu,0xu)..{dir 80}(0xu,-2xu) dashed evenly pensize(2pt)withcolor ecolor;
  	savepic:=currentpicture;
  	currentpicture:= nullpicture;
 
  	draw (-2xu,0xu) ..controls (-2xu,0.75xu) and (2xu,0.75xu) .. ( 2xu,0xu) dashed evenly pensize (0.8pt) withcolor ecolor;
  	draw (0xu,-0.15xu) .. controls (0.2xu,-0.18xu) and (0.2xu,-0.92xu) .. ( 0xu,-0.95xu) dashed evenly pensize (0.8pt) withcolor bcolor;
  	draw fullcircle xscaled 4xu yscaled 1.9xu pensize (1pt);
  	draw (-1xu,0.03xu){dir -20}..(1xu,0.03xu) pensize (0.8pt);
  	draw (-0.8xu,-0.02xu){dir 25}..(0.8xu,-0.02xu) pensize (0.8pt);
  	draw (-2xu,0xu) ..controls (-2xu,-0.75xu) and (2xu,-0.75xu) .. ( 2xu,0xu) pensize (0.8pt) withcolor ecolor;
  	draw (0xu,-0.15xu) .. controls (-0.2xu,-0.18xu) and (-0.2xu,-0.92xu) .. ( 0xu,-0.95xu) pensize (0.8pt) withcolor bcolor;

	draw (-1.05xu,-0.42xu)..(-1.05xu,-0.55xu) pensize(1pt);
	label(btex $z_3$ etex,(-1.05xu,-1.12xu));
	draw (-0.6xu,-0.48xu)..(-0.6xu,-0.61xu) pensize(1pt);
	label(btex $z_2$ etex,(-0.6xu,-1.18xu));
	draw (-0.15xu,-0.5xu)..(-0.15xu,-0.63xu) pensize(1pt);
	label(btex $z_1$ etex,(-0.15xu,-1.2xu));
	draw (0.3xu,-0.5xu)..(0.3xu,-0.63xu) pensize(1pt);
	label(btex $z_L$ etex,(0.3xu,-1.2xu));
	draw (0.75xu,-0.46xu)..(0.75xu,-0.59xu) pensize(1pt);
	label(btex $z_{L-1}$ etex,(0.98xu,-1.16xu));
        
	label(btex $\cdots$ etex rotated -20,(-1.45 xu,-0.95xu));
	label(btex $\cdots$ etex rotated 20,(1.5 xu,-0.95xu));

  	draw pic shifted (-8xu,0.0xu);

	draw savepic scaled 0.5 shifted (-3.8xu,2xu);

	label(btex $x_1$ etex,(-3.6xu,0.8xu));
	label(btex $x_{L+1}$ etex,(-3.42xu,3.2xu));

	label(btex $x_1\equiv x_{L+1}\to z_1$ etex,(-3.8xu,-0.78xu));
	label(btex $x_i=z_i/z_2$ etex,(-3.8xu,-1.28xu));
	drawarrow (-5.1xu,-1xu)--(-2.5xu,-1xu);

	label(btex $z_2\to 0$ etex,(-3.8xu,0xu));
	drawarrow (-2.5xu,-.22xu)--(-5.1xu,-.22xu);
\end{mpostfig}

\begin{mpostfig}[label=fundamentaldomain]
  picture pic;
  draw (-2xu,0xu) ..controls (-2xu,0.75xu) and (2xu,0.75xu) .. ( 2xu,0xu) dashed evenly pensize (0.8pt) withcolor ecolor;
  draw (0xu,-0.15xu) .. controls (0.2xu,-0.18xu) and (0.2xu,-0.92xu) .. ( 0xu,-0.95xu) dashed evenly pensize (0.8pt) withcolor bcolor;
  draw fullcircle xscaled 4xu yscaled 1.9xu pensize (1pt);
  draw (-1xu,0.03xu){dir -20}..(1xu,0.03xu) pensize (0.8pt);
  draw (-0.8xu,-0.02xu){dir 25}..(0.8xu,-0.02xu) pensize (0.8pt);
  draw (-2xu,0xu) ..controls (-2xu,-0.75xu) and (2xu,-0.75xu) .. ( 2xu,0xu) pensize (0.8pt) withcolor ecolor;
  draw (0xu,-0.15xu) .. controls (-0.2xu,-0.18xu) and (-0.2xu,-0.92xu) .. ( 0xu,-0.95xu) pensize (0.8pt) withcolor bcolor;
  pic:=currentpicture;
  currentpicture := nullpicture;
  pickup pencircle scaled 1pt;
  drawarrow (-0.15xu,0xu)--(5.2xu,0xu) withcolor black;
  drawarrow (0xu,-0.15xu)--(0xu,2.6xu) withcolor black;
  draw (0xu,0xu)--(3.9xu, 0xu) withcolor ecolor;
  draw (4.9 xu, 2 xu)--(1 xu, 2 xu) withcolor ecolor;
  draw (3.9 xu, 0 xu)--(4.9 xu, 2 xu) withcolor bcolor;
  draw (0xu, 0xu)--(1 xu, 2 xu) withcolor bcolor;
  label.bot (btex $0$ etex, (0,-.1xu));
  label.top (btex $\tau$ etex, (0.9xu,2xu));
  label.top (btex $\tau+1$ etex, (4.8xu,2xu));
  label.bot(btex $1$ etex, (3.9xu,-.1xu));
  label.lft(btex Im$(z)$ etex, (0xu,2.6xu));
  label.top(btex Re$(z)$ etex, (5.2xu,0xu));
  draw pic shifted (-3.8xu,0.8xu) scaled (1.2);

\end{mpostfig}

\begin{mpostfig}[label=genusonegeneral]
	draw fullcircle scaled 1.8xu pensize(1.3pt) withcolor ecolor;
	drawbdry(11,30);
	drawbdrydashed(30,36);
	drawbdrydashed(0,11);
	vlabel(27,1.2,btex $z_1=0\equiv 1$ etex);
	vlabel(23,1.2,btex $z_2$ etex);
	vlabel(19,1.2,btex $z_3$ etex);
	vlabel(15.5,1.2,btex $z_4$ etex);
	vlabel(11,1.2,btex $z_5$ etex);
	vlabel(30,1.2,btex $z_L$ etex);
	hpath:= subpath (0,11) of cpath scaled 1.15;
	draw ppath(hpath,0,0.9) dashed withdots scaled 0.7 pensize(1.3pt);
	hpath:= subpath (30,36) of cpath scaled 1.15;
	draw ppath(hpath,0.07,1) dashed withdots scaled 0.7 pensize(1.3pt);
\end{mpostfig}

\begin{mpostfig}[label=genusonez2to1]
	draw fullcircle scaled 1.8xu pensize(1.3pt) withcolor ecolor;
	drawbdry(11,30);
	drawbdrydashed(30,36);
	drawbdrydashed(0,11);
	vlabel(27,1.2,btex $z_1$ etex);
	vlabelcol(23, 1.2, faint, faint, btex $z_2$ etex);
	vlabelcol(25, 1.2, gcolor, gcolor, btex $z_2$ etex);
	vlabel(19,1.2,btex $z_3$ etex);
	vlabel(15.5,1.2,btex $z_4$ etex);
	vlabel(11,1.2,btex $z_5$ etex);
	vlabel(30,1.2,btex $z_L$ etex);
	hpath:= subpath (0,11) of cpath scaled 1.15;
	draw ppath(hpath,0,0.9) dashed withdots scaled 0.7 pensize(1.3pt);
	hpath:= subpath (30,36) of cpath scaled 1.15;
	draw ppath(hpath,0.07,1) dashed withdots scaled 0.7 pensize(1.3pt);
	hpath:= subpath (23,27) of cpath scaled 1.15;
	draw ppath(hpath,0.12,0.38) pensize(1.3pt) withcolor gcolor;
	drawarrow ppath(hpath,0.60,0.88) pensize(1.3pt) withcolor gcolor;
\end{mpostfig}

\begin{mpostfig}[label=genusonez2to0]
	draw fullcircle scaled 1.8xu pensize(1.3pt) withcolor ecolor;
	drawbdry(0,28);
	drawbdry(30,36);
	drawbdrydashed(28,30);
	vlabel(27,1.2,btex $z_1$ etex);
	vlabelcol(23, 1.2, faint, faint, btex $z_2$ etex);
	vlabelcol(33, 1.2, gcolor, gcolor, btex $z_2$ etex);
	vlabel(32,0.8,btex $z_3$ etex);
	vlabel(31,0.8,btex $z_4$ etex);
	vlabel(30,0.8,btex $z_5$ etex);
	vlabel(27.7,0.8,btex $z_L$ etex);
	hpath:= subpath (28,30) of cpath scaled 0.8;
	draw ppath(hpath,0.2,0.6) dashed withdots scaled 0.7 pensize(1.3pt);
	hpath:= subpath (23,0) of cpath scaled 1.15;
	draw ppath(hpath,0.02,1) pensize(1.3pt) withcolor gcolor;
	hpath:= subpath (36,33) of cpath scaled 1.15;
	draw ppath(hpath,0,0.8) pensize(1.3pt) withcolor gcolor;
	hpath:= subpath (33,27) of cpath scaled 1.15;
	drawarrow ppath(hpath,0.09,0.92) pensize(1.3pt) withcolor gcolor;
\end{mpostfig}

\begin{mpostfig}[label=momentumrelation]
	drawbdry(0,9);
	drawbdry(25,36);
	drawbdrydashed(9,25);
	vlabel( 8,1.24,btex $k_{L-1}^\mathrm{tree}$ etex);
	vlabel( 5,1.24,btex $k_L^\mathrm{tree}$ etex);
	vlabel( 2,1.24,btex $k_{L+1}^\mathrm{tree}$ etex);
	vlabel(35,1.24,btex $k_1^\mathrm{tree}$ etex);
	vlabel(32,1.24,btex $k_2^\mathrm{tree}$ etex);
	vlabel(29,1.24,btex $k_3^\mathrm{tree}$ etex);
	vlabel(26,1.20,btex $k_4^\mathrm{tree}$ etex);
	hpath:= subpath (8,26) of cpath scaled 1.2;
	draw ppath(hpath,0.07,0.93) dashed withdots scaled 0.7 pensize(1.3pt);
	label(btex $\mathbf{C}^\mathrm{E}_0$ etex,(0xu,0xu));
	pic:=currentpicture;
  	currentpicture := nullpicture;

	draw fullcircle scaled 1.8xu pensize(1.3pt) withcolor ecolor;
	drawbdry(0,7);
	drawbdry(29,36);
	drawbdrydashed(7,29);

	vlabel( 6,1.24,btex ~ etex);
	label(btex $\tilde{k}_{L-1}^\mathrm{1\mbox{-}loop}=k_{L-1}^\mathrm{tree}$ etex,(2.2xu,2xu));
	vlabel( 3,1.24,btex ~ etex);
	label(btex $\tilde{k}_L^\mathrm{1\mbox{-}loop}=k_L^\mathrm{tree}$ etex,(3.1xu,1.1xu));
	vlabel( 0,1.24,btex ~ etex);
	label(btex \begin{minipage}{3.2cm}$\tilde{k}_1^\mathrm{1\mbox{-}loop}=$ \\ \phantom{wi}$k_1^\mathrm{tree}+k_2^\mathrm{tree}+k_{L+1}^\mathrm{tree}$ \end{minipage} etex,(3.8xu,-0.1xu));
	vlabel(33,1.24,btex ~ etex);
	label(btex $\tilde{k}_3^\mathrm{1\mbox{-}loop}=k_3^\mathrm{tree}$ etex,(3.1xu,-1.1xu));
	vlabel(30,1.24,btex ~ etex);
	label(btex $\tilde{k}_4^\mathrm{1\mbox{-}loop}=k_4^\mathrm{tree}$ etex,(2.2xu,-2xu));

	hpath:= subpath (6,30) of cpath scaled 1.2;
	draw ppath(hpath,0.03,0.97) dashed withdots scaled 0.7 pensize(1.3pt);

        draw pic shifted (-6xu,0.0xu) scaled (1.0);
	label(btex $\mathbf{C}^\mathrm{E}_1$ etex,(0xu,0xu));
\end{mpostfig}